\newif\ifcqg\cqgtrue
\ifcqg
\documentclass[]{iopart}
\usepackage{iopams}
\usepackage{setstack}
\eqnobysec
\def\flamoi{\fl}
\else
\documentclass[aps,letter,prd,thightenlines,nofootinbib,showpacs,eqsecnum,amsfonts,amssymb,amsmath]{revtex4}
\usepackage{dcolumn}
\usepackage{bm}           

\def\flamoi{{}}

\fi

\usepackage[dvips,usenames]{color}
\usepackage{dsfont}
\ifcqg
\usepackage{hyperref}
\else
\fi
\usepackage{graphicx,epsfig,graphics,rotate}

\ifcqg
\newcommand{\abc}{\numparts}
\newcommand{\eabc}{\endnumparts}
\else
\newcommand{\abc}{{}}
\newcommand{\eabc}{{}}
\fi

\newcommand{\myref}[1]{Eq.~(\ref{#1})}

\newcommand{\be}{\begin{equation}}
\newcommand{\ee}{\end{equation}}
\newcommand{\bea}{\begin{eqnarray}}
\newcommand{\eea}{\end{eqnarray}}
\newcommand{\gr}[1]{{\mathbf{#1}}}

\newcommand{\demi}{\frac{1}{2}}

\newcommand{\vk}{{\bf k}}
\newcommand{\vx}{{\bf x}}

\newcommand{\ii}{\mathrm{i}}
\newcommand{\init}{\mathrm{init}}

\newcommand{\dirac}[1]{\delta_{D}^{#1}}
\newcommand{\Dirac}[1]{\delta_{D}^{#1}}
\newcommand{\para}{\parallel}

\def\st{\sigma_{\mathrm T}}

\def\dd{{\rm d}}
\def\HH{\mathcal{H}}

\definecolor{Myblue}{rgb}{0,0,1}
\definecolor{Myred}{rgb}{1,0,0}

\def\iB{{\color{Myred}I}}
\def\jB{{\color{Myred}J}}
\def\kB{{\color{Myred}K}}

\def\zT{{{\color{Myblue}o}}}
\def\iT{{{\color{Myblue}i}}}
\def\jT{{{\color{Myblue}j}}}

\def\ib{{\mathrm b}}
\def\ir{{\mathrm r}}
\def\ie{{\mathrm e}}
\def\ic{{\mathrm c}}
\def\im{{\mathrm m}}

\def\hashamoi{\#}

\newcommand{\troisj}[6]{\left(\begin{array}{ccc}
      #1 & #2 & #3 \\
      #4 & #5 & #6\end{array}\right)}

\newcommand{\eq}{{\rm eq}}

\newcommand{\lss}{{\rm LSS}}

\newcommand{\kPlmsn}[4]{{}^{#4}_{#3} {\uparrow \!\!\!K}^{#2}_{#1}}
\newcommand{\kMlmsn}[4]{{}^{#4}_{#3} {\downarrow \!\!\!K}^{#2}_{#1}}

\newcommand{\llmn}[3]{{}^{#3} \lambda^{#2}_{#1}\,}

\newcommand{\fracr}{f_{_{\rm r}}}

\newcommand{\sym}{{\hbox{sym.}}}
\newcommand{\fNL}{f_{_{\rm NL}}}
\newcommand{\nohat}{}

\newcommand{\dbi}{\partial_{\iB}}
\newcommand{\dbj}{\partial_{\jB}}
\newcommand{\dhi}{\partial^{\iB}}
\newcommand{\dhj}{\partial^{\jB}}

\newcommand{\etalss}{\eta_{\rm LSS}}

\ifcqg

\else\fi

\begin{document}
\title[The CMB bispectrum from the cosmological perturbations]{The cosmic microwave background bispectrum from the non-linear evolution of the cosmological perturbations}
\author{Cyril Pitrou}
\ifcqg
\ead{cyril.pitrou@port.ac.uk}
\address{Institute of Cosmology and Gravitation,
        Dennis Sciama Building, Burnaby Road, 
        Portsmouth, PO1 3FX, United Kingdom,}

\author{Jean-Philippe Uzan}
\ead{uzan@iap.fr}
\address{Institut d'Astrophysique de Paris, UMR7095 CNRS,
         Universit\'e Pierre~\&~Marie Curie - Paris,
         98 bis bd Arago, 75014 Paris, France,}

\author{Francis Bernardeau}
\ead{francis.bernardeau@cea.fr}
\address{CEA, IPhT, 91191 Gif-sur-Yvette c{\'e}dex, France,\\
         CNRS, URA-2306, 91191 Gif-sur-Yvette c{\'e}dex, France.}
\else

\author{Cyril Pitrou}
\email{cyril.pitrou@port.ac.uk}
\affiliation{Institute of Cosmology and Gravitation,
        Dennis Sciama Building, Burnaby Road, 
        Portsmouth, PO1 3FX, United Kingdom,}

\author{Jean-Philippe Uzan}
\email{uzan@iap.fr}
\affiliation{Institut d'Astrophysique de Paris, UMR7095 CNRS,
         Universit\'e Pierre~\&~Marie Curie - Paris,
         98 bis bd Arago, 75014 Paris, France,}

\author{Francis Bernardeau}
\email{francis.bernardeau@cea.fr}
\affiliation{CEA, IPhT, 91191 Gif-sur-Yvette c{\'e}dex, France,\\
         CNRS, URA-2306, 91191 Gif-sur-Yvette c{\'e}dex, France.}

\fi

\date{March 3, 2010}
\begin{abstract}
This article presents the first computation of the complete bispectrum of the cosmic microwave background temperature anisotropies arising from the evolution of all cosmic fluids up to second order, including neutrinos. Gravitational couplings, electron density fluctuations and the second order Boltzmann equation are fully taken into account. Comparison to limiting cases that appeared previously in the literature are provided. These are regimes for which analytical insights can be given. The final results are expressed in terms of equivalent $\fNL$ for different configurations. It is found that for moments up to $\ell_{\rm max}=2000$, the signal generated by non-linear effects is equivalent to $\fNL\simeq 5$ for both local-type and equilateral-type primordial non-Gaussianity.
%
\end{abstract}


\ifcqg
\else
\maketitle
\fi

\section{Introduction}

The Cosmic Microwave Background (CMB) anisotropies are now observed with a high
precision and have become a key observation of modern cosmology. They are in particular very precious
to constrain the theories of the primordial Universe~\cite{WMAP7}. So far the temperature anisotropies have been 
found to have statistical properties that are compatible with Gaussian statistics~\cite{WMAP7}. The CMB data can therefore entirely be captured in its
power spectrum and the latter has been used to set constraints on the cosmological parameters and on the shape
of the inflationary potential. There is hope however that future observations such
as Planck~\cite{PLANCK}, that will provide better data on the statistical properties of the temperature and polarization fields, open a new window on the physics of the early Universe with the use of higher-order statistical properties of the CMB sky such as its bispectrum.

In the analysis of those data, one should keep in mind though that the properties of the CMB temperature and polarization anisotropies depend both on the properties of the initial conditions and on their evolution. As long as measurements are restricted to second-order statistics such as the angular correlation function or the power spectra, a linear perturbation theory suffices, for the required precision, to relate the (2-dimensional) angular power spectrum to the (3-dimensional) initial power spectrum of the metric
perturbations at the end of inflation. It has thus been understood that
the characteristic features observed in the CMB temperature spectrum originate from the
developments of acoustic oscillations encoded in the linear transfer~\cite{Hu1994b,Hu1995}
function while its overall amplitude and its scale
dependence are fixed by the initial power spectrum, the shape
of which agrees with the predictions from an inflationary era~\cite{Mukhanov1992,Linde2007}.
At this level of description, all aspects are now fully understood and is part of textbooks~\cite{2003moco.book.....D,2005pfc..book.....M,FrancisLivre,PeterUzanTrans,2008cmb..book.....D,2008cosm.book.....W} 
and since the metric and matter perturbations are linearized, any model that predicts
Gaussian initial conditions, as standard single field inflation, is expected to
produce Gaussian statistical properties for the CMB temperature
field.

In general bispectra arise whenever non-linear mode couplings are at play during the cosmological evolution and 
general relativity being in essence a non-linear field theory, deviations from
Gaussianity are expected to be ubiquitous, arising either from the inflationary era
(and thus called primordial non-Gaussianity) or from the post-inflationary
evolution. Gravity mediated couplings are however generally small and in full agreement with the current data 
that clearly favors only mild non-Gaussianities if any~\cite{WMAP7}. In particular, it is now widely accepted that standard single field inflation cannot
produce significant non-Gaussianities since its amplitude is mostly
dictated by gravity induced couplings~\cite{Maldacena2003}. On the other hand, significant
deviations from Gaussianity can arise from non-standard kinetic terms, for which models based on the Dirac-Born-Infeld action are typical models \cite{2004PhRvD..70l3505A} or in the context of 
of multiple-field inflation specially when non-gravity type couplings are at play~\cite{Bernardeau2002,Komatsu2002,Bartolo2002,Bernardeau2003,Bernardeau2006,Rigopoulos2006,2008PThPh.120..159S,2009PhRvL.103g1301B}.
It is therefore generally admitted that non-Gaussianity searches can open a
window on the details of the inflationary mechanism at work in the
early Universe (see Refs.~\cite{Bartolo2004,2009astro2010S.158K,Bartolo2010revue} for general reviews).

The level with which primordial non-Gaussianity could actually be detected is however still largely debated,
mainly because this source of non-Gaussianity is in ``competition'' with the couplings induced by non-linear effects throughout the whole recombination and photon propagation processes.
This has motivated a
series of general studies aiming at characterizing the bispectrum~\cite{Bartolo2004b,Pyne1996,Mollerach1997,Goldberg1999,Komatsu2001,Babich2004,Bartolo2004a,Komatsu2005,Liguori2006,Bartolo2008,Khatri2008,Senatore2008b,Nitta2009,Hanson2009,Boubekeur2009,Mangilli2009,Liguori2010}  and even the trispectrum~\cite{Komatsu2002,Okamoto2002,Kogo2006} to be expected from
the observation made by CMB experiments. While the identification of the mode
couplings (at the quantum level) during the inflationary phase has been set on
secure grounds~\cite{Maldacena2003,Bernardeau2004a,Weinberg2005,Weinberg2006,2008PhRvD..78f3534W}, the evolution of the cosmological perturbations during the post-inflationary era
is still largely unexplored, the primordial statistical properties being often related to the observed statistical properties 
only through a linear transfer function. However, in order to relate the 
angular bispectrum of the observed CMB to the spatial bispectrum of the metric perturbations at the end of inflation, 
one needs to derive transfer functions up to second order that incorporate all type of couplings.

The work we present in this article is the end result of a task which was initiated in Ref.~\cite{PUB2008}, that was 
followed by some partial reanalysis~\cite{Bartolo2008,Senatore2008a,Khatri:2008kb}, In this article we described what we thought is the main mechanism at play for the generation of the bispectra at small angular scales and for which
we could give some physical insights.
The present article extends this analysis using the theoretical developments of the non-linear 
perturbation theory described in Refs.~\cite{Nakamura2007,Pitrou2008,Pitrou2008letter}, and in particular
concerning the second order Boltzmann equation that we need in order to describe the evolution of radiation and neutrinos. Its goal is three-fold: 
(i) to support the predictions and analytic understanding obtained on small scales in Ref.~\cite{PUB2008} by a full numerical integration of the second order Boltzmann equation, {\em without neglecting any term}, 
(ii) to compute the expected bispectrum on intermediate scales, and 
(iii) to revisit the large-scale behaviour of the bispectrum which has been presented in the previous literature~\cite{Bartolo2004a,Boubekeur2009,Bartolo2005}.
We emphasize that the amplitude of this non-Gaussianity is completely fixed once the amplitude and
power spectrum of the first order scalar perturbations are constrained so that this offers a definite
prediction of the minimal amount of non-Gaussianity expected in any CMB observation, that is provided
Einstein theory remains a good description of gravity.

Let us mention that our present analysis refines our previous descriptions by classifying the evolution effects, linear  or not, 
in two categories. 
First, the  \emph{primary} or \emph{early} effects arise from the evolution of the perturbations in the radiation dominated era after horizon crossing, during recombination between protons and electrons and until the potentials are constant in their linear evolution. Then  \emph{secondary} or \emph{late} effects arise at late time when the potential starts to evolve due the late time acceleration  of the Universe or due to reionisation effects (see Ref.~\cite{2008RPPh...71f6902A} and references therein). This separation is unambiguous for the linear evolution and is therefore pertaining to the power spectrum computation but also to the calculation of the bispectrum that mix linear and second order source terms. This is actually clear from Fig. \ref{FigSources1} below, and it is summarized in table~\ref{Tableclassement}.
\begin{table}
\begin{tabular}{|l|l|l|}
\hline
Effect & {\bf Linear evolution} &  {\bf Non-linear evolution} \\
\hline
Primordial & Primordial power spectrum $P(k)$ &  Primordial $\fNL$ \\
\hline
Primary (early) & Sachs-Wolfe and Doppler effects &  This article \\
\hline
Secondary (late) & Integrated Sachs-Wolfe effect, &  Lensing-ISW correlation\\
&reionization& \\
\hline
\end{tabular}
\caption{Classification of the linear and non-linear effects.
\label{Tableclassement}}
\end{table}
This present work focuses on the non-linear evolution of the field during the early period and the resulting bispectra it induces. In particular we compute the shape, amplitude and bispectrum of the temperature anisotropies\footnote{Note that these results actually depend on the actual definition of the temperature one uses (see the discussion of \S~\ref{subsec30}). This is due to the spectral distortion that second order effects necessarily induce as described in Ref.~\cite{Pitrou2009ysky}.} due to these effects and eventually compare them to secondary sources, and specifically the ISW-lensing effects, and to primordial coupling effects (through standard parameterizations). In this work the  non-linear evolution of the fluids, including neutrinos, is therefore treated exactly up to second-order in the linear perturbations until
reionization is complete. That includes of course the use of the second order Boltzmann equation. Adiabatic initial conditions are also assumed. All the results described here can actually be reproduced from a {\it Mathematica} code which is freely available -- with its documentation -- on the webpage~\cite{CMBquick}.

The article is organized as follows. Section~2 describes the main concepts of the non-linear
cosmological perturbation theory while all equations are gathered in~\ref{app_equations}.
Then, in Section~3 we determine the initial conditions for both the metric variables
and cosmic fluids (baryon, cold dark matter, photons, neutrinos). In Section~4
we describe, after a careful definition of the temperature in \S~\ref{subsec30},  the numerical integration based on the flat sky
approximation presented in Section~3. The numerical results concerning
the bispectrum are discussed in Section~5 while in Section~6 we provide an
analytical understanding of these results in various limiting cases.

%

\section{Non-linear Perturbation Theory}

We follow Ref.~\cite{Pitrou2008} for the notation. This section summarizes our main choices 
and conventions and we refer the reader to the latter reference for any further details.
We first detail the description of the metric in \S~\ref{subsec11}, of the matter fields
in \S~\ref{subsec12}  and then describe the structure of the perturbation equations
in \S~\ref{sec_struct_equation}.

\subsection{Metric perturbations}\label{subsec11}

At lowest order, we assume that the Universe is well described by a Friedmann-Lema\^{\i}tre (FL) space-time 
with Euclidean spatial sections and with scale factor $a$.
The Universe is then described by a perturbed space-time around this 
FL background. In the Newtonian gauge (often also named Poisson gauge), the form of the perturbed metric is then given by
\begin{eqnarray}\label{metric}
\flamoi \dd s^2 &=& g_{\mu\nu}\dd x^\mu\dd x^\nu \\*
\flamoi &=& a(\eta)^2 \big\{-(1 + 2\Phi )\dd\eta^2 + 2
 B_{\iB} \dd x^{\iB}\dd\eta + [(1-2 \Psi)\delta_{\iB\jB} + 2 H_{\iB\jB}]\dd x^{\iB}\dd x^{\jB}\big\},\nonumber
\end{eqnarray}
with $\partial^{\iB} B_{\iB}=\partial^{\iB} H_{\iB \jB}=H^{\iB}_{\phantom{\iB}\iB}=0$. 
Here, indices of the type $\iB,\jB,\kB\dots$ are spatial coordinates indices running from $1$ to $3$, and $\eta$ is the conformal time. 
We also define the comoving Hubble factor $\HH=a'/a$ where a prime denotes a derivative with respect to $\eta$.
Each of the perturbation variables has an order by order expansion of the type~\cite{Bruni1997}
\begin{equation}\label{decomposition-ordre2}
 X=\bar{X}+ X^{(1)}+\frac{1}{2}X^{(2)}\,.
\end{equation}
In the case of metric perturbations, the definition is made such that their background value vanishes. 
Additionally, since vector modes fail to be generated at first order in the
standard models of inflation~\cite{Mukhanov1992,FrancisLivre,PeterUzanTrans}, we only consider the second order vector modes $B_\iB^{(2)}$.
Similarly, the amplitude of first order tensor modes (i.e. gravitational waves) is expected to be sufficiently 
low~\cite{WMAP5} so that we can safely neglect $H_{\iB \jB}^{(1)}$, and we will consider only $H_{\iB \jB}^{(2)}$.

\subsection{Describing the matter content}\label{subsec12}

In the concordance model of cosmology, the matter content of the Universe includes relativistic particles (radiation (photons), neutrinos) and non relativistic particules, which fall either in the category of baryons or cold dark matter.
Once the equation of state for each species is specified, the background energy density is sufficient to characterize its state.
They are denoted $\bar \rho_\nu$, $\bar \rho_\ir$, $\bar \rho_\ic$ and $\bar \rho_\ib$ for neutrinos, radiation, cold dark matter and baryons, respectively. 
Usually the (perfect) fluids contained in the Universe are characterized in the perturbed Universe by their energy density but also by their velocity.
However this is not possible for relativistic particules and, in order to describe photons and neutrinos, we need to rely 
on a kinetic description based on the moments of their {\em distribution function}. We will thus describe the fluids as a special case of this statistical description when only the two first moments, which are related to the energy density and the velocity~\cite{Pitrou2008}, are needed.

\subsubsection{The moments of species}

In general the moments of the distribution function are defined according to the spatial part of a local Minkowski space-time which itself is defined at any point of the space-time manifold by the value of a tetrad field. They are also taken with respect to an azimuthal direction that we arbitrarily
align with a given Fourier mode when working out their spatial dependence in Fourier space.
We thus describe the radiation with the brightness moments ${\cal I}_{\ell}^{m(1)}(\gr{k},\eta)$ and ${\cal I}_{\ell}^{m(2)}(\gr{k},\eta)$, but also with the moments describing its polarization which are separated into the electric-type moments $\nohat{\cal E}_{\ell}^{m(1)}(\gr{k},\eta)$ and $\nohat{\cal E}_{\ell}^{m(2)}(\gr{k},\eta)$, and the magnetic type moments $\nohat{\cal B}_{\ell}^{m(2)}(\gr{k},\eta)$ (which appear only at second order since we have assumed that the first order vector and tensor modes can be neglected).
The polarization of radiation is generated through the interactions with baryons while the neutrinos, being collisionless,
develop no polarization. As a consequence, we only need the brightness moments ${\cal N}_{\ell}^{m(1)}(\gr{k},\eta)$ and ${\cal N}_{\ell}^{m(2)}(\gr{k},\eta)$ to describe them (we also assume that the three species of neutrinos are massless and
we describe them altogether though their mass splitting can have some interesting cosmological 
implications~\cite{Lesgourgues2006,Tereno:2008mm}).

The matter species, i.e. baryons and cold dark matter, are described  by perfect fluids. The perturbation of their energy densities is defined according to the decomposition~(\ref{decomposition-ordre2}). The spatial moments of the velocities of these fluids are noted $v_{m}$ with $m=-1,0,1$ and their perturbation is also defined according to Eq.~(\ref{decomposition-ordre2}) with no background value thanks to the symmetries of the background space-time, which imply that all fluids have the four-velocity $\bar u^\mu = \frac{1}{a} \delta^{\mu}_{0}$ since no preferred direction exists.
Furthermore, since the first order vector modes are not generated at first order, we have $v_{\pm 1}^{(1)}=0$.

Additionaly, it will prove more convenient to work also with the moments of the vector and tensor perturbations, that
is with $\Phi_{\pm 1}^{(2)}$ and $H_{\pm 1}^{(2)}$ instead of $B_I$ and $H_{IJ}$. 
Again, we refer to Ref.~\cite{Pitrou2008} for the exact definitions which are inspired from Ref.~\cite{Hu1997}.

\subsubsection{Stress-energy tensor}

For a perfect fluid, the stress-energy tensor is given by
\begin{equation}
 T_{\mu\nu} = \rho u_\mu u_\nu + P\left( g_{\mu\nu} + u_\mu u_\nu\right),
\end{equation}
where the pressure is related to the energy density by an equation of state $P= w \rho$. We can see on this expression of the stress-energy tensor that we neglect the anisotropic stress and it thus applies for baryons and cold dark matter only. For radiation and neutrinos, the stress-energy tensor can be constructed from the distribution function and it is thus possible to relate its expression to the moments of the brightness (see Ref.~\cite{Pitrou2008} for instance). 

\subsection{Field equations}\label{sec_struct_equation}

\subsubsection{Structure of the equations}

Three equations are essential to understand the evolution of perturbations of the radiation. First, we need the Einstein equation $G_{\mu \nu}= \kappa T_{\mu\nu}$ with $\kappa=8 \pi G$, in order to relate the perturbations of the metric to the ones of the different matter species.
Then, the evolution of the distribution functions is dictated by the Boltzmann equation from which one can extract the evolution hierarchy of its moments.
Finally, since the collision rate between photons and baryons is proportional to the fraction of free electrons, $x_{\rm e}$, which is governing the recombination process, we need to derive the evolution equation of $x_{\rm e}$ up to the first order in the perturbations. 
These three sets of equations are detailed in~\ref{app_equations}.

At linear order, the evolution equations are partial differential equations which are 
by construction linear in the perturbation variables. They formally take the form
\begin{equation}
{\cal D}_i[X^{(1)}_j]=0\,,
\end{equation}
where the $X^{(1)}_j$ is the set of first order perturbation variables and ${\cal D}_i$ the set of differential operators acting on them.
At second order the purely second order perturbation variables satisfy the same differential equations, but supplemented with source terms which are quadratic in the first order variables, i.e. they are of the form
\begin{equation}
{\cal D}_i[X^{(2)}_j]={\cal S}_i[X^{(1)}_j],
\end{equation}
where the terms ${\cal S}_i$ are operators (possibly differential) quadratic
in their arguments. It is thus sufficient to write down the second order
equations, that is the linear operators ${\cal D}_j$ and the source terms
${\cal S}_i$, in order to fully specify both the first and the second order equations. 

\subsubsection{Structure of the solutions}

At first order, the space of solutions is spanned by the linear combinations of
independent solutions since the operators ${\cal D}_i$ are linear. Then, the
choice of the initial conditions selects a particular solution in this vector space of
solutions. However, at second order, the space of solutions is spanned by the
linear combinations of the same independent solutions to which a particular
solution has to be added, thanks to the non vanishing source terms ${\cal S}_i$ 
which break the linearity, so that it belongs to
an affine space instead of a vector space. The choice of this particular solution is
arbitrary so that we can decide to take the one which vanishes deep in the
radiation dominated era. This particular solution will then be physically
interpreted as the cumulative effect of the sources and this is actually what we mean
when we compare the effects of non-linear evolution to those of the linear
evolution, the latter being the solutions of the homogeneous equation. 
As explained in the introduction, the goal of this article is to assess the importance of the non-linear evolution effects on the temperature bispectrum with respect to the linear evolution effects which are now well under control from a theoretical and numerical point of view.

\subsubsection{Fourier space and convention}

From a technical point of view, we will solve numerically the evolution equations in Fourier space.
We define the Fourier transform of any function $X(\vx,\eta)$ as
\begin{equation}
 X(\vk,\eta) = \int \frac{\dd^3\vx}{(2 \pi)^{3/2}}X(\vx,\eta)\hbox{e}^{-\ii \vk\cdot\vx}
\end{equation}
and we  will use the expression of the Fourier transform of a product as the convolution
\begin{equation}\label{Convolution_Notation}
[X Y](\vk)={\cal K}\left\{X(\vk_1)Y(\vk_2)\right\}\,,
\end{equation}
hence introducing the short hand notation
\begin{equation}\label{Convolution}
 \mathcal{K}\{\dots\} \equiv \int \frac{\dd^3\gr{k}_1
 \dd^3\gr{k}_2}{(2 \pi)^{3/2}}\, \Dirac{3}(\gr{k}_1+\gr{k}_2-\gr{k})\dots\,,
\end{equation}
where $\Dirac{3}$ is the 3-dimensional Dirac distribution. 

\section{Initial conditions}\label{sec3}

Since both radiation and neutrinos are relativistic, the ratio of their energy densities is conserved
during the cosmological evolution, and we define the fraction of radiation among the relativistic species by 
$\fracr \equiv \bar \rho_\ir/(\bar \rho_\ir+\bar \rho_\nu)$. 
Similarly, cold dark matter and baryons being pressureless and coupled to
gravity universally~\cite{Coc:2008yu}, the ratio of their energy densities is also conserved.
We also define
\begin{equation}\label{e.defR}
 R=\frac{3}{4}\frac{\bar \rho_\ib}{\bar\rho_\ir}\,.
\end{equation}
In this section focusing on the initial conditions, all expressions are meant to hold only at
an initial time $\eta_\init$. We thus omit to specify it in order to simplify the notation.
We now give the initial conditions for the matter perturbation variable in \S~\ref{subsec21}
and the geometry perturbation variables in \S~\ref{subsec22}, both at first and
second orders.

\subsection{Initial conditions for the matter variables}\label{subsec21} 

We only consider in this article adiabatic initial conditions since isocurvature modes
are highly constrained observationnally~\cite{Bucher:2000cd} but also because
the single field models of inflation~\cite{FrancisLivre,PeterUzanTrans} predict
adiabatic initial conditions. The link between isocurvature modes and
multiple-field models remains to be investigated in depth~\cite{Bernardeau2003,Linde1997}.

For adiabatic initial conditions, the velocity of all matter species is negligible both at the first 
and at second orders. It is indeed suppressed by a factor $k \eta_\init$ compared to $\Phi_{\rm init}$.
We thus take the initial condition 
\begin{equation}
v_m^{(1)} = v_m^{(2)} = 0,\quad \hbox{with}\quad m=-1,0,1
\end{equation} 
for baryons and cold dark matter and we take
\begin{equation}
{\cal I}_{1}^{m(1)} = {\cal N}_{1}^{m(1)} ={\cal I}_{1}^{m(2)} = {\cal N}_{1}^{m(2)} =0,
\end{equation}
for radiation and neutrinos. Adiabaticity also implies that the ratios between $\rho_\ib^{1/3}$, $\rho_\ic^{1/3}$, $\rho_\ir^{1/4}$ and $\rho_\nu^{1/4}$ remain constant. We deduce the following initial conditions\footnote{${\cal I}_0^0$ is not exactly the energy density since the frame used to define the multipoles is not necessarily comoving with the species. However the difference are quadratic in the velocity field, mainly because it arises from boost factors, and it is thus not important for the initial conditions. See Refs.~\cite{Pitrou2008,Pitrou2007} for the link between these two quantities.}
\abc
\begin{equation}\label{CI1abiab}
\frac{1}{3}\frac{\rho_\ic^{(1)}}{ \bar \rho_\ic}=\frac{1}{3}\frac{\rho_\ib^{(1)}}{\bar \rho_\ib}=\frac{1}{4}{\cal I}_{0}^{0(1)}\,\,,
\end{equation}
\begin{equation}\label{CI2abiab}
\frac{1}{3}\frac{\rho_\ic^{(2)}}{\bar \rho_\ic}=\frac{1}{3}\frac{\rho_\ib^{(2)}}{\bar \rho_\ib}=\frac{1}{4}{\cal I}_{0}^{0(2)}-\left[\frac{1}{4}{\cal I}_{0}^{0(1)}\right]^2\,,
\end{equation}
\begin{eqnarray}
{\cal I}_{0}^{0(1)}(k)&=&{\cal N}_{0}^{0(1)}(k)\,,\\
{\cal I}_{0}^{0(2)}(k)&=&{\cal N}_{0}^{0(2)}(k)\,.
\end{eqnarray}
\eabc
Now, using the Poisson equation~(\ref{eq00}) deep in the radiation era, we can relate these initial conditions to those of the metric perturbations as\abc
\begin{eqnarray}
{\cal I}_0^{0(1)}&=& -2 \Phi^{(1)}\,,\\
{\cal I}_0^{0(2)}&=& -2 \Phi^{(2)}+ 8 \Phi^{(1)2}\label{e35b}\,.
\end{eqnarray}
\eabc
Additionally, the presence of neutrinos implies that $\frac{\HH^2}{k^2} {\cal N}_2^{0}$\footnote{Note that the quadrupole moments, that is the moments with $\ell=2$ (${\cal N}^m_2$ and ${\cal I}^m_2$) are different from the anisotropic stress of neutrinos and photons~\cite{Pitrou2007}. In fact they match only if the frame in which the moments are taken is aligned with the velocity of the species considered.} does not vanish and this term is required in order to determine the initial conditions for the perturbations of the two gravitational
potentials since $\Phi-\Psi$ is proportional to ${\cal N}_2^{0}+{\cal I}_2^{0}$; 
see Eq.~(\ref{eqijst}). \ref{App_quadrupoleconditions} details the initial conditions satisfied by the 
quadrupoles ${\cal I}_{2}^{0(2)}$, ${\cal N}_{2}^{0(1)}$ and ${\cal N}_{2}^{0(2)}$ while we always have ${\cal I}_{2}^{0(1)}=0$.

\subsection{Initial conditions for the metric variables}\label{subsec22} 

As previously, we fix the initial conditions deep in the radiation dominated era for super-Hubble modes,
so that we can expand all our equations in terms of $k \eta_\init \ll 1$. For adiabatic perturbations,
we recall that on super-Hubble scales the comoving curvature perturbation is conserved. Its expressions at first and second orders are~\cite{Malik2004,Vernizzi2005}
\abc
\begin{equation}
\flamoi \mathcal{R}^{(1)} =\Psi^{(1)} +\frac{2}{3(1+w)\HH}\left[\Psi'^{(1)} + \HH\Phi^{(1)} \right]\,,
\end{equation} 
\begin{eqnarray}
\flamoi \mathcal{R}^{(2)}&=&\Psi^{(2)} +\frac{2}{3(1+w)\HH}\left[\Psi'^{(2)} +
  \HH\Phi^{(2)} - 4 \HH \Phi^{(1)2} -  \frac{\Psi^{(1)'2}}{\HH}-4(\Phi^{(1)}-\Psi^{(1)})\Psi^{(1)'}\right]\nonumber\\
\flamoi &&+ \left(1 + 3 c_s^2\right)\left[\frac{\rho^{(1)}}{ 3(1+w)\bar \rho}\right]^2
+\frac{4}{3(1+w)}\frac{\rho^{(1)}}{\bar \rho} \Psi^{(1)}\ ,
\label{Rsecondorder}
\end{eqnarray} 
\eabc
where $w$ and $c_s^2$ are the equation of state and the adiabatic speed of sound of the total fluid.
Deep in the radiation era the properties of the total fluid are close to the  radiation+neutrinos fluid since $\bar \rho_\im \ll \bar \rho_\ir$ and $w=c_s^2=1/3$. 
At first order, we obtain the initial conditions~\cite{Ma1995}
\abc
\begin{equation}
 \left[1+\frac{4}{15}\left(1-\fracr \right)\right]\Phi^{(1)} =\frac{2}{3}
 \mathcal{R}^{(1)}\ ,
\end{equation}
\begin{equation}
\Psi^{(1)}-\Phi^{(1)}=\frac{2}{5}(1-\fracr)\Phi^{(1)}\,,
\end{equation}
which in particular implies $\mathcal{R}^{(1)}=\Psi^{(1)}+\Phi^{(1)}/2$.
\eabc
At second order, we obtain
\abc
\begin{equation}
\Psi^{(2)} =\frac{2}{3}\left(\frac{3}{2}\Phi^{(1)2} +2\Phi^{(1)}\Psi^{(1)} \right)+\frac{1}{3}S_3+\frac{2}{3}{\cal R}^{(2)}\,,
\end{equation}
\begin{equation}
\Phi^{(2)} =\frac{2}{3}\left(\frac{3}{2}\Phi^{(1)2} +2\Phi^{(1)}\Psi^{(1)} \right)-\frac{2}{3}S_3+\frac{2}{3}{\cal R}^{(2)}\,\
\end{equation}
\eabc
where
\begin{equation}\label{EqS3vsS3tilde}
\flamoi S_3=\tilde S_3- \frac{3}{5}{ \HH^2}\Delta^{-1}\left[\fracr {\cal I}_2^{0(2)}+(1-\fracr){\cal N}_2^{0(2)} \right]\,,
\end{equation}
 
\begin{eqnarray}
\flamoi \tilde S_{3} &=& -4 \Psi^{(1)2}-2 \Phi^{(1)2}\nonumber \\
\flamoi&&-\Delta^{-1}\left[ \dbi (\Psi^{(1)}+\Psi^{(1)}) \dhi (\Psi^{(1)}+\Phi^{(1)}) +2\Psi^{(1)}\Delta \Phi^{(1)} \right]\nonumber \\
\flamoi   && + 3(\Delta \Delta)^{-1}\dbi \dbj\left[ \dhi (\Psi^{(1)}+\Phi^{(1)}) \dhj (\Psi^{(1)}+\Phi^{(1)})+2 \Psi^{(1)} \dbi \dbj \Phi^{(1)} \right]\,.
\end{eqnarray}
It can also be recast in the following set of two initial conditions
\abc
\begin{equation}
\left[1+\left(1-\fracr \right)\frac{4}{15}\right]\Phi^{(2)} =\frac{2}{3}\left\{ \frac{3}{2}\Phi^2 +2\Phi\Psi+ \mathcal{R}^{(2)}-S_3  \right\}\,,\label{Phi2functionofR}
\end{equation}
\begin{equation}
\Psi^{(2)}-\Phi^{(2)}=S_3\,.
\end{equation}
\eabc
In order to make link with previous works~\cite{Bartolo2004b}, note that in the case where there are no neutrinos ($\fracr =1$) we obtain
\abc
\begin{eqnarray}
\flamoi \Phi^{(1)} &=&\Psi^{(1)}=\frac{2}{3} \mathcal{R}^{(1)}\,,\\
\flamoi \Psi^{(2)} &=& \frac{2}{3} \mathcal{R}^{(2)}
+  \Psi^{(1)2} - \Delta^{-1}\left[ \dbi \Psi^{(1)} \dhi
  \Psi^{(1)} \right] +  3 \Delta^{-2}\dhj \dbi
   \left[ \dbj \Psi^{(1)}\dhi \Psi^{(1)} \right]\,,\label{psiI}\\
\flamoi \Phi^{(2)}&=&\Psi^{(2)} + 4 \Psi^{(1)2} +
3 \Delta^{-1}\left[\dbi \Psi^{(1)} \dhi
\Psi^{(1)}\right]
 - 9 \Delta^{-2}\dhj \dbi \left[\dbj
\Psi^{(1)} \dhi \Psi^{(1)}\right]\
.\label{Eq_lienPhiPsi_superHubble}
\end{eqnarray}
\eabc

For the vector and tensor modes, we have argued that they are respectively purely decaying
modes or of negligible amplitude at first order, so that
\begin{equation}
 \Phi_{\pm1}^{(1)} = H_{\pm1}^{(1)} = 0.
\end{equation}
At second order, vector and tensor type perturbations exclusively arise from the integrated contribution 
of the quadratic sources so that they also vanish at sufficiently early times. We thus have
\begin{equation}
 \Phi_{\pm1}^{(2)} = H_{\pm1}^{(2)} = 0.
\end{equation}

\subsection{Primordial non-Gaussianity and $\fNL$}

In the previous section, we have not specified the value of $\mathcal{R}^{(2)}$.
Indeed this is what we would like to constrain by measuring the level of non-Gaussianity since different models of inflation 
have different predictions.
In general, theoretical predictions are given in terms of the spectrum and of the bispectrum of ${\cal R}$ related
to the 2- and 3-point functions by
\begin{eqnarray}
\langle {\cal R}(\vk_{1}){\cal R}(\vk_{2})\rangle &=&\langle {\cal R}^{(1)}(\vk_{1}){\cal R}^{(1)}(\vk_{2})\rangle
=\Dirac{3} \left(\vk_{1}+\vk_{2}\right) P_{\cal R}(k_1)
 \,,
\end{eqnarray}
and
\begin{eqnarray}
\langle {\cal R}(\vk_{1}){\cal R}(\vk_{2}){\cal R}(\vk_{3})\rangle &=&\demi \langle {\cal R}^{(1)}(\vk_{1}){\cal R}^{(1)}(\vk_{2}){\cal R}^{(2)}(\vk_{3})\rangle+\sym\nonumber\\
&=&\Dirac{3} \left(\vk_{1}+\vk_{2}+\vk_{3}\right)\
\fNL^{{\cal R}}\ F(\vk_{1},\vk_{2},\vk_{3}) \,,
\end{eqnarray}
where $P_{\cal R}(k)$ is the primordial power spectrum and
$F(\vk_{1},\vk_{2},\vk_{3})$ is a function which specifies the shape of the
primordial non-Gaussianity and $\fNL^{{\cal R}}$ is its amplitude. Each possible type of primordial non-Gaussianity is characterized by a function $F(\vk_{1},\vk_{2},\vk_{3})$, and these functions are normalized such that they all have the same value in the Fourier configuration $k_1=k_2=k_3$.

Different limiting cases can be found in the literature~\cite{Creminelli2006}, the main one being the non-Gaussianity of the \emph{local type} for which 
\begin{equation}
F_{\rm loc}(\vk_{1},\vk_{2},\vk_{3})=2 P_{{\cal R}}(k_{1})P_{{\cal R}}(k_{2})+ 2\,\sym
\end{equation}
In this case, it is equivalent to specify that 
\begin{equation}
\mathcal{R}^{(2)}(\vk)= 2 {\cal K}\left\{  \fNL^{{\cal R}} {\cal R}^{(1)}(\vk_1) {\cal R}^{(1)}(\vk_2)\right\}
\end{equation} 
in Fourier space.
In particular, it has been shown~\cite{Maldacena2003} that for slow-rolling single field inflationary models
\begin{equation}
\mathcal{R}^{(2)} \simeq -2\mathcal{R}^{(1)2}\,,
\end{equation}
that is $\fNL^{\cal R} \simeq -1$. Note that this is equivalent to impose that $\fNL^{{\cal R}_M} \simeq 0$
with $\exp(-2{\cal R}_M)\equiv 1- 2{\cal R}$~\cite{Maldacena2003}.
This is the initial condition that we shall assume in this
work since we want to study the effects of evolution in the pursuit of non-Gaussianity and want to consider the minimum amount of primordial non-Gaussianity (see however the discussion of \S~\ref{subsec340} below). 

Note also that most expressions are given in terms of the power-spectrum of the gravitational potential $\Phi^{(1)}$ rather than ${\cal R}^{(1)}$. The two power spectra are easily related thanks to
\begin{equation}
P_\Phi(k)=\frac{4}{9}\frac{1}{\left[1+ \frac{4}{15}\left(1-\fracr \right)\right]^2}P_{\cal R}(k)\,.
\end{equation}
Similarly, the primordial non-Gaussianity is more often expressed in terms of a $\fNL$ parameter which is related to $\Phi$ rather than ${\cal R}$. In order to be consistent with previous literature in which neutrinos were not included at second order, we define $\fNL^{\Phi}$ through
\be\label{deffnlphi}
\flamoi \Phi^{(2)}(\vk)=-2{\cal K} \left\{\fNL^{\Phi}  \frac{3}{5} \left[\frac{3}{2}\left(1+\frac{4}{15}\left(1-\fracr \right)\right)\right]\Phi^{(1)}(\vk_1)\Phi^{(1)}(\vk_2)\right\}\,,
\ee
which is inspired from Eq.~(\ref{Phi2functionofR}) when quadratic terms are ignored, that is when the relation between $\Phi^{(2)}$ and ${\cal R}^{(2)}$ is approximated to be linear, which is a valid approximation for large primordial non-Gaussinities. This is the non-Gaussianity parameter that we will discuss in the rest of this article. Historically, the definition (\ref{deffnlphi}) was made for $\fracr=1$, and since the matter dominated era potential is related to the primordial one by $\Phi^{(1)}_{\rm mat}=\frac{9}{10}\Phi^{(1)}_{\rm init}$ on large scales, then we would obtain
\be\label{DeffNLPhi}
\Phi_{\rm mat}^{(2)}(\vk)=-2{\cal K} \left\{\fNL^{\Phi}\Phi^{(1)}_{\rm mat}(\vk_1)\Phi^{(1)}_{\rm mat}(\vk_2)\right\}\,.
\ee
It is useful to notice that when we consider the neutrinos  ($\fracr \neq 1$) for the concordance model~\cite{WMAP7}
\be
\frac{3}{5}\left[\frac{3}{2}\left(1+\frac{4}{15}\left(1-\fracr \right)\right)\right]\left(\Phi^{(1)}\right)^2\simeq 0.99\left(\Phi^{(1)}\right)^2\,,
\ee
and thus, the same type of relation holds when considering neutrinos, but deep in the radiation dominated era.

Another interesting family of non-Gaussian initial conditions, motivated by models such DBI inflation, is of the \emph{equilateral type} where the function $F$ peaks when $k_1=k_2=k_3$. The expression for such type of non-Gaussianity is given in Ref.~\cite{Creminelli2006b} and reads
\begin{eqnarray}
\flamoi F_{\rm equi}(\vk_{1},\vk_{2},\vk_{3})&=&6\left\{-P_{{\cal R}}(k_{1})P_{{\cal R}}(k_{2})+ 2\,\sym-2 \left[P_{{\cal R}}(k_{1})P_{{\cal R}}(k_{2})P_{{\cal R}}(k_{3})\right]^{2/3}\right.\nonumber\\
&&\left.\quad +P^{1/3}_{{\cal R}}(k_{1})P^{2/3}_{{\cal R}}(k_{2})P_{{\cal R}}(k_{3})+ 5\,\sym\right\}.
\end{eqnarray}
Similarly to what has been done in Eq.~(\ref{deffnlphi}) we can also define a $\fNL^{\Phi}$ for the equilateral configuration.
The local and equilateral type non-Gaussianities are interesting from a technical point a view, since there analytical expression in function of $k_1$, $k_2$ and $k_3$ is factorizable in powers of these quantities. This implies that the optimal estimator, that we need to build to constrain this type of primordial non-Gaussianity, is faster than for a general shape of the primordial non-Gaussianity~\cite{Babich2004}. Even if the true primordial non-Gaussianity is not strictly of the local or equilateral type, it can be approximated by these ideal cases in order to produce constraints on the primordial non-Gaussianity in a fast and efficient manner~\cite{Liguori2010}.
 
\subsection{Why we can concentrate on $\fNL^{\cal R}=-1$ and still be general}\label{subsec340}

Let us now stress that as long as one considers the bispectrum, the
result is linear in the second order solution. This linearity
shall not be confused with the non-linearity of the perturbation equations.
Indeed, as long as the initial conditions at first order are identical, the
second order initial conditions (\ref{CI2abiab}, \ref{e35b}, 
\ref{psiI}, \ref{Eq_lienPhiPsi_superHubble}) for the matter and metric
perturbations are linear in ${\cal R}^{(2)}$. Since the solutions of the 
second order perturbation equation belong
to an affine space, we can always write $\fNL^{\cal R}
\equiv{\fNL^{\cal R}}_{\rm prim} + {\fNL^{\cal R}}_{\rm 1field}$, where the first
quantity refers to the extra primordial non-Gaussianity compared
to the standard one-field inflation prediction ${\fNL^{\cal R}}_{\rm 1field}$.
It follows that the initial condition for ${\cal R}^{(2)}$ satisfies
$$
\mathcal{R}^{(2)}(\vk)= 2 {\cal K}\left\{{\fNL^{\cal R}}_{\rm prim} {\cal R}^{(1)}(\vk_1) {\cal R}^{(1)}(\vk_2)\right\}
 + 2{\cal K}\left\{{\fNL^{\cal R}}_{\rm 1field} {\cal R}^{(1)}(\vk_1) {\cal R}^{(1)}(\vk_2)\right\}\,,
$$
that is  $\mathcal{R}^{(2)}(\vk)=\mathcal{R}_{\rm prim}^{(2)}(\vk)+\mathcal{R}_{\rm NL~evolution}^{(2)}(\vk)$.
All the purely second order variables will thus be decomposed 
as $X^{(2)}(\vk)=X_{\rm prim}^{(2)}(\vk)+X_{\rm NL~evolution}^{(2)}(\vk)$ and
the final bispectrum will simply be the superposition of a primordial bispectrum and of the
bispectrum induced by the non-linear evolution that is computed in this article.

\section{Computation of the spectrum and bispectrum}

We adopt the line of sight approach~\cite{Hu1997,Ma1995,Zaldarriaga1997,Hu1997a,Hu1998} in the resolution of the Boltzmann equation. 
This consists in considering that the observed brightness in a given direction of the CMB is the sum of the
brightness of the emitting sources. It simplifies the numerical implementation since,
at least at the first order in perturbations, only lower multipoles contribute to the sources.

In this section, we first discuss the definition of the temperaturre in \S~\ref{subsec30} and
then describe the
expression of the emitting sources in \S~\ref{subsec31} (after having
defined our choice of definition for the temperature in \S~\ref{subsec30}). Then we
use a flat sky approximation (to be distinguished from the Limber
approximation which is an extra approximation, which can be made or not)
in order to simplify the numerical integration both for
the spectrum in \S~\ref{subsec31}  and for the bispectrum in \S~\ref{subsec33}. 
We refer to Ref.~\cite{PUBenpreparation} for the detailed derivation of the expressions
summarized below and for a discussion of their accuracy.

\subsection{Defining the temperature}\label{subsec30} 


The primary physical quantity related to the distribution function is the brightness, 
and we need to define the CMB temperature from this quantity. First we define
the {\em brightness temperature} fractional perturbation by
\be\label{defTbrightness}
\Theta \equiv \frac{1}{4}\left(\frac{{\cal I}}{\bar{\cal I}}-1\right)\,,
\ee
where $\bar{\cal I}=\bar\rho_\gamma$ is the background value of the brightness, from which we can define the {\em brightness temperature} by $T_{\cal I}\equiv \bar T (1+\Theta)$.
This definition is inspired from the more general {\em bolometric temperature} $T$ which is defined by
\be\label{defTbolometrique}
\left(\frac{T}{\bar T}\right)^4\equiv \frac{{\cal I}}{\bar{\cal I}}\,\,.
\ee 
The brightness temperature and the bolometric temperature agree at the background and first order level. Since the Boltzmann equation is derived in function of the brightness, the brightness temperature is the most straightforward quantity to use, but the bolometric temperature carries more meaning since it is the temperature of the black-body distribution which would have the same energy density as the actual distribution.
However, it has been shown~\cite{Pitrou2009ysky} that the non-linear dynamics sources a $y$-type spectral distortion and this would affect the brightness and thus both the brightness and the bolometric temperatures. Following our previous analysis~\cite{Pitrou2009ysky}, it would be more natural to use the {\em occupation number temperature} defined as
the temperature of a black-body which would have the same number density of photons
as the observed distribution (to be contrasted with the bolometric temperature which is the temperature
of the black body which carries the same energy density as the observed distribution). It is thus defined  by
\begin{equation}\label{defToccupation}
 T_N\equiv \frac{T}{(1+4 y)^{\frac{1}{4}}}\,\,.
\end{equation}
Defined like this, this temperature is not affected by spectral distortions since Compton scattering conserves the number of photons in collisions, and it has a non-ambiguous signification with the type of distortions induced by the non-linear collision term~\cite{Pitrou2009ysky}. 

These definitions agree at the background level and at first order in the
perturbation so that $C_\ell^{T_{\cal I}T_{\cal I}}=C_\ell^{TT}=C_\ell^{T_NT_N}\equiv \bar T^2 C_\ell^{\Theta\Theta}$.
Since they differ at second order in perturbations, we can define three different
bispectra, depending on the choice of definition for the temperature. As we shall
see in \S~\ref{subsec54}, since at lowest order in the spectral distortions~\cite{Pitrou2009ysky}
\begin{equation}
 T = T_N(1+ y),
\end{equation}
the three bispectra are related to one another so that one only needs
to compute one of them. In order for our results to be easily
compared with the existing literature, we will use the bolometric temperature.
although the occupation number temperature is, at least from a theoretical point of view, more natural to use.


\subsection{Emitting sources and transfer functions}\label{subsec31} 

Given the definition of the multipoles used in Ref.~\cite{Pitrou2008}, the brightness temperature 
emitted by a multipole ${\cal I}_\ell^m(\vk,\eta)$ is given by
\begin{equation}
\frac{1}{4}Q_\ell^m(\vk) {\cal I}_\ell^m(\vk,\eta) \,,
\end{equation}
where we have introduced
\begin{equation}\label{defQl}
Q_\ell^m(\vk) \equiv N_\ell^{-1} Y_\ell^m(\hat \vk)\,,
\end{equation}
$\hat \vk$ being the unit vector in the direction of $\vk$ and where
\begin{equation}
N_\ell \equiv \ii^\ell \sqrt{\frac{2 \ell+1}{4 \pi}}\,.
\end{equation}
It follows that the total expression of the emitting sources is given by
\begin{equation}
W(\vk,\eta) \equiv \exp(-\bar \tau) \sum_{\ell,m} S_\ell^m(\vk,\eta) Q_\ell^m(\vk)\,,
\end{equation}
where $\bar \tau$ is the optical thickness which satisfies the integral equation
\be
\bar \tau = \int_{\eta}^{\eta_0} \dd \eta' \bar \tau'(\eta')\,,
\ee
that is $\dd \bar \tau/ \dd \eta = - \bar \tau'$, and where the emitting sources multipoles $S_\ell^m$ are defined in~\ref{AppEmittingSources}. 
This expression is valid both at first and at second order, depending on whether we consider the first order or the second order sources. 
Since we want to determine how the primordial perturbations of the gravitational potential $\Phi$ translate 
into the observed temperature, we are rather interested in the transfer functions defined by
\abc
\be
 w^{(1)}(\vk,\eta) \Phi(\vk,\eta_\init)\equiv W^{(1)}(\vk,\eta)\,,
\ee
\be
\label{NLkernel}
{\cal K}\left\{w^{(2)}(\vk_1,\vk_2,\eta)\Phi^{(1)}(\vk_1,\eta_\init)\Phi^{(1)}(\vk_2,\eta_\init)\right\}\equiv W^{(2)}(\vk,\eta)\,.
\ee
\eabc
As we shall see, $w^{(1)}(\vk,\eta)$ and $w^{(2)}(\vk_1,\vk_2,\eta)$ will be computed numerically. 
In principle the ${\cal I}_\ell^m(\vk)$ are complex functions, but numerically we prefer to manipulate 
real numbers and real differential equations. At first order this is not an issue since 
only the scalar perturbations are non vanishing and  ${\cal I}^{0(1)}_\ell(\vk)$ 
remains unchanged under a rotation of the coordinates system around $\vk$, that is
about the azimuthal direction used for the spherical harmonics. However, at second order, this
is no more the case since for a given $\ell$ all the allowed $m$ are to be taken into account. In
order to perform the integration with real numbers, we thus choose the
orientations of $\vk_1$ and $\vk_2$ such that $\varphi_{\vk_1}=\varphi_{\vk_2}=0$, that is $k_{1y}=k_{2y}=0$. When the
azimuthal direction of the spherical harmonics is $\vk$, ${\cal I}^{m(1)}_\ell(\vk_1)$ and
${\cal I}^{m(1)}_\ell(\vk_2)$ are then computed according to Eq.~(\ref{Rotatemode}), which ensures that they remain real. 
In the computation of the bispectrum, we will then need a general orientation of $\vk_1$ and $\vk_2$. It 
can be obtained by rotating this configuration around the azimuthal direction $\vk$ by an angle $\alpha(\vk_1,\vk_2)$.
Defining ${\cal I}_{\ell}^{m(2)}$ in a similar way as Eq.~(\ref{NLkernel}),
\begin{equation}\label{DeftransferI}
{\cal K}\left\{{\cal I}_{\ell}^{m(2)}(\vk_1,\vk_2,\eta) \Phi^{(1)}(\vk_1,\eta_\init)\Phi^{(1)}(\vk_2,\eta_\init) \right\}\equiv {\cal I}^{m(2)}_\ell(\vk,\eta)\,,
\end{equation}
it can be related to the result of the (real) numerical integration ${\cal I}_{\ell\,\rm real}^m (\vk_1,\vk_2)$ by
\begin{equation}\label{usealpha}
{\cal I}_{\ell}^m (\vk_1,\vk_2)=e^{\ii m \alpha(\vk_1,\vk_2)} {\cal I}_{\ell\,\rm real}^m (\vk_1,\vk_2)\,.
\end{equation}
Our notation $\alpha$ refers to the Euler angles. 

\subsection{Flat sky approximation}

In the flat sky approximation, the sky is expanded around a reference line of sight 
with direction $\hat {\bf r}$. Any Fourier mode $\vk$ can then be split into a component $k_\parallel=k\cdot \hat {\bf r}$ 
parallel to the line of sight and a 2-dimensional projection, $\vk_\perp=\vk - k_\parallel\hat {\bf r}$, 
in the plane orthogonal to this line of sight. 

While the orientation of the Fourier mode with respect to the line of sight
does not matter for the monopole of the brightness ($\ell=0$), it plays an important role for
higher order multipoles ($\ell \ge 1$). Indeed their contribution to the radiation that we receive
depends on the orientation of $\vk$ with respect to the direction of the flat sky. 

The flat sky approximation is independent from the resolution of the Boltzmann equation. We start by solving the first few moments of the emitting sources by integrating a truncated Boltzmann hierarchy, and only after this we use the line of sight method to obtain all the multipoles of the CMB. The flat sky approximation is only used in this final step and proves to be a very good approximation for $\ell > 10$, since the error is much smaller than the cosmic variance, and reaches a percent accuracy beyond $\ell=100$. Two subcases of the flat sky approximation exist~\cite{PUBenpreparation}. The Limber approximation corresponds to cases where the sources vary slowly in  time and over a wide range of distances along the line of sight. This case is not well suited for CMB except for effects like the late ISW. The second regime corresponds to the thin shell approximation. It relies on the hypothesis that the sources are located in a narrow range of distances. This is the one that we use in this article since it appears to be well suited for primary CMB calculations as illustrated in Fig.~\ref{FigSources1}. 

Let us also note that the procedure described in the previous section in
order to consider a general
orientation of $\vk$, $\vk_1$ and $\vk_2$ with respect to the flat sky can be
obtained as follows. First, we align the azimuthal direction $\vk$
with the direction $\hat {\bf r}$ of the flat sky which is chosen as the ${\bf z}$
direction of a Cartesian coordinate system. Two other axes ${\bf x}$ and ${\bf y}$ 
perpendicular ${\bf z}$ can then be defined. 
We choose $\vk_1$ and $\vk_2$ to lie in the plane spanned by
${\bf z}$ and ${\bf x}$. From this particular configuration, we can reach a
general configuration by first rotating around ${\bf z}$ by an angle $\alpha$,
then by rotating around ${\bf y}$ by an angle $\beta$, and finally by rotating
around ${\bf z}$ by an angle $\gamma$. These last two rotations enable to
reach $\vk$ from the direction ${\bf z}$. In the notation $Y_\ell^m(\hat \vk)$
of Eq.~(\ref{defQl}), we thus meant $Y_\ell^m(\alpha,\beta)$. 

\subsection{Spectrum}\label{subsec32} 

As usual, the temperature anisotropies can be decomposed in spherical harmonics as
\begin{equation}
 \Theta(n^i)=\sum_{\ell m}a_{\ell m}Y^\ell_m(n^i),
\end{equation}
hence defining the coefficients $a_{\ell m}$. The global isotropy
of the Universe implies~\cite{Abramo:2010gk,Prunet:2004zy} that the 2-point correlation function of $\Theta$ can be
decomposed as
\begin{equation}
 \langle \Theta(n_1^i)\Theta(n_2^i) \rangle = \sum_\ell \frac{2\ell+1}{4\pi}C^{\Theta\Theta}_\ell P_\ell(\cos\theta)
\end{equation}
where $n_1^in_{2i}=\cos\theta$ and the brackets refer to an ensemble average
or to an average on the sky if $\Theta$ is the observed temperature. The angular power
spectrum can then be related to the ensemble average of the $a_{\ell m}$ as
\begin{equation}
 \langle a_{\ell m}a_{\ell'm'}^* \rangle = C^{\Theta\Theta}_\ell \delta_{\ell\ell'}\delta_{mm'}.
\end{equation}
The angular power spectrum can be expressed in terms of the initial power spectrum
and linear transfer functions~\cite{FrancisLivre,PeterUzanTrans}. 
At lowest order in the flat-sky expansion, the angular power spectrum of the CMB anisotropies 
takes the form~\cite{Bond1996} (see also~\cite{PUBenpreparation})
\begin{equation}\label{Cl_flat-sky1}
\flamoi C^{\Theta\Theta}_\ell = \frac{1}{2 \pi}\int \dd r \dd r' \dd k_\parallel w^{(1)}(\vk,r)  w^{\star(1)}(\vk,r') e^{-\ii k_\parallel (r-r')} \frac{P_\Phi(k)}{[(r+r')/2]^2}
\end{equation}
where $r=\eta_0-\eta$, $k_\perp \equiv (\ell+ 1/2)/r$ and $k^2=k_\parallel^2+k_\perp^2$, and where a star denotes the complex conjugate. 
If we neglect the integrated effects which occur at low $z$ and affect only
the largest scales, we can then approximate that all the signal in
$w(\vk,\eta)$ arises from the last scattering surface whose thickness is defined
from the visibility function $\bar g(\eta)\equiv -\bar \tau' \exp(-\bar \tau)$. This visiblity
function peaks at $r_{\lss}$ which defines the center of the last scattering
surface. In this thin shell approximation, we thus obtain that
\begin{equation}\label{Eq_Cl_approx}
C^{\Theta\Theta}_\ell \simeq \frac{1}{2 \pi} \int \dd k_\parallel \left| \int \dd r \,w^{(1)}(\vk,r) e^{-\ii k_{\para} r} \right|^2\frac{P_\Phi(k)}{r_{\lss}^2}\,\,,
\end{equation}
now with $k_\perp = (\ell +1/2)/r_{\lss}$.
Additionally, since modes of different $m$ are in general not statistically correlated\footnote{Note that in our case this is also 
the case because we consider scalar modes at first order in the perturbations.}, 
the phases in the spherical harmonics of~(\ref{defQl}) are unimportant and we can take for practical purposes
\begin{equation}\label{Qlmreal}
Q_\ell^m (\hat \vk) \equiv (-\ii)^\ell \sqrt{\frac{(\ell-m)!}{(\ell+m)!}}P_\ell^m\left(k_\parallel/k\right) \,,
\end{equation}
where $P_\ell^m$ are the associated Legendre polynomials.

\subsection{Bispectrum}\label{subsec33} 

\begin{figure}[htb]
\center
\includegraphics[width=6cm]{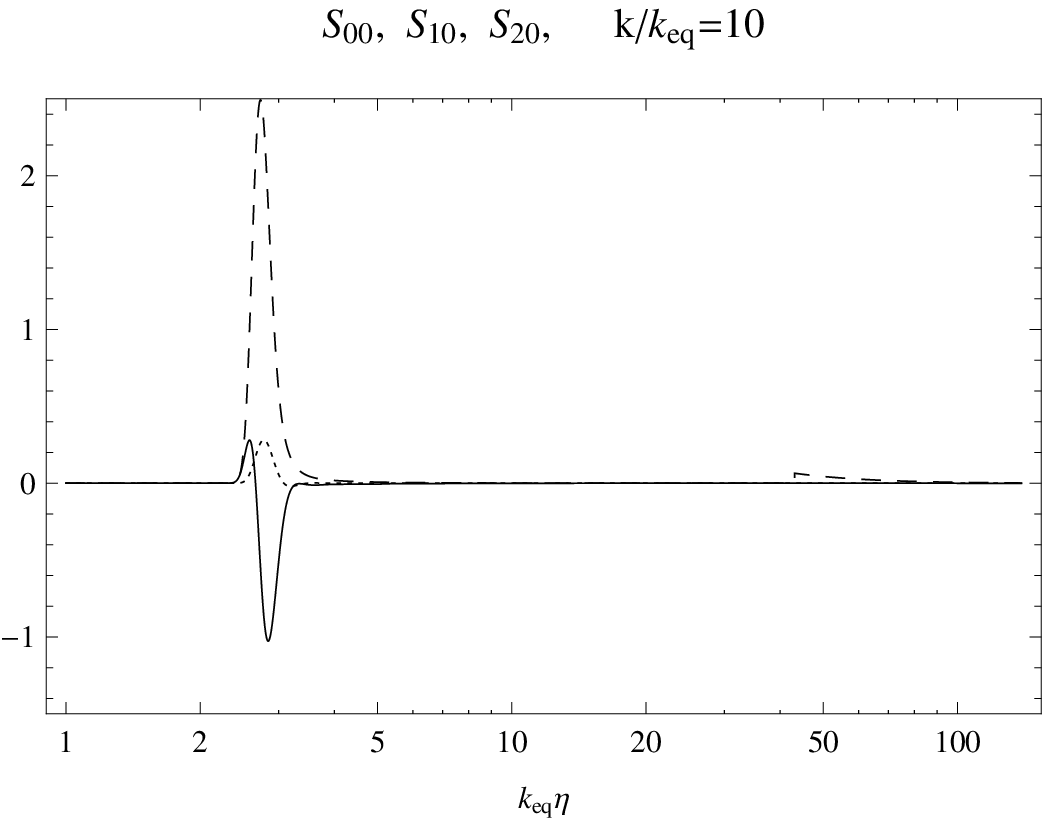}
\includegraphics[width=6cm]{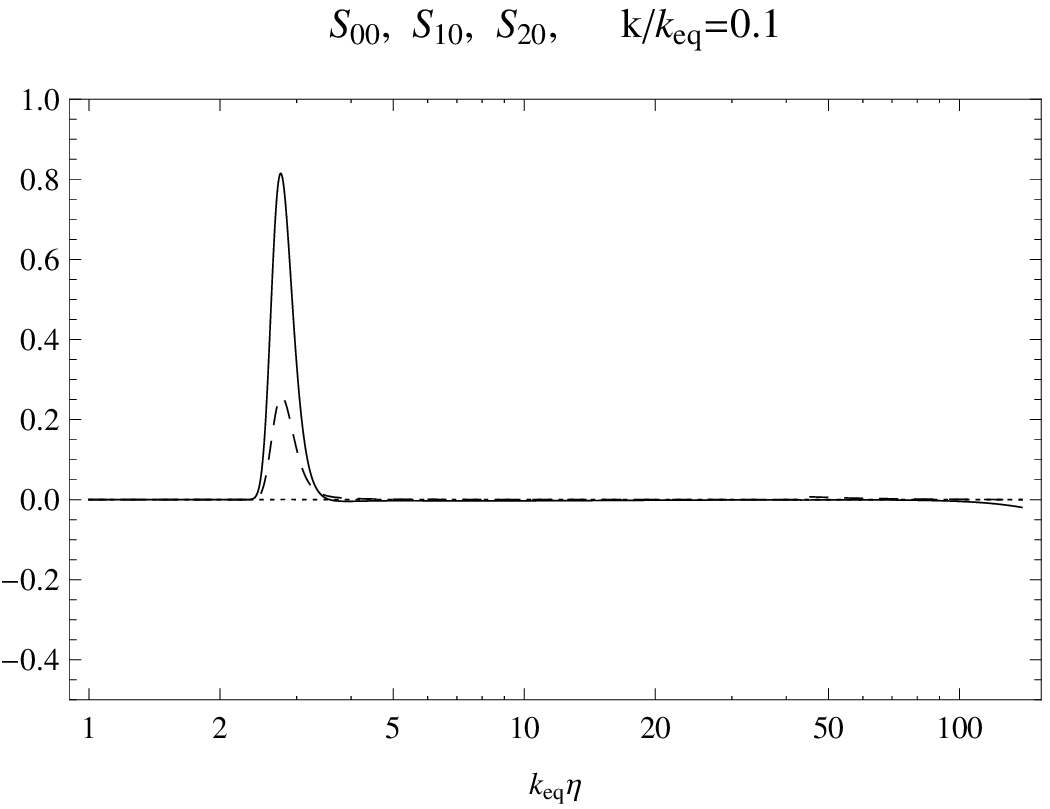}
\caption{First order scalar emitting sources $S_\ell^0$ with $\ell=0,1,2$ in respectively solid, dashed and dotted lines. 
The plot on the left corresponds to a mode which is well inside the Hubble radius around recombination time, whereas the plot on the right corresponds to a mode which is still super-Hubble around recombination time. In both cases the contributions 
are sharply peaked and this defines the last-scattering surface and validates the flat-sky approximation.
 }
\label{FigSources1}
\end{figure}

In a similar way as the angular spectrum, one defines the bispectrum which, again under the assumption of isotropy, is related to the 3-point function of the $a_{\ell m}$ by
\begin{equation}
 \langle a_{\ell_1 m_1}a_{\ell_2m_2}a_{\ell_3m_3}  \rangle = B^{\Theta\Theta\Theta}_{\ell_1\ell_2\ell_3}
  \troisj{\ell_1}{\ell_2}{\ell_3}{m_1}{m_2}{m_3},
\end{equation}
where the matrix denotes a Wigner-$3j$ symbol and $B^{\Theta\Theta\Theta}_{\ell_1\ell_2\ell_3}$ is the bispectrum. It has become more usual to use  the reduced bispectrum $b^{\Theta\Theta\Theta}_{\ell_1 \ell_2 \ell_3}$~\cite{Komatsu2002,Komatsu2001} which is defined from the bispectrum $B^{\Theta\Theta\Theta}_{\ell_1 \ell_2 \ell_3}$ by
\be
B^{\Theta\Theta\Theta}_{\ell_1 \ell_2 \ell_3}
 \equiv \troisj{\ell_1}{\ell_2}{\ell_3}{0}{0}{0} \sqrt{\frac{(2 \ell_1+1)(2 \ell_3+1)(2 \ell_3+1)}{4 \pi}}b^{\Theta\Theta\Theta}_{\ell_1 \ell_2 \ell_3}\,.
\ee
In the flat-sky approximation~\cite{PUBenpreparation}, it takes the form
\begin{eqnarray}\label{bl1l2l3_fullFS}
\flamoi b^{\Theta\Theta\Theta}_{\ell_1 \ell_2 \ell_3}&=&\frac{1}{2 \pi}\int \dd k_{1 \parallel} \dd k_{2 \para} \dd r_1 \dd r_2 \dd r_3 e^{-\ii k_{1 \para}(r_1-r_3)} e^{-\ii k_{2 \para}(r_2-r_3)}\\
\flamoi && \times w^{(1)}(\vk_1,r_1) w^{(1)}(\vk_2,r_2) w_{\rm NL}^{\star(2)}(\vk_1,\vk_2,r_3)\frac{P_\Phi(k_1)}{r_1^2}\frac{P_\Phi(k_2)}{r_2^2}+ 2 \,\, \sym \nonumber\,,
\end{eqnarray}
where
\abc
\begin{equation}
k_{1 \perp} \equiv \ell_1/r_1\,,\quad k_{2 \perp} \equiv \ell_2/r_2\,, 
\end{equation}
\begin{equation}
(k_{3 \perp})^2\equiv \frac{\ell_3^2}{r_1 r_2} + \frac{\ell_1^2}{r_1 r_2}\left(\frac{r_2}{r_1}-1 \right)+\frac{\ell_2^2}{r_1 r_2}\left(\frac{r_1}{r_2}-1 \right)\,.
\end{equation}
\eabc
We also remind that, since $\vk_3=\vk_1+\vk_2$, the parallel components are related according to $k_{3 \para} =k_{1 \para} + k_{2 \para}$.
Similarly to the spectrum, this expression can be avantageously simplified in the case 
where the emitting sources peak in a narrow range of distances. 
More precisely, provided that the first order emitting sources are peaked in a narrow range of distance, which is the case if we neglect the late ISW effect, the following approximation holds out of the largest scales.
In order to grasp how sharply peaked the first order  emitting sources are, we plot them in Fig.~\ref{FigSources1} for $\ell=0,1,2$ for scalar perturbations ($m=0$). 
In this thin shell approximation, the expression~(\ref{bl1l2l3_fullFS}) can be simplified to give
\begin{eqnarray}\label{bl1l2l3_FS}
\flamoi b^{\Theta\Theta\Theta}_{\ell_1 \ell_2 \ell_3}&=&\frac{1}{2 \pi} \int \dd k_{1 \parallel} \dd k_{2 \para} \left[\left(\int\dd r_1 w^{(1)}(\vk_1,r_1) e^{-\ii k_{1 \para} r_1}\right)\left( \int \dd r_2 w^{(1)}(\vk_2,r_2) e^{-\ii k_{2 \para} r_2} \right)\right.\nonumber\\*
\flamoi && \,\,\left.\times\left(\int \dd r_3 e^{\ii (k_{1 \para} +k_{2 \para})r_3} w_{\rm NL}^{\star (2)}(\vk_1,\vk_2,r_3) \right)\right]\frac{P_\Phi(k_1)P_\Phi(k_2)}{r_{\lss}^4} + {\rm 2\,\, perm.} \,
\end{eqnarray}
now with the simpler flat-sky relations
\begin{equation}
k_{1 \perp} \equiv \ell_1/r_\lss\,,\quad k_{2 \perp} \equiv \ell_2/r_\lss\,,\quad k_{3 \perp} \equiv \ell_3/r_\lss\,. 
\end{equation}
From the rotational invariance of the spectrum, we can also use the
expression~(\ref{Qlmreal}) in the computation of $w^{(1)}$ and $w^{(2)}$.
Furthermore, given the definition of the multipoles with the factor $N_\ell$~\cite{Pitrou2008}, the symmetries 
$k_{1 \para} \rightarrow -k_{1 \para}$ and  $k_{2 \para}
\rightarrow -k_{2 \para}$ correspond to $\alpha \rightarrow \pi -\alpha$, and since $b_{\ell_1 \ell_2 \ell_3}^{\Theta\Theta\Theta}$ is real, we can replace $\exp[\ii m \alpha(\vk_1,\vk_2)]$ by $\cos[m \alpha(\vk_1,\vk_2)]$ in Eq.~(\ref{usealpha}).
 
\section{Numerical implementation}

\subsection{Width of correlations}\label{Sec_width_correlation}

In the expression~(\ref{Cl_flat-sky1}) of the spectrum,
the term  multiplying $\hbox{e}^{-\ii k_\parallel (r-r')}$ is an almost constant in the regime $k_\para \ll k_\perp$ and is 
damped, at least by the power spectrum
$P_\Phi\left(\sqrt{k_\para^2+k_\perp^2}\right)$, when $k_\para \gg k_\perp$. It follows that the integration over $k_\para$ leads to a window function ~\cite{Boubekeur2009} 
of argument $(r-r')$ whose width is of the order $1/k_\perp \simeq
r_{\lss}/ \ell$. A similar argument holds for the bispectrum and the
integration over $k_{1 \para}$ and $k_{2 \para}$ in Eq.~(\ref{bl1l2l3_fullFS})
leads to two window functions, one for $(r_1-r_3)$ with a typical width
$1/k_{1 \perp} \simeq r_{\lss}/\ell_1$, and a second for $(r_2-r_3)$ with a typical width
$1/k_{2 \perp} \simeq r_{\lss}/\ell_2$. 
We thus conclude that for a given mode $\ell$, the typical scale of correlation
is given by $r_{\lss}/\ell$. This justifies why we can safely split the effects between primary 
and secondary effects, that is between early and late time effects. 
Indeed, at first order, the gravitational potential 
reaches a constant value until the Universe starts to be dominated by the cosmological constant. The 
integrated effects are then separated into an early effect, that is before the potential freezes, and a late effect when it 
starts evolving again. 
Since the correlations of physical effects must occur at approximately the same distance, then even if integrated effects occur everywhere at second order as the potential grows, they would only contribute in the CMB bispectrum by correlation with the first order early and late effects. As a result, when we are interested in the bispectrum, the second order effects can still be split into primary and secondary effects. In the numerical implementation of Eq.~(\ref{bl1l2l3_fullFS}),
this implies that, as long as we neglect the late ISW, we only need to integrate over $r_3$ in a range of $r_{\lss}$ plus  or minus a few times $r_{\lss}/\ell$ to encompass the early ISW.

\subsection{Numerical integration}

The numerical integration can be sketched as follows.
\begin{enumerate}
 \item[(I)]  We first integrate numerically the sources. Using the initial conditions defined in \S~\ref{sec3}, we integrate numerically in Fourier space the  coupled system constituted of
 (1) the Einstein equation, (2) the Boltzmann equation both up to second order in the perturbations
 and (3) the recombination equation up to first order in the perturbations.
 We first need to integrate the first order equations and then use their results to
 determine the sources of the second order integration.
 \item[(II)] We then compute the line of sight expressions~(\ref{Eq_Cl_approx}) and~(\ref{bl1l2l3_FS}) 
 respectively for the spectrum and for the bispectrum, with their flat sky approximation.
\end{enumerate}

Let us emphasize the following details:
\begin{itemize}
\item All our numerics assume the cosmological parameters as determined by WMAP-5~\cite{WMAP5}.
\item The numerics requires to sample the Fourier space and for each Fourier configuration to integrate numerically in time. 
At first order, we only need to sample the norm $k$ of the mode $\vk$, since we aligned the mode with the azimuthal direction. 
At second order, we also align $\vk$ with the azimuthal direction. We only need to sample $k_1$ and $k_2$ and 
the cosinus  of the angle ($\mu_{12}$) between $\vk_1$ and $\vk_2$. 
\item The Boltzmann hierarchy has to be truncated in order to avoid spurious reflections. We use the closing 
relations defined in Ref.~\cite{Ma1995} for both the first and the second order. The latter being defined only for scalar modes, we need to supplement them with the appropriate closing relations for the modes with $m\ge1$ that appear
at the second order. These can be found in Ref.~\cite{Riazuelo} and we report them as well in \ref{App_closure}.
\item We use $\ell_{\rm max}=8$ at first order and $\ell_{\rm max}=5$ at second order for the
truncation of the Boltzmann hierarchy.
\item The line of sight integrals for the spectrum and for the bispectrum are then sampled only up
to 3 times the caracteristic scale of the correlation (i.e. $r_\lss/\ell$) from the end of the last scattering surface. 
\item We do not include the late ISW and our results hold only for $\ell \gtrsim  10$.
\item  The code used runs in {\it Mathematica} and is freely available with documentation on the webpage~\cite{CMBquick}. 
The package {\tt CMBquick1} contains all the functions used for the first order integrations including the transfer functions 
and the spectra, while the package {\tt CMBquick2} contains in the same way all the functions used in the second order integrations. 
\end{itemize}

\subsection{Transfer functions}

\begin{figure}[htb]
\center
\includegraphics[width=6cm]{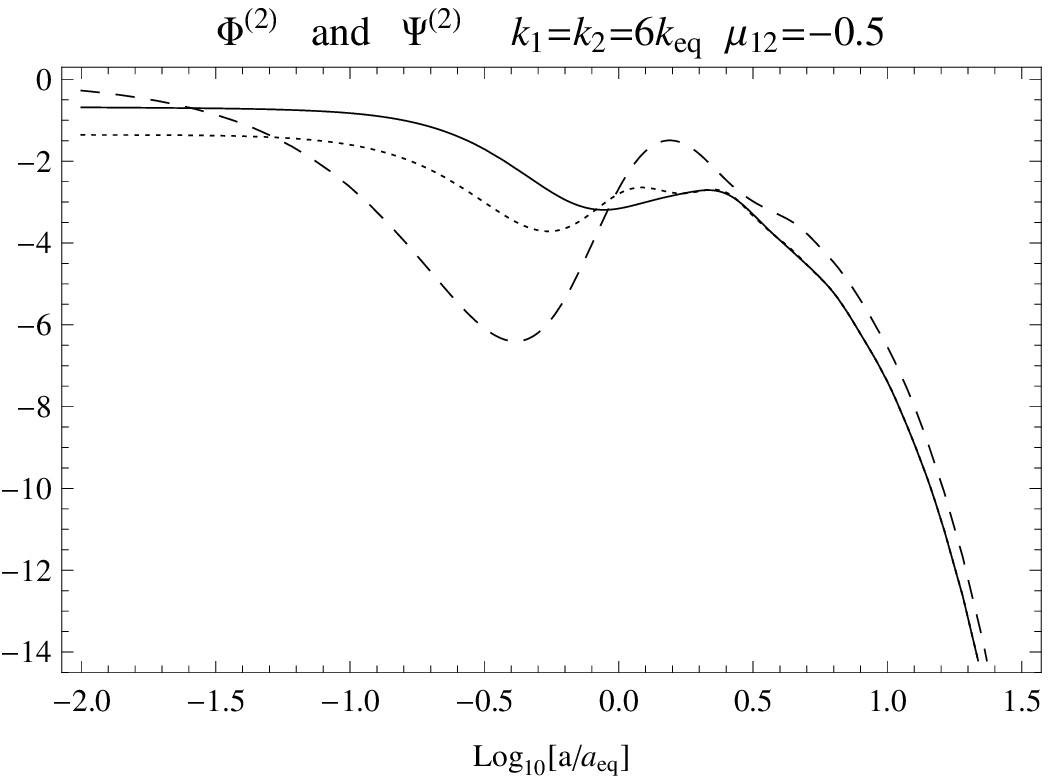}
\includegraphics[width=6cm]{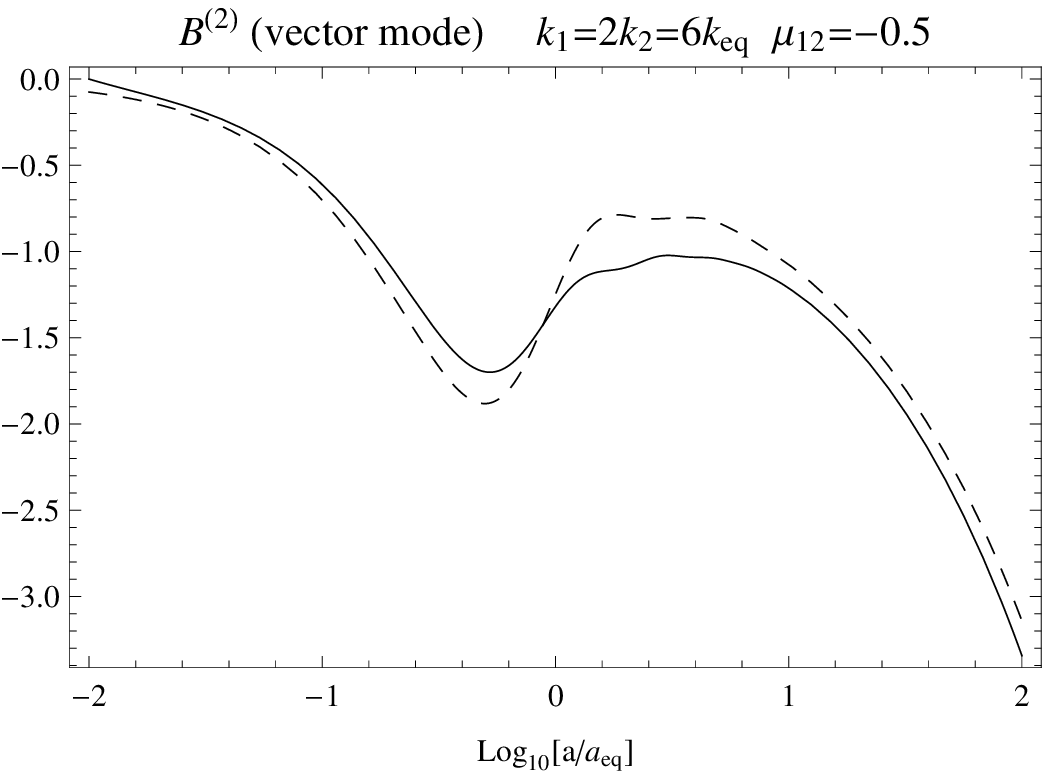}\\
\includegraphics[width=6cm]{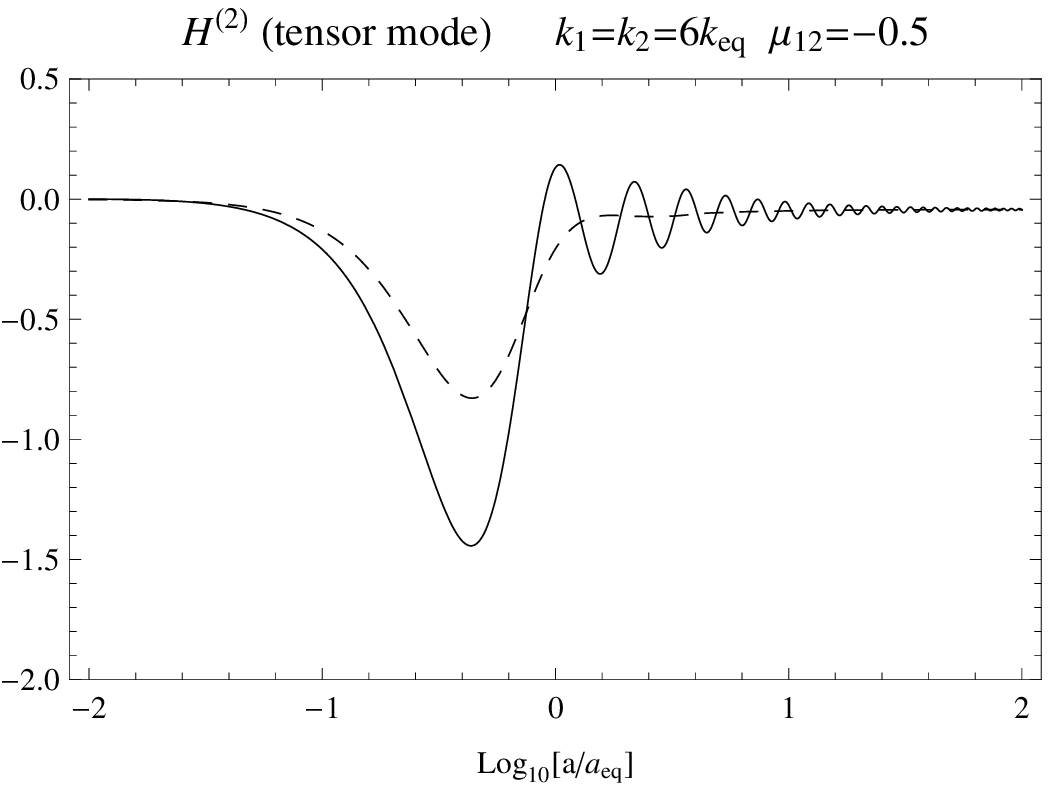}
\includegraphics[width=6cm]{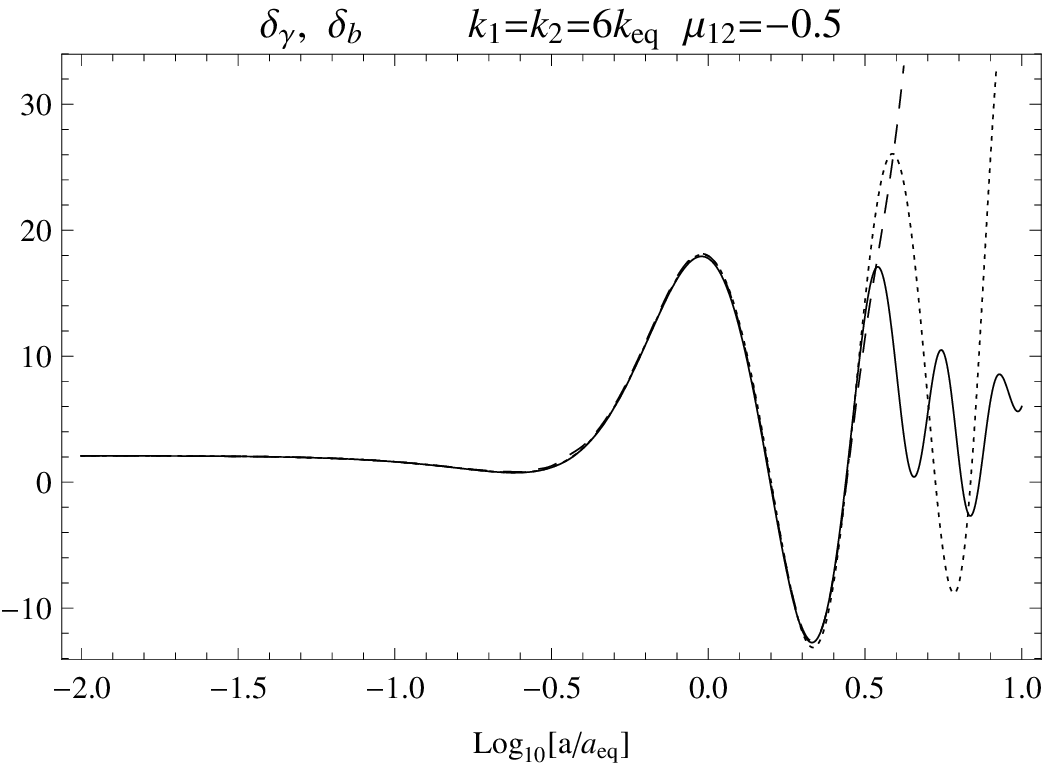}
\caption{Top left: $\Phi^{(2)}$ and $\Psi^{(2)}$ are in plotted in solid and dotted lines respectively. 
Top right: Vector perturbation $B^{(2)}$ (solid line); Bottom left: Tensor perturbation $H^{(2)}$ (solid line).
For all these panels, the asymptotic transfer functions in the limit $a/a_{\rm eq} \gg 1$ are depicted in dashed line.
Bottom right: Second order energy density contrasts for the radiation, $\delta_\gamma$ (solid), and baryons;
$\delta_b$ (long dashed). The tight coupling approximation which holds as 
long as $\tau'/k \gg 1$, is drawn in dotted line.}
\label{figtransfers}
\end{figure}

Fig.~\ref{figtransfers} depicts the transfer functions at second order, using
a definition similar to Eq.~(\ref{DeftransferI}), for the scalar, vector
and tensor degrees of freedom of the metric, restricting to the particular configuration $k_1=k_2=6 k_\eq$ and $\mu_{12}=-0.5$.
These transfer functions, computed numerically and without any approximation are perfectly
compatible with the ones reported  in Ref.~\cite{PUB2008} obtained from an approximation
of the Boltzmann equation. This gives an a posteriori confirmation of the physical understanding
we have provided in Ref.~\cite{PUB2008} and we refer to this article for further discussion.

Fig.~\ref{figtransfers}  also compares the transfer functions to their asymptotic behaviours in the limit $a/a_{\rm eq} \gg 1$.
These asymptotic behaviours correspond to the transfer functions computed in a purely matter dominated universe, and 
we have checked that they match with their previous estimations~\cite{Boubekeur2009}.

Note that the vector modes transfer function almost vanishes when $k_1=k_2$ for all $\mu_{12}$  since in that case 
the only contribution arises from the coupling of the anisotropic stress contracted 
with gradients~\cite{RoyElisa}. Indeed in the harmonic 
decomposition of terms like $X\partial_\iB Y$, the terms for $m=1$ and $m=-1$ vanish 
for symmetry reasons when $k_1=k_2$. For such configurations, the contributions to vector production then arise solely from terms of the form $X^{\iT \jT}\partial_\jB Y$. As a consequence, we have plotted a Fourier configuration which escapes this restriction.

\subsection{Reduced bispectra}\label{subsec54}

\subsubsection{Relation between bispectra}

In \S~\ref{subsec30}, we emphasized the different possible definitions of temperature. Each definition defines
a bispectrum while it does not affect the spectrum.

While the most straightforward would be to use the bispectrum of the temperature brightness $\Theta$ defined in Eq.~(\ref{defTbrightness}),
the previous literature~\cite{Nitta2009} has focused on the bispectrum of the  bolometric temperature $T$ 
defined in Eq.~(\ref{defTbolometrique}). And, as we emphasized, to avoid the spectral distortion that affect
the bolometric temperature, it may be more robust to use the occupation number temperature $T_N$. Let
us now discuss the relation between the bispectra of these 3 quantities.

First, it is obvious from Eqs.~(\ref{defTbrightness}) and~(\ref{defTbolometrique}) that
\be
 b^{TTT}_{\ell_1 \ell_2 \ell_3}=b^{T_{\cal I}T_{\cal I}T_{\cal I}}_{\ell_1 \ell_2 \ell_3}-3 \left(C_{\ell_1} C_{\ell_2}
 +C_{\ell_2} C_{\ell_3}+C_{\ell_1} C_{\ell_3}\right)\,.
\ee
Then, we deduce from Eq.~(\ref{defToccupation}) that
the total second order information is carried by the combinattion of $b^{T_{\rm N}T_{\rm N}T_{\rm N}}_{\ell_1 \ell_2 \ell_3}$ 
and $b^{yT_{\rm N}T_{\rm N}}_{\ell_1 \ell_2 \ell_3}$. It follows that
\be
b^{T_{\rm N}T_{\rm N}T_{\rm N}}_{\ell_1 \ell_2 \ell_3}=b^{TTT}_{\ell_1 \ell_2 \ell_3}-b^{yT_{\rm N}T_{\rm N}}_{\ell_1 \ell_2 \ell_3}-b^{yT_{\rm N}T_{\rm N}}_{\ell_2 \ell_3 \ell_1}-b^{yT_{\rm N}T_{\rm N}}_{\ell_3 \ell_1 \ell_2}\,.
\ee

In conclusion, it is an easy translation to switch from one bispectrum to
the other (once the evolution of $y$ is known but this has been achieved
in Ref.~\cite{Pitrou2009ysky}). In order for our results to be easily compared
to the previous literature, we will only consider the bolometric temperature.

\subsubsection{Examples of numerical results}
%

\begin{figure}[htb]
\center
\includegraphics[width=7cm]{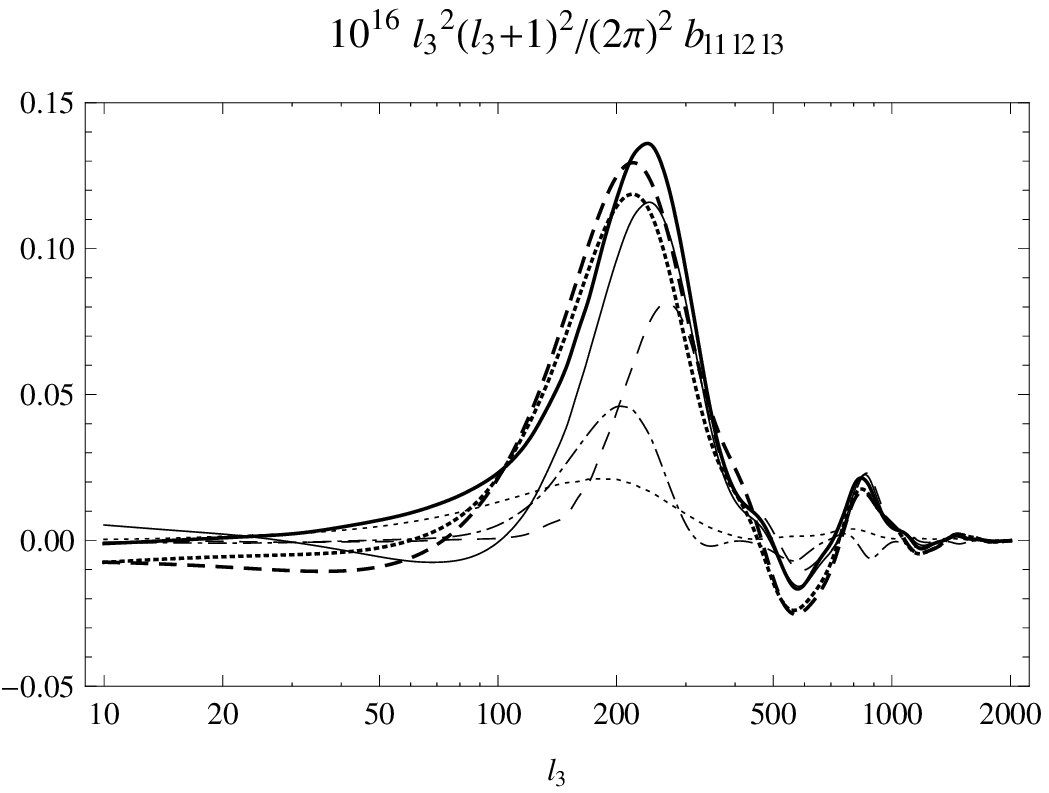}\\
\includegraphics[width=7cm]{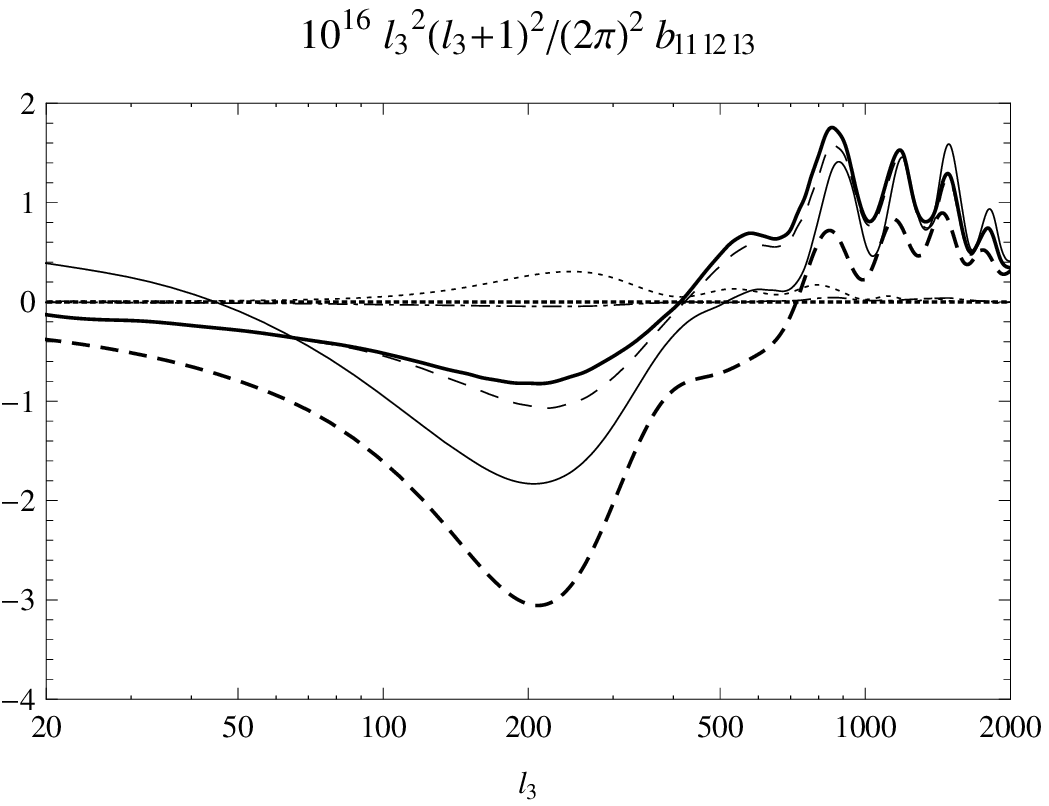}\\
\includegraphics[width=7cm]{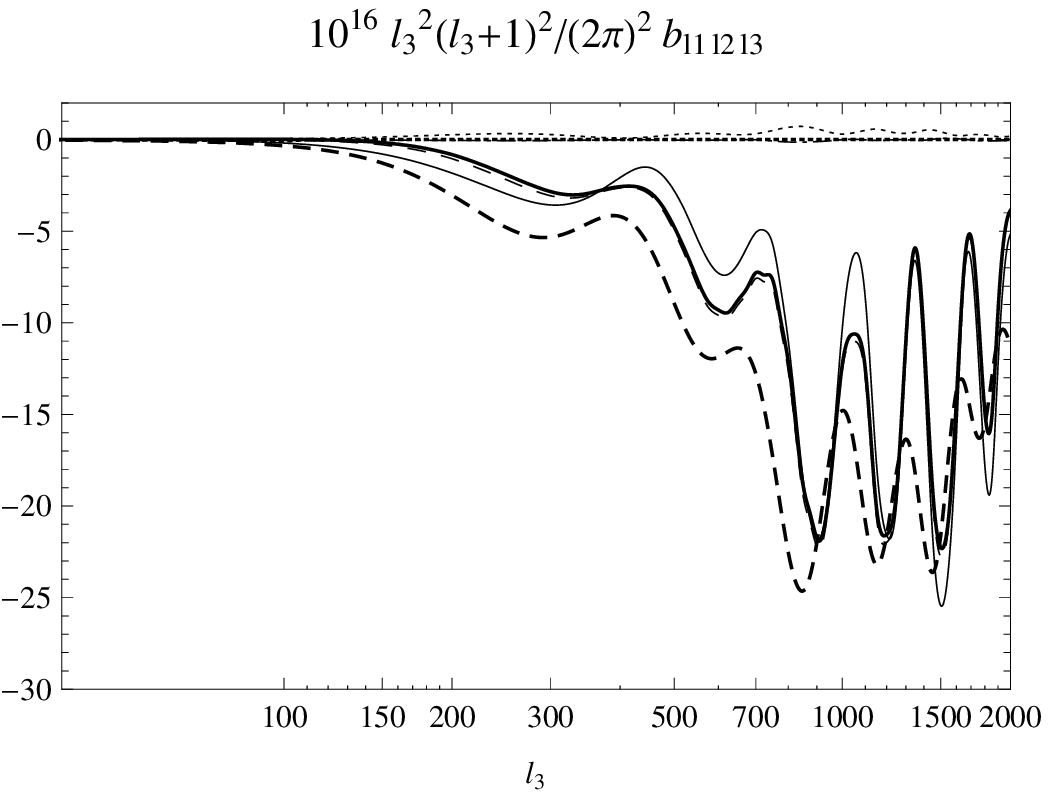}
\caption{Comparison of the bispectrum induced by the non-linear dynamics (thick solid line) to a primordial bispectrum with $\fNL^{\Phi}=5$ (thick dashed line for local type, thick dotted line for the equilateral type). We also plot the approximation of Ref.~\cite{PUB2008} that is considering only purely second order scalar sources in thin solid line, and the contribution from sources for $m=0,1,2$ in respectively thin dashed, dotted, and dot-dashed lines. We considered three different configurations in $\ell$ space, from top to bottom: an equilateral configuration with $\ell_1=\ell_2=\ell_3$ and then two squeezed configurations with $10 \ell_1=\ell_2=\ell_3$ and  $\ell_1=20$ with $\ell_2=\ell_3$. }
\label{figbllls2}
\end{figure}

Representing the bispectrum completely would require a four dimensional plot. For the sake
of clarity, we restrict our presentation to a few specific configurations. But once again any other configuration is accessible via
the use of the packages made available in \cite{CMBquick}. On Fig.~\ref{figbllls2}, we show the explicit shape of the bispectrum that we have found and show the three contributions coming from the sources with $m=0$, $m=1$ and $m=2$ respectively\footnote{Note that the meaning of $\ell$ for a source term is indeed ambiguous since for instance if we consider the gradient of a scalar, it contributes in the source with $\ell=1$ and $m=0$ in an integrated effect. However if we perform an integration by part of
the form $n^\iT \partial_\iB X = \dd X/\dd \eta -\partial X/\partial \eta$ then the total derivative contributes 
in the source term $\ell=m=0$ in an effect on the last scattering surface, and the partial derivative contributes as well in $\ell=m=0$ in an integrated effect. Actually, at first order, this is precisely what happens to the term $n^\iT \partial_\iB \Phi^{(1)}$ and that is why after an integration by parts it contributes to the Sachs-Wolfe effect and as an integrated effect. Furthermore, there is no physical meaning 
in splitting the contributions of the sources into purely second order terms and quadratic terms as in Ref.~\cite{Nitta2009} 
since this decomposition is not gauge invariant. Additionally, there can be large contributions of opposite sign which thus nearly cancel as for instance on large scales (see 
\S~\ref{Sec_analytic_large} or Refs.~\cite{Bartolo2004a,Boubekeur2009}) so that picking up
a single contribution may be misleading and give an overestimation of the signal. With the same argument, 
the coupling of the first order collision term with the perturbed ionization fraction which has been studied in 
Refs.~\cite{Khatri2008,Senatore2008b,Khatri2009} is also not a gauge invariant quantity and we do not try 
to reproduce these results for this reason.}. We see that the scalar contribution, that is from sources with $m=0$ is dominant, especially on small scales. The top panel corresponds to equilateral configurations (with a varying size) and the bottom panels
to squeezed cases (isosceles triangles) with length ratio of 10 for the middle panel and a fixed value $l_{1}=20$ for the bottom. The curves are given as a function of $l_{3}$.

\section{Implications for the detection of the bispectrum}

The previous sections have allowed us to numerically compute different bispectra taking
into account the full non-linear evolution of the cosmological perturbations and compare
it to different types of primordial non-Gaussianities. We now turn, in \S~\ref{subsec61} to the question
of defining an equivalent $\fNL$ and then to the estimation of the signal to noise ratio in \S~\ref{subsec62}. 

\subsection{Equivalent $\fNL$}\label{subsec61}
\newcommand\fNLeq{\widehat{f}_{_{\rm NL}}}

\begin{figure}[htb]
\center
\includegraphics[width=6cm]{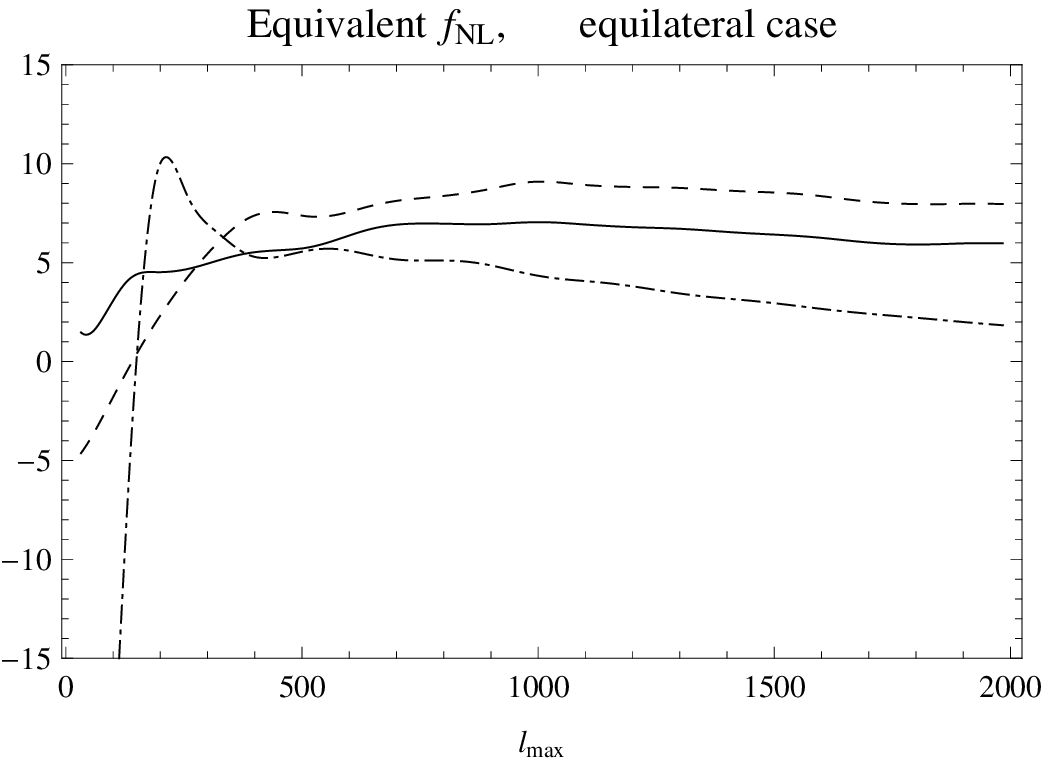}
\includegraphics[width=6cm]{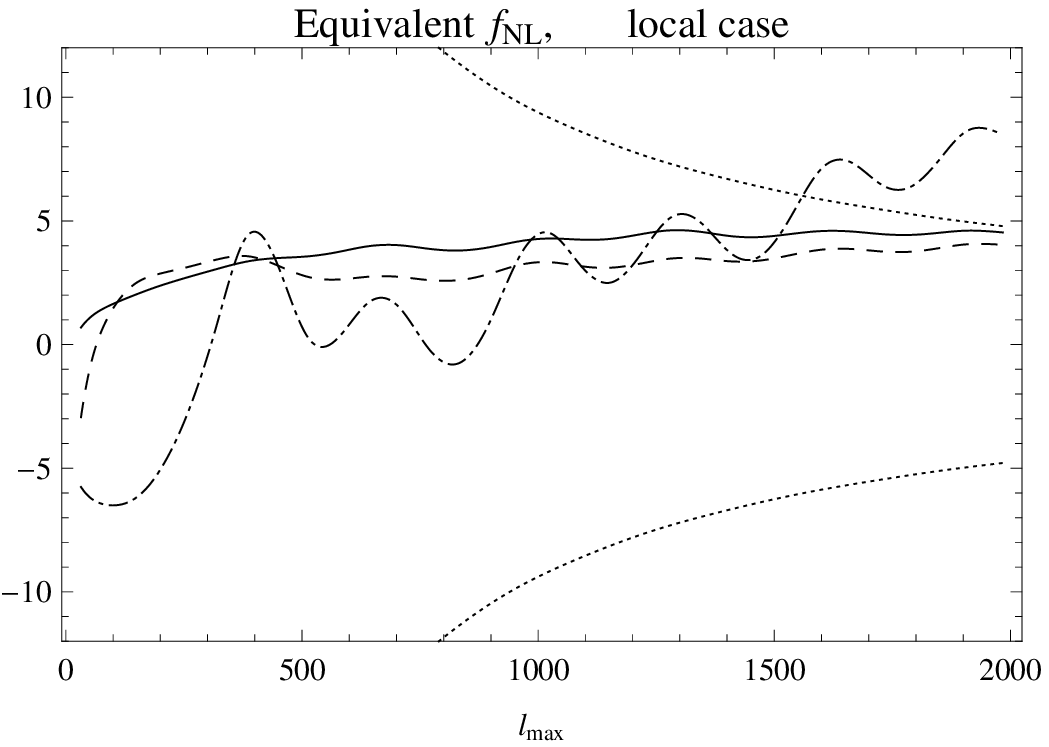}
\caption{Comparison of $\fNLeq$ (solid line) induced by the non-linear dynamics
to  $\fNLeq$ obtained by including only the scalar ($m=0$) modes (dashed line) of the purely second order sources, which corresponds to the approximation of our previous work~\cite{PUB2008}.  We also plot $\fNLeq$ (dot-dashed lined) induced by lensing-ISW secondary effect. Right panel corresponds to local type couplings and the left panel to equilateral type couplings and the $\pm \sigma$ detection limit of $\fNL^\Phi$ is depicted in dotted lines for the local type couplings case.}
\label{fig1}
\end{figure}

We follow the extensive previous 
works~\cite{Bartolo2010revue,Bartolo2008,Nitta2009,Hanson2009,Boubekeur2009,Creminelli2006} in order to define what is the equivalent $\fNL$, noted $\fNLeq$, that the primary non-linear effects would induce.

More precisely, let us consider an ideal experiment, that would be only cosmic variance limited, and a 
statistical estimator designed to detect the primordial non-Gaussianity of the type $\fNL^{\Phi}$ assuming
a linear evolution of the cosmological fields. Then $\fNLeq$ is precisely what such an estimator 
would actually measure because of the non-linear dynamics. In summary, for a given choice of model for the
primordial couplings, $\fNL^{\Phi}$ is such that 
\begin{eqnarray}
\flamoi \hbox{Signal}\left(\begin{array}{c}
  \fNL^{\Phi} = \fNLeq  \\ 
  \hbox{with}\\
  \hbox{linear evolution}
  \end{array}\right)
=
  \hbox{Signal}\left(\begin{array}{c}
  \fNL^{{\cal R}_M} = 0 \\ 
  \hbox{with}\\
  \hbox{non-linear evolution}
  \end{array}\right).
\end{eqnarray}
According to this point of view, the signal coming from the second order evolution represents a spurious signal, a noise, and $\fNLeq$ 
is the amplitude of that noise. Of course from an observational point of view, this is not a genuine noise and is not related to any 
experimental imperfections. It may also have some interests on its own! In the future, it will have to be incorporated consistently in the design of the estimators of primordial non-Gaussianity.

Being mostly interested in the equilateral and the local configurations, we report
in Fig.~\ref{fig1}  $\fNLeq$ for these two configurations as a function of the maximum multipole $\ell_{\rm max}$ used.

We have also plotted on Fig.~\ref{fig1} the $\fNLeq$ arising only from the scalar ($m=0$) purely second order source. 
As expected from our previous analysis~\cite{PUB2008} and from the numerical results of the previous section, these are clearly the dominant contributions. When comparing with the $\fNLeq$ coming from the secondary non-linear effects as reported in 
Ref.~\cite{Hanson2009}, we observe that the primary effects are of slighlty less importance for the local case while they are far more important for the equilateral case. For the local case, the contributions to $\fNLeq$ from primary and secondary effects have the same sign.
The sum of the primary and secondary effects is then beyond the $2 \sigma$ limit of detection when $\ell_{\rm max}>1500$. This is certainly above the detection limit of \emph{Planck} since it uses multipoles up to $\ell_{\rm max}\simeq 2000$ and it should thus be taken into account. 
The primary effects alone would bias the measurement of the local type non-Gaussianity by $\Delta \fNL^{\Phi}\simeq 5$.

\subsection{Signal to noise}\label{subsec62}

\begin{figure}[htb]
\center
\includegraphics[width=8cm]{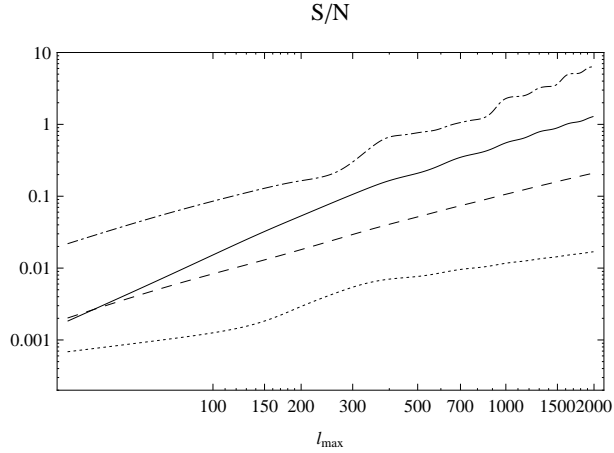}
\caption{The signal to noise ratio as a function of the maximum multipole $\ell_{\rm max}$ for an ideal experiment. The signal to noise ratio of the total bispectrum generated by non-linear effects (solid line), of the local type primordial bispectrum 
(dashed line) and of the equilateral type (dotted line) 
when $\fNL^{\Phi}=1$. We also plot the signal to noise ratio of the bispectrum due to the
 ISW-lensing correlation in dot-dashed line. 
 }
\label{FigSN}
\end{figure}

We can now report the signal to noise ratio (S/N) as a function of the maximum 
multipole $\ell_{\rm max}$ used in the estimator described in Refs.~\cite{Hanson2009,Hu2000}. 
Fig.~\ref{FigSN} represents the signal to noise ratio of the primordial bispectra, both for the local and equilateral cases,
as well as the signal to noise ratio of the bispectrum generated by non-linear effects. 
We also superimpose the signal to noise ratio of the bispectrum generated by ISW-lensing correlations 
(see \S~\ref{subsec72}). We found that the signal to noise ratio of the bispectrum 
generated by the non-linear dynamics reaches unity for  $\ell_{\rm max}=2000$. 

\section{Analytical understanding of the limiting cases}

The aim of the section is to derive analytical formulae, a priori valid in asymptotic regimes, that
offer theoretical insights on the underlying physics. We first  ignore the contribution of neutrinos, and thus set $\fracr=1$,
since they play only a marginal role it will greatly simplify the expressions  obtained. 


After rewriting the Boltzmann equation in \S~\ref{subsec71}, we first focus on secondary
anisotropies in \S~\ref{subsec72}. We then consider the large angular scale limit in
\S~\ref{Sec_analytic_large} (considering separately local and equilateral configurations), 
the small angular scale limit in \S~\ref{Sec74}, and finally a mixing between small and large scales in \S~\ref{Sec75}

\subsection{Rewriting the Boltzmann equation non perturbatively}\label{subsec71}

Since the Boltzmann equation is solved numerically by using the line of sight approach, it is
convenient to rewrite it in terms of the total time derivative along geodesics. It has been noted that 
defining 
\be
\exp[2 \Phi_{\rm BRM}] \equiv 1+2\Phi\,,\qquad \exp[- 2 \Psi_{\rm BRM}] \equiv 1-2\Psi\,,
\ee
simplifies its expression~\cite{Bartolo2006}. In particular,
the Boltzmann equation, written non perturbatively (that is in such a way that it is valid up to any order in perturbations), can be recast as
\begin{equation}
e^{\bar \tau} \frac{\dd \left(e^{-\bar \tau} \tilde {\cal I}\right)}{\dd \eta}+\frac{\partial \tilde {\cal I}}{\partial n^\iT}\frac{\dd n^\iT}{\dd \eta} -4 \frac{\tilde {\cal I}}{E}\frac{\dd E}{\dd \eta}= e^{\Phi_{\rm BRM}}{\cal C}[\tilde {\cal I}]+\bar \tau' \tilde {\cal I}\,.
\end{equation}
In this expression, $E$ stands for the scale factor multiplied by the energy\footnote{In Ref.~\cite{Pitrou2008} $E$ is noted $a p^\zT$ and in Ref.~\cite{Bartolo2008} it is noted $a p$. Since this is the energy of photons, we use this more obvious notation.} of a photon\footnote{See 
Ref.~\cite{Pitrou2008} for details about the observer which actually measures this energy.}, 
$\tilde {\cal I }(\gr{x},\eta,n^\iT)$ is the reduced brightness of the radiation, that is the brightness divided by its average 
value (this implies that $\bar {\tilde{\cal I}}=\bar {\tilde{\cal I}}_0^0=1$). 
Note also that $\dd/\dd \eta$ is the total derivative on the perturbed space-time which includes the convective derivative, but not lensing, i.e. defined on the perturbed space-time by 
\be
\frac{\dd}{\dd \eta} \equiv\frac{\partial }{\partial \eta}+\frac{\partial x^\iB}{\partial \eta}\frac{\partial}{\partial x^\iB}\,\,.
\ee 
While being very general, $E$ cannot be specified so easily in a non perturbative manner, 
and neither can we obtain a non perturbative expression of the collision term. 
However, up to second order we obtain
\be\label{EqdE}
\frac{1}{E}\frac{\dd E}{\dd \eta}=-\left[\frac{\dd \Phi_{\rm
      BRM}}{\dd \eta}-\Phi_{\rm BRM}'-\Psi_{\rm BRM}' + \left(H'_{\iB \jB}-\partial_\iB B_\jB  \right)n^\iT n^\jT \right]\,,
\ee
where we must bear in mind that the vector and tensor modes are only second order quantities.
As for the collision term, it is reported in~\ref{App_Boltzmann} up to second order as well. 

However, we can go a few steps further in our non-perturbative reformulation of the Boltzmann equation. Indeed, we can integrate by parts $\tilde {\cal I}\dd (\ln E)/\dd \eta$, and use the Boltzmann equation itself to replace $(\ln E) \dd \tilde {\cal I}/ \dd \eta$. We then use 
that $(\ln E) \dd (\ln E)/\dd \eta=1/2 \dd (\ln E)^2/ \dd \eta$ and iterate this method and finally resum all the terms to obtain
\be
 e^{\bar \tau} \frac{\dd \left[e^{-\bar \tau}\tilde {\cal I}E^{-4}\right]}{\dd \eta}+E^{-4}\frac{\partial \tilde {\cal I}}{\partial n^\iT}\frac{\dd n^\iT}{\dd \eta}= E^{-4}\left[e^{\Phi_{\rm BRM}}{\cal C}[\tilde {\cal I}]+\bar \tau' \tilde {\cal I}\right]\,.
\ee
If we expand the photon geodesic equation in perturbations according to
\be
\frac{\partial x^\iB}{\partial \eta} = \overline{\frac{\partial x^\iB}{\partial \eta}} + \delta\left(\frac{\partial x^\iB}{\partial \eta} \right)\,,
\ee
we can expand accordingly the total derivative as
\be
\frac{\dd }{\dd \eta} = \overline{\frac{\dd }{\dd \eta}}+ \delta \left( \frac{\dd }{\dd \eta} \right)\,,\quad \overline{\frac{\dd }{\dd \eta}}\equiv \frac{\partial }{\partial \eta} +\overline{\frac{\partial x^\iB}{\partial \eta}}\frac{\partial}{\partial x^\iB}\,\,.
\ee

An integral solution of Eq.~(\ref{EqdE}) can be found immediatly by integrating 
the Boltzmann equation over the background geodesics. We obtain finally that the 
observed brightness today in the direction of observation $-n^\iT$ is
\bea\label{GeneralIntegralForm}
\flamoi E^{-4}(\eta_0)\tilde {\cal I}(\eta_0,-n^\iT)&=&\int^{\eta_0} \dd \eta' \bar g(\eta')E^{-4}\left({\cal C}[\tilde {\cal I}]\frac{e^{\Phi_{\rm BRM}}}{\bar \tau'}+ \tilde {\cal I} \right) -\int^{\eta_0}e^{-\bar \tau }E^{-4} \frac{\partial \tilde {\cal I}}{\partial  n^\iT}\frac{\dd n^\iT}{\dd \eta}\nonumber\\
&&-\int ^{\eta_0}e^{-\bar \tau }\delta\left(\frac{\partial
    x^\iB}{\partial \eta} \right) \partial_\iB \left(\tilde {\cal I} E^{-4} \right)\dd \eta'\,.
\eea
This expression is valid up to any order in the perturbations, but the difficulties are hidden in the computation 
of $E$ and in the expression of the collision term ${\cal C}[\tilde {\cal I}]$. 
The first integral on the right hand side encodes the intrinsic anisotropy. As expected, it is weighted by the visibility function 
and adds up the contributions from the last scattering surface. The second integral contains the lensing effect while 
the last one is the time delay. This latter contribution has already been computed~\cite{Hu2000}
for the spectrum and for the bispectrum, and its effect is by far negligible for geometric reasons. 
However, the lensing effect is both a primary effect and a secondary effect since it contributes 
from the last scattering surface up to now. Since $\dd n^\iT/ \dd \eta$ vanishes at the background level, 
we can use the general integral equation~(\ref{GeneralIntegralForm}) to express it in terms of the intrinsic 
anisotropies by iteration. We detail in the next section how this gives rise to a bispectrum when coupled 
to secondary linear effects.

\subsection{Secondary effects}\label{subsec72}
  
After having reached a constant value in the matter dominated era, the first order potentials 
remain constant until the cosmological constant eventually starts to dominate the matter content of the Universe. 
This gives rise to a late integrated effect (which enters in $E$) known as late Integrated Sachs-Wolfe effect (late ISW). 
Given the arguments of \S~\ref{Sec_width_correlation} on the width of the correlation in the bispectrum, 
the intrinsic second order anisotropy will not correlate to the late ISW. However the lensing term can correlate to the late
ISW since it is also an integrated effect. The lensing term is at least a second order term, and if we make use 
of the first order Boltzmann equation to replace for $ \partial \tilde {\cal I}^{(1)} / \partial n^\iT$, it can be recast in 
(see next section as well as Refs.~\cite{Hu2000,Zaldarriaga2000,Lewis2006})
\be\label{lensingterm}
\nabla^{n^\iT} \phi^{(1)} \nabla_{n^\iT}\tilde {\cal I}(\eta_0,-n^\iT)\,,
\ee
where $\phi^{(1)}$ is the lensing potential whose expression we give further in Eq.~(\ref{EqDeflensingpotential}). 
Note that in order to obtain this expression, we have to neglect the lensing inside the last scattering surface
as well as the lensing of the late ISW. Both approximations are justified from the geometrical properties of the lensing effect since 
it vanishes when the lens is  either at the source or at the observer position. 
Under these approximations, the lensing term is separable into a lensing term $\phi^{(1)}$ and a lensed term ${\cal I}^{(1)}$.
The computation of separable terms is straightforward since the lensing potential correlates with the late ISW effect and the lensed source correlates with the temperature emitted on the last scattering surface. Strictly speaking the bispectrum arising from the lensing term is thus a combination of primary and secondary effects\footnote{Pure secondary effects appear only at large scale and are thus less interesting since the cosmic variance limits the information extracted from the observations.}.
The reduced bispectrum obtained is given for instance in Refs.~\cite{Goldberg1999,Hanson2009,Mangilli2009,Hu2000,Zaldarriaga2000,Lewis2006,Spergel1999} and reads
\be
\flamoi b_{\ell_1 \ell_2 \ell_3}= \demi \left[\ell_2(\ell_2+1)+\ell_3(\ell_3+1) -\ell_1(\ell_1+1)\right] C^{ \phi \Theta_{\rm ISW}}_{\ell_2} C^{\Theta \Theta}_{\ell_3}+ 5 \,\,\sym,
\ee
where the prefactor comes from the angular gradients on $\phi^{(1)}$ and ${\cal I}^{(1)}$.
This method to derive the correlation of the lensing term with first-order secondary effects is general~\cite{Spergel1999,Goldberg1999}. Reionization for instance will create a late Doppler effect, and the effect of the non-linear structures formation (Rees-Sciama, Sunyaev-Zel'dovich...) can also be mapped to additional secondary linear effects (in an effective manner) in order to apply this method~\cite{Mangilli2009,Verde2002} and compute their bispectrum.
\subsection{Large scales}\label{Sec_analytic_large}

\subsubsection{Generalities}

On large scales, that is for modes such that $k \etalss \ll 1$ where $\etalss$
is the average time of recombination, we want to derive the equivalent of the
formula $\Theta^{(1)}=\frac{1}{3}\Phi_{\lss}$ that allows to
explain the Sachs-Wolfe plateau of the angular power spectrum. This
was already obtained by following the photons geodesics in
Refs.~\cite{Bartolo2004a,Bartolo2005,Boubekeur2009}. 

We assume that the last scattering surface is far in the matter dominated era, that is long after the matter-radiation
equivalence. We also ignore the effect of the cosmological constant. The Bardeen potentials $\Phi^{(1)}$ and
$\Psi^{(1)}$ have then reached a constant value that we label by ${\lss}$. On large scales, this is even true at any order
in perturbations. The width of the visibility function can also be neglected on large scales, and 
we approximate $\bar g(\eta)$ by a Dirac distribution centered around $\etalss$, and $e^{-\bar \tau}$ 
by a step function which is unity for $\eta>\etalss$. We can also neglect the velocity fields, that is the velocity 
moments $v_m$ of baryons and cold dark matter, and the first moment of radiation and
neutrinos, that is ${\cal I}_1^m$ and ${\cal N}_1^m$ since they are suppressed
by $k \etalss$. Higher order moments can also be neglected since they are suppressed by higher powers of $k \etalss$. 

A first simplification in \myref{GeneralIntegralForm} lies in the fact that the collision term
has no monopole on large scales. We further assume that we can neglect the vector and tensor 
modes and the integrated terms in the derivative of $E$ given by \myref{EqdE}. In that simplified case we have
\be
E = e^{-\Phi_{\rm BRM}}\,.
\ee
The general solution obtained in \myref{GeneralIntegralForm} then reduces to
\bea
\flamoi \tilde {\cal I}(\eta_0,-n^\iT)e^{4 \Phi_{\rm BRM}(\eta_0)}&=&\tilde {\cal I}(\etalss)e^{4
  \Phi_{\rm BRM}(\etalss)}-\int_{\etalss}^{\eta_0}e^{4 \Phi_{\rm BRM} } \frac{\partial \tilde {\cal I}}{\partial
  n^\iT}\frac{\dd n^\iT}{\dd \eta}\nonumber\\*
&&-\int_{\etalss}^{\eta_0}\delta\left(\frac{\partial
    x^\iB}{\partial \eta} \right) \partial_\iB \left(\tilde {\cal I} e^{ 4\Phi_{\rm BRM}} \right)\dd \eta'\,,
\eea
which is also valid non perturbatively (that is up to any order in perturbations) provided that the approximations hold. We will discard from now on the last integral coming from the time-delay as it is negligible~\cite{Hu2000}.
Restricting to an expression valid only up to second order, we obtain (restoring the potential $\Phi$)
\be
\flamoi \tilde {\cal I}(\eta_0,-n^\iT)=4 \Phi_{\lss}+\tilde{\cal I}_{0\,{\lss}}^0+4(\tilde{\cal I}_{\lss}+ \Phi_{\lss})\Phi_{\lss}-\int_{\etalss}^{\eta_0} \frac{\partial \tilde {\cal I}}{\partial n^\iT}\frac{\dd n^\iT}{\dd \eta}\dd \eta'\,.
\ee
Note that in the previous expression, following standard practice, we have omitted the terms which depend on 
the observer space-time location as they are not observable.

At first order, we obtain the conservation relation
\be\label{Eq_consradiation_ordre1}
\frac{\dd \left(\Theta^{(1)} + \Phi^{(1)} \right)}{\dd \eta}=\dirac{1}(\eta-\etalss)\left(\Theta^{(1)}+\Phi^{(1)}\right)\,,
\ee
where $\dirac{1}$ is the one-dimensional Dirac distribution. We thus recover the textbook result 
\begin{equation}
\Theta^{(1)}(\eta_0,-n^\iT)=\left[\Theta_0^{0(1)}+\Phi^{(1)}\right](\etalss)\,,
\end{equation}
where we recall that $\Theta$ is defined in Eq.~(\ref{defTbrightness}).
We also recall that the right hand side is evaluated at $\etalss$ on background
geodesics defined by the direction $n^\iT$ and that we have dropped the potential evaluated today since it is not observable. 

At second order,  we have
\bea\label{Eq_largescales_final_ordre2}
\Theta^{(2)}(\eta_0,-n^\iT)&=&\Theta_{0\,\lss}^{0(2)}+\Phi_{\lss}^{(2)}+2 \left[4 \Theta^{(1)}_{\lss}+\Phi^{(1)}_{\lss} \right]\Phi^{(1)}_{\lss}\nonumber\\
&&-2\int_{\etalss}^{\eta_0} \frac{\partial \Theta^{(1)}}{\partial n^\iT}\left(\frac{\dd n^\iT}{\dd \eta'}\right)^{(1)}\dd \eta'\,.
\eea
The first line of the right hand side represents the intrinsic anisotropy while the last line contains the lensing effect.
Using Eq.~(\ref{Eq_consradiation_ordre1}) and the fact that the gravitational
potential does not depend on the direction of propagation but only on the
space-time coordinates, the lensing term can be recast as
\be
 -\int_{\etalss}^{\eta_0} \frac{\partial \Theta^{(1)}}{\partial n^\iT}\left(\frac{\partial
  n^\iT}{\partial \eta'}\right)^{(1)}\dd \eta'=\nabla^{n^\iT} \phi^{(1)} \nabla_{n^\iT} \left(\Theta^{(1)}_{\lss}+\Phi^{(1)}_{\lss}\right),
\ee
where we have used the first order expression for $\dd n^\iT /\dd \eta$ 
(given in our notation in Ref.~\cite{Pitrou2008}), and where we have defined the lensing potential by
\be\label{EqDeflensingpotential}
\phi^{(1)}\equiv-\int_{\etalss}^{\eta_0} \frac{\left[\Phi^{(1)}(\eta')+\Psi^{(1)}(\eta')\right](\eta'-\etalss)}{(\eta_0-\eta')(\eta_0-\etalss)}\dd \eta'\,,
\ee
the geometic factor being simply the usual combination of angular distances in a spatially Euclidean universe.

For adiabatic initial conditions, and assuming that the Universe is completely
matter dominated during the recombination, we obtain the following
initial conditions for the temperature anisotropies
\abc
\be
\Theta^{0(1)}_{0,\lss}=-\frac{2}{3}\Phi^{(1)}_\lss\,,
\ee
\be
\Theta^{0(2)}_{0,\lss}=-\frac{2}{3}\Phi^{(2)}_\lss+\frac{28}{9}\Phi^{(1)}_\lss \Phi^{(1)}_\lss\,,
\ee
\eabc
which can be obtained easily~\cite{Boubekeur2009,Bartolo2005} using $\tilde {\cal
  I}_{\lss} = \exp \left(-8 \Phi_{\rm BRM}^{\lss}/3\right)$.
Furthermore, we can use that in the matter dominated era the second
order gravitational potential is given on large scales by
\be\label{Phi2_large-scales}
\flamoi \Phi^{(2)}= 2\Phi^{(1)2}+2 \Delta^{-1}(\partial_\iB \Phi^{(1)} \partial^\iB \Phi^{(1)} )-6
\Delta^{-2}\partial^\iB \partial^\jB(\partial_\iB \Phi^{(1)} \partial_\jB \Phi^{(1)} )\,,
\ee
when $\fNL^{{\cal R}}=-1$. This enables us to obtain a complete expression of the second order temperature anisotropies on large scales in the case where the approximations made hold. 

\subsubsection{Local configuration}

In the local configuration, the primary non-Gaussianity contributes mostly to
squeezed configurations of the bispectrum, that is when one of the modes is much smaller than
the other two, for instance if $k_1 \ll k_2$. We are thus interested in the contribution of evolution in that same configuration in the $\ell$ space in order to assess the contribution of non-linear effects to the total CMB bispectrum. The contributions of the purely second order vector and tensor modes are suppressed by $k_1/k_2$ and can thus be
ignored. The same happens for the integrated effects involving $\Phi^{(2)'}$
and $\Psi^{(2)'}$. The equation~(\ref{Eq_largescales_final_ordre2}) is thus valid in that case.
Furthermore, the lensing term~\cite{Boubekeur2009,Hu2001} can also be neglected in this large scale limit, and we are thus left
with the intrinsic anisotropy. For the local configuration we read from~(\ref{Phi2_large-scales}) 
that $\Phi^{(2)}=2\Phi^{(1)2}$, and we thus obtain
\be
\Theta^{(1)}=\frac{1}{3}\Phi^{(1)}\,,\qquad \Theta^{(2)}=\frac{4}{9}\Phi^{(1)2}\,.
\ee 
We recover the results of Ref.~\cite{Boubekeur2009} for the bolometric temperature $T$,
since the expression obtained reads precisely up to second order 
\be\label{IdeallargescaleT}
(T/\bar T)^4 \equiv \frac{{\cal  I}}{\bar{\cal I}}=(1+ 4\Theta) =\exp[4 \Phi^{(1)}/3]\,.  
\ee
Should we need to compute the temperature with a non-vanishing
$\fNL^{{\cal R}_M}$, that is with $\fNL^\Phi\neq-1$, the previous result would be modified by replacement
of $\Phi^{(1)}$ by $\Phi^{1} -(\fNL^{\Phi}+1) \left( \Phi^{(1)2}-\langle
  \Phi^{(1)2} \rangle \right)$ and the arguments exposed in
Ref.~\cite{Boubekeur2009} for the interpretation of the result would remain unchanged.
Note also that this ideal regime is never reached since the last scattering surface is not deep in the matter dominated era. Since the first and second order potentials undergo a change of value between the radiation dominated era and the matter dominated era, the relation $\Phi^{(2)}=2\Phi^{(1)2}$ is violated around equivalence and in particular on the last scattering surface. This fact implies that the limit in which Eq.~(\ref{IdeallargescaleT}) was obtained cannot be seen from our numerical results.



\subsubsection{Equilateral configuration}

In the equilateral case, the previous approximations do not hold. Indeed the
lensing term and the integrated contributions of the scalar, vector and tensor
modes cannot be ignored since terms such as
$\Delta^{-1} \partial^\iB \Phi \partial_\iB \Phi $ are negligible in the squeezed limit but not in an
equilateral configuration. Since these effects do not correlate only at the last
scattering surface, a proper numerical integration of the bispectrum
expression~(\ref{bl1l2l3_FS}) has to be performed. 
They are computed in details in Ref.~\cite{Boubekeur2009}.

\subsection{Small scales: Equilateral configuration}\label{Sec74}

On small scales, another approximation scheme can be used in order to assess
the order of magnitude of the evolutionary bispectrum. This has been presented in
detail in our previous analysis~\cite{PUB2008}, the validity of which has
been confirmed by the numerics of the present work. It relies on the assumption that all modes are on sub-Hubble scales, and it thus holds only for non-squeezed configurations, that is for configurations of the equilateral type.
To summarize, it is based on the fact that any perturbation mode of the cold dark matter starts collapsing as soon as it 
becomes sub-Hubble. The non-linear gravitational potential is thus predominently shaped by cold dark matter
component. The photon-baryon plasma then develops acoustic oscillations, that are imprinted on the last scattering surface, 
with a forcing term given by the non-linear gravitational potential since on very small scales the purely second order Sachs-Wolfe effect is driven towards
\be
\Theta_0^{0(2)}+\Phi^{(2)} \simeq -R\Phi^{(2)}\,,
\ee
and other effects can be neglected. The convergence of this approximation onto the complete computation is shown on 
Fig.~\ref{fig1}, and we see that it is responsible for most of signal on small scales. In
particular, we conclude that the approximation is good as soon as $\ell>500$. Note also that it is in very good agreement with the full analytic estimation of Ref.~\cite{Bartolo2008} which also find $\fNLeq \simeq 5$ for equilateral type non-Gaussianity when $\ell_{\rm max}=2000$.


\subsection{Mixing large and small scales}\label{Sec75}

\begin{figure}[htb]
\center
\includegraphics[width=6cm]{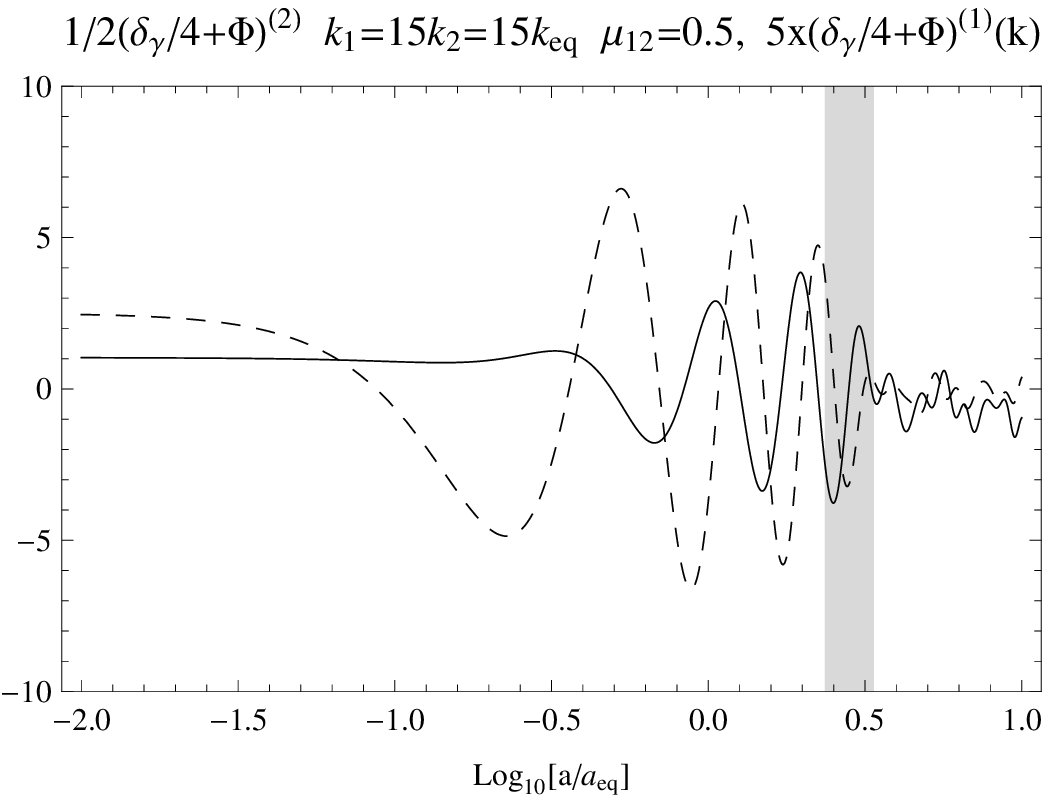}
\includegraphics[width=6cm]{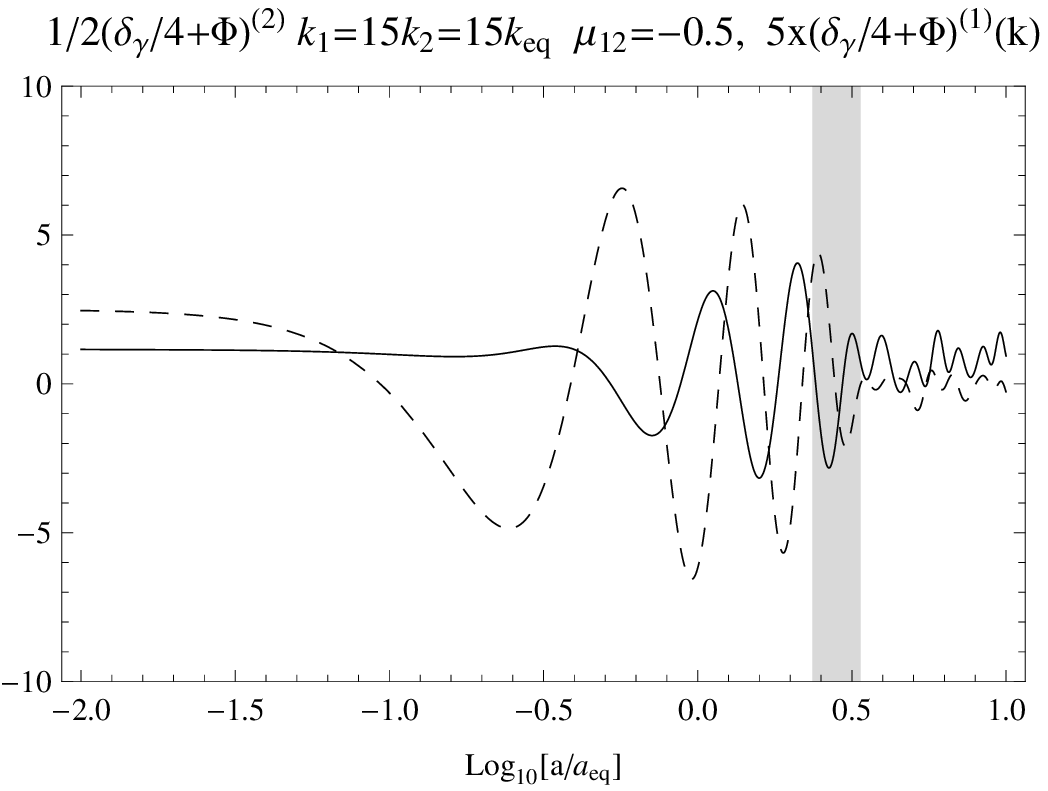}
\caption{In continous line we depict $\demi\left(\Theta_0^{0(2)}+\Phi^{(2)}\right)({\gr k}_1,{\gr k}_2)$ and in dashed line $5 \times \left(\Theta_0^{0(1)}+\Phi^{(1)}\right)(k)$. The vertical shaded area corresponds to the last-scattering surface, that is to the period during which the vast majority of photons are emitted (99\%), except for those emitted during the reionized era. The ratios between the modes $k_1$, $k_2$ and $k_\eq$ is kept fixed and the angle $\mu_{12}$ takes  the values $0.5,\,-0.5$ on respectively the left and right plots. We notice that the purely second order Sachs-Wolfe effect is nearly in phase with opposite sign with respect to the first order one during the last-scattering surface, and approximately $5$ times larger in magnitude.}
\label{figsqueezedintuition}
\end{figure}

The fact that the convolution~(\ref{Convolution}) performed on the source terms is invariant under the exchange $k_1 \leftrightarrow k_2$ makes it very difficult to find analytic solutions to the evolution of perturbations in that case. To see this, let us consider a contribution to the second order potential $\Phi^{(2)}$  which would be quadratic in first order variables, that is of the form $[X Y](\vk)$ [using the notation of ~(\ref{Convolution_Notation})]. The first order quantities $X$ and $Y$ can be related to the primordial first-order potential $\Phi^{(1)}$ through a transfer function, these functions being defined for instance by $X(k_1) \equiv T_X(k_1) \Phi^{(1)}(k_1)$ and $Y(k_2) \equiv T_Y(k_2) \Phi^{(1)}(k_2)$. 
The contribution to the bispectrum in $\Phi$ of such quadratic term is then found to be
\be
\flamoi\langle [XY](\vk)\Phi(\vk_1)\Phi(\vk_2)\rangle = \Dirac{3}(\vk+\vk_1+\vk_2)\left[T_X(k_1) T_Y(k_2)+k_1 \leftrightarrow k_2\right]P_\Phi(k_1)P_\Phi(k_2)\,.
\ee
We thus notice that if we want to find an approximation in a configuration $k_1 \ll k_2$, we have to consider an approximation in that regime for both $T_X(k_1) T_Y(k_2)$ and $T_X(k_2) T_Y(k_1)$. For such configurations, there is thus no simple  approximation scheme that could be used and one has to rely entirely on the numerical results. 
This is typically an issue in order to estimate the non-Gaussian signal for local configurations. 
Indeed most of the signal in the convolution comes from configurations where one of the mode is much smaller than the other (for instance $k_1 \ll k_2$). And for this type of configuration, $k_2$ might corresponds to sub-Hubble scales if $k_2 \gg k_\eq$, but $k_1$ must also be much smaller and thus super-Hubble.
One can however estimate the non-Gaussian signal for local configuration by comparing numerically the purely second order Sachs-Wolfe effect $\demi\left(\Theta_0^{0(2)}+\Phi^{(2)}\right)({\gr k}_1,{\gr k}_2)$ with $k_1\ll k_2$, to the first order one $\left(\Theta_0^{0(1)}+\Phi^{(1)}\right)(k)$, where we recall that ${\gr k}={\gr k}_1+{\gr k}_2$. For two types of squeezed configurations where the large mode $k_1 = 15 k_\eq$ corresponds approximately to $\ell\simeq2000$, we plot these two effects on Fig.~\ref{figsqueezedintuition}. In order to stress that the purely second order effect is larger than the first order one, we multiply the first order effect by a factor $5$. We note that the first and second order Sachs-Wolfe effects are oscillating nearly in phase with opposite sign. 
This result suggests that, in such squeezed configurations, most of the non-Gaussian signal is carried by $\Theta_0^{0(2)}$. And, given the sign definition of $\fNL^{\Phi}$ in Eq.~(\ref{DeffNLPhi}), the second-order acoustic oscillations have an amplitude such that it leads to $\fNLeq \simeq 5$.

\section{Conclusion}

This article presents a complete investigation of the imprint of the non-linear
dynamics on the CMB bispectrum. The calculations were carried with all the matter fields
of the standard cosmological $\Lambda$CDM model included with three families of massless neutrinos.
The numerical calculations make use of a full numerical integration
of the coupled system of the second order Boltzmann and Einstein equations for both the photons, with their
polarization, and the neutrinos. Furthermore, line of sight integrations include first order effects in the recombination history. 
The initial conditions correspond to adiabatic initial conditions with a vanishing intrinsic primordial non-Gaussianity (in the context of standard single-field  inflation, that implies that a contribution of the order of the slow-roll parameters has been neglected).

The numerical integrations were done with the cosmological parameters of the best fit model derived from the WMAP data~\cite{WMAP5}. This article is focused on the bispectrum of the temperature anisotropies. We have been forced however to define
the temperature we use, namely in this article the bolometric temperature, since second order effects are bound to induce spectral distortions. This effect has been described in details in Ref.~\cite{Pitrou2009ysky}, but we do not expect though that a change of definition for the temperature, like the occupation number temperature, would change significantly the conclusions we have reached.
This work also demonstrates that the second order Boltzmann equation can be exactly integrated numerically and be used to produce bispectra. The resulting shape and amplitude of those bispectra is the result of intricate phenomena. It is possible though to obtain theoretical insights into peculiar cases, at small or large scale for instance. Those results confirm in particular that, at small scales, the major mechanism at play is the impact of the gravitational coupling of the dark matter 
potential during the matter dominated era as it had been put forward in Ref.~\cite{PUB2008}.

It is obviously difficult to grasp those results in details. In order to be able to compare the amplitude of those effects to
primordial couplings, we have defined and computed equivalent
$\fNL$ parameters, $\fNLeq$. They are defined in such a way that it is the signal a statistical indicator designed to measure primordial $\fNL$ would get from the amplitude of the temperature bispectrum. We have found that for both primordial non-Gaussianity of local or equilateral types we have  $\fNLeq\simeq5$ for $\ell$ above 500. 
When compared to secondary effects, namely ISW-lensing couplings, primary effects are found to be of comparable amplitude. The former are however more efficient in producing a $\fNLeq$ for the local type ; the situation is reverse for the equilateral type for which the signal comes predominantly from the primary effects. 


Evaluations of signal to noise ratio however show that the non-Gaussianity induced by the primary second order effects in the temperature field alone  can only be marginally detected by the \emph{Planck} mission. Secondary effects are more likely to be detected. This is however the first ever explicit and complete computation of these effects and even though the concordant model does not offer a good chance of detection, it might be a good way to put constraints on alternative cosmological models.
%


We remind that the numerical tools used in this article are freely available and
can be downloaded at~\cite{CMBquick} so that
many other bispectrum configurations and transfer functions can be investigated at will.
The code makes uses of a flat sky approximation. It is accurate enough
for our purpose and this approximation does not interfere with the resolution
of the Boltzmann equation. The code should also be complemented with
secondary non-linear effects, that have been investigated in Refs.~\cite{Hanson2009,Mangilli2009,Spergel1999}.

\ack
C.P. is supported by STFC and would like to thank Institut d'Astrophysique de Paris for its kind hospitality during part of this project.

\appendix

\section{Governing equations}\label{app_equations}

\subsection{Einstein equations}

We use the standard definition that for a species labelled by $a$, $\Omega_a \equiv \bar \rho_a/(3 \HH^2)$. 
When unspecified, $a$ runs on $r,\nu,b,c$ that is on photons, neutrinos, baryons and cold dark matter. 
We also use the definition $\kappa \equiv 8 \pi G$. In the expressions of the quadratic sources, 
it is implicitly meant that the two perturbation variables in each quadratic term are first order variables. 
We thus omit the order superscript in that case to alleviate the notation. 
We have a set of four scalar equations (see Refs.~\cite{Bartolo2004b,Nakamura2007,Acquaviva2003})
\begin{eqnarray}
 \flamoi &&\Delta \Psi^{(2)} -3 \HH \Psi^{(2)'} -3 \HH^2 \Phi^{(2)} -\frac{\kappa}{2} \sum_{a}
  \rho^{(2)}_a = S_1\ ,\label{eq00}\\
\flamoi &&\Psi^{(2)''} + \HH^2 \Phi^{(2)} + \frac{1}{3}\Delta (\Phi^{(2)}-\Psi^{(2)}) + \HH \Phi^{(2)'} + 2 \HH \Psi^{(2)'} \nonumber\\
 &&\qquad\qquad+2 \HH'\Phi^{(2)} -  \frac{\kappa}{6}  \sum_{a=\ir,\nu} \rho^{(2)}_a  = S_2\ ,\label{eqii}\\
\flamoi  && \Psi^{(2)} - \Phi^{(2)} +\frac{\kappa}{5}\Delta^{-1}\left(\bar \rho_\ir {\cal I}_2^{2(2)}+\bar \rho_\nu {\cal N}_2^{2(2)}\right) =S_3\,,\label{eqijst}\\
 \flamoi  && \Psi^{(2)'} + \HH \Phi^{(2)} + \frac{\kappa}{2} \sum_{a} \bar \rho_a (1+w_a)V^{(2)}_a  = S_4\label{eq0i}\ .
\end{eqnarray}
Their quadratic source terms read respectively
\begin{eqnarray}
\flamoi S_{1} &=& -8 \Psi \Delta \Psi -3 \dbi \Psi \dhi \Psi -3 \Psi'^{2}
       + \kappa \sum_{a} \bar \rho_a (1+w_a)\dbi V_{a} \dhi V_{a}  \, ,\label{e.S1}\nonumber\\
\flamoi &&-12 \HH (\Phi-\Psi)\Psi'-12 \HH^2 \Phi^2\\
\flamoi S_{2} &=&  4 \HH^2 \Phi^2  + \frac{8}{3} \Psi \Delta \Psi + 8 \HH (\Phi-\Psi) \Psi'
    + 8 \HH' \Phi^2 - \Psi'^2 +2 \Phi' \Psi'    \nonumber\\
\flamoi &&+\frac{2}{3} \dbi \Phi \dhi \Phi+\frac{2}{3} \dbi \Phi \dhi \Psi+\dbi \Psi \dhi \Psi+ \frac{\kappa}{3} \sum_{a} \bar \rho_a (1+w_a)\dbi V_{a} \dhi V_{a}\nonumber\\
\flamoi &&+\frac{4}{3}(\Phi-\Psi)\Delta \Phi+8\HH(\Phi-\Psi)\Psi'+4 (\Phi-\Psi)\Psi''\,,\\
\flamoi S_{3} &=& -4 \Psi^2-2 \Phi^2\nonumber \\
\flamoi &&-\Delta^{-1}\left[ \dbi (\Psi+\Psi) \dhi (\Psi+\Phi) +2\Psi\Delta \Phi + \sum_{a} \kappa \bar \rho_a (1+w_a) \dbi V_{a} \dhi V_{a} \right]\nonumber \\
 \flamoi   && + 3(\Delta \Delta)^{-1}\dbi \dbj\left[ \dhi (\Psi+\Phi) \dhj (\Psi+\Phi)+2 \Psi \dbi \dbj \Phi  \right. \nonumber\\
\flamoi && \qquad \qquad \left.+  \kappa\sum_{a} \bar \rho_a (1+w_a) \dhi V_a \dhj V_a \right]\ ,\\
\flamoi S_4 &=& 2 \HH \Phi^2 - 4 \Psi \Psi' + 2\dbi^{-1}(\Psi'\dbi \Psi)\nonumber\\
 \flamoi    && + \kappa \sum_{a} \bar \rho_a \dbi^{-1}\left[(1+w_a)\Psi\dbi V_a
    -(1+c_{s,a}^2)\delta_a \dbi V_a \right]\ .\label{e.S4}
\end{eqnarray}
These expressions are easily computed in Fourier space using that~\cite{Pitrou2008} 
\begin{eqnarray}
[\dbi V^{(1)} \dhi V^{(1)}](\vk) &=&-{\cal K}\left\{ \vk_1.\vk_2 V^{(1)}(k_1) V^{(1)}(k_2)\right\}\\
&=&-{\cal K}\left\{\sum_{n=-1}^{1} (-1)^n v^{(1)}_{n}(\vk_1)v^{(1)}_{-n}(\vk_2)\right\}\\
&=&-{\cal K}\left\{ \hat{\vk}_1.\hat{\vk}_2 v^{(1)}_0(k_1) v^{(1)}_0(k_2)\right\}
\end{eqnarray}
since
\be
[\partial_\iB V^{(1)}](\vk)= \ii \hat k_\iB v^{(1)}_0(k) \,, 
\ee
\be
v^{(1)}_m(\vk)= -\hat k^{(m)} v^{(1)}_0(k)\,,
\ee
\be
\vk_1.\vk_2=k_1^{\iB} {k_{2}}_{\iB}=\sum_{n=-1}^{1} (-1)^n k_1^{(n)} k_2^{(-n)}\,,
\ee
where $\hat k_\iB\equiv k_\iB/k$ and $k^2 \equiv k_\iB k^\iB$. The notation $V$ denotes here the scalar part of the velocity field in the coordinate frame, whereas the moments $v_m$ are taken in a local Minkowski frame. See Ref.~\cite{Pitrou2008} for the
precise definitions of these decompositions.

Note also that in order to express terms like
\begin{equation}
 3(\Delta \Delta)^{-1}\dbi \dbj \left(\dhi \Phi \dhj \Phi \right)-\Delta^{-1} \dbi \Phi \dhi \Phi\,,
\end{equation}
in function of the multipolar components of the modes $k_1^{(n)}$ and $k_2^{(-n)}$, we can use the identity (see~\ref{App_Boltzmann} for definitions)
\begin{equation}
\sum_{n=-1}^1 \kMlmsn{2}{0}{0}{n} \,k_1^{(n)} k_2^{(-n)}=3(\hat \vk.\vk_1)(\hat \vk .\vk_2)-\vk_1.\vk_2\,.
\end{equation}
As for the second order metric vector modes, we can determine them from the constraint (that we only report here in Fourier space)
\begin{equation}
-\frac{k^2}{2}B^{(2)}_{m}=\kappa \sum_a \bar \rho_a (1+w_a) v_{a,m}^{(2)} +S_{\rm V}
\end{equation}
with
\begin{eqnarray}
\flamoi S_{\rm V}&=&  {\cal K} \left\{4 \Psi^{(1)'}(k_1)k_2^{(m)}\Phi^{(1)}(k_2)\right.\\
\flamoi &&\qquad \left.+2 \kappa \sum_a \bar \rho_a (1+w_a)\left(\frac{\rho^{(1)}(k_1)}{\bar \rho_a}-\Phi^{(1)}(k_1)-\Psi^{(1)}(k_1)\right)v^{(1)}_{a,m}(k_2)\right\}\,.\nonumber
\end{eqnarray}

Finally the second order tensor modes are determined by a second order differential equation which describes how they are sourced by quadratic terms,
\begin{eqnarray}
\flamoi&&H_m^{(2)''}(k)+2 \HH H_m^{(2)'}(k)+k^2H_m^{(2)}= \frac{2 \kappa}{15}\left(\bar \rho_\ir {\cal I}_2^{m(2)}+\bar \rho_\nu {\cal N}_2^{m(2)}\right)\\
\flamoi&&+\kMlmsn{2}{2}{0}{1}\frac{2}{3}{\cal K}\left\{k_2^{(1)}k_1^{(1)} [\Phi^{(1)}(k_1)\Phi^{(1)}(k_2)+\Psi^{(1)}(k_1)\Psi^{(1)}(k_2)] +\sum_{a=\ib,\ic} \kappa \bar \rho_a v^{(1)}_{a,1}(k_1)v^{(1)}_{a,1}(k_2)\right\}.\nonumber
\end{eqnarray}


\subsection{Boltzmann equation for radiation}\label{App_Boltzmann}

\subsubsection{General remarks}

As mentionned in \S~\ref{sec_struct_equation} (see also Ref.~\cite{Pitrou2008}) it is sufficient to specify 
the purely second order part and the quadratic part of an equation to fully specify it at first and second order.
The Boltzmann equation, which is formally written
\begin{equation}
{\cal L}[{\cal X}]={\cal C}[{\cal X}],
\end{equation}
where ${\cal X}$ stands for ${\cal I}$, ${\cal E}$ and ${\cal B}$, and where ${\cal L}$ and ${\cal C}$ are respectively the Liouville operator and the collision operator. We decompose the operators in the form 
\be\label{DecL}
\flamoi {\cal L}[\bar {\cal X},{\cal X}^{(1)},{\cal X}^{(2)}]=\bar {\cal L}[\bar {\cal X}]+{\cal L}^{(1)}[\bar
{\cal X},{\cal X}^{(1)}]+\frac{1}{2}\left({\cal L}^{(2)}[\bar {\cal X},{\cal X}^{(2)}]+{\cal L}^{(1)(1)}[\bar {\cal X},{\cal X}^{(1)}] \right)\,,
\ee
and similarly for ${\cal C}$.
The details of the derivation with our present notation can be found in Ref.~\cite{Pitrou2008} (see also Refs.~\cite{Bartolo2006,Bartolo2007}). Contrary to what has been performed in 
Ref.~\cite{Pitrou2008} we choose to report here the Boltzmann equation in the form ${\cal L}[]={\cal C}[]$ rather than in the form ${\cal L}^{\hashamoi}[]={\cal C}^{\hashamoi}[]$, in order to facilitate the comparison with existing literature.
We will use the following definitions that will simplify the notation
\bea
\kPlmsn{\ell}{m}{s}{0}&\equiv& \kMlmsn{\ell}{m}{s}{0}\equiv
\sqrt{\frac{(\ell^2-m^2)(\ell^2-s^2)}{\ell^2}}\,,\\
\kPlmsn{\ell}{m}{s}{\pm1}&\equiv& -\sqrt{\frac{(\ell \pm m)(\ell\pm  m
    +1)(\ell^2-s^2)}{2 \ell^2}}\,,\\*
\kMlmsn{\ell}{m}{s}{\pm1}&\equiv& \sqrt{\frac{(\ell\pm m)(\ell\pm m
    -1)(\ell^2-s^2)}{2 \ell^2}}\,,
\eea
\be
\llmn{\ell}{m}{0}\equiv-\frac{m}{\ell}\,,\qquad\llmn{\ell}{m}{\pm 1}\equiv\pm
\frac{1}{\ell}\sqrt{\frac{(\ell+1\pm m)(\ell\mp m)}{2}}\,.
\ee

We report first the link between the moments of radiation and the fluid description involving the energy density and the velocity of radiation. The link is similar for neutrinos. More details can be found in Refs.~\cite{Pitrou2008,Pitrou2007}. Note that we define the multipoles ${\cal I}_\ell^m$ from ${\cal I}/\bar {\cal I}$ so that it is dimensionless, and we do the same for the polarization and neutrinos multipoles.
At first order we have
\begin{equation}
{\cal I}_{0}^{0(1)}(k) = \frac{\rho_\ir^{(1)}(k)}{\bar \rho_\ir}\,, \quad {\cal I}_{1}^{m(1)}(k) = 4 v_{\ir,m}^{(1)}(k)\,,
\end{equation}
and at second order we obtain
\begin{eqnarray}
{\cal I}_{0}^{0(2)}(\vk) &=& \frac{\rho_\ir^{(2)}(\vk)}{\bar \rho_\ir} - 2 {\cal K}\left\{\frac{4}{3} \sum_{n=-1}^{1}(-1)^n v_{\ir,n}^{(1)}(\vk_1) v_{\ir,-n}^{(1)}(\vk_2)\right\}\,,\\
{\cal I}_{1}^{m(2)}(\vk) &=& v_{\ir,m}^{(2)}(\vk) + 2 {\cal K}\left\{\frac{4}{3} \frac{\rho_{\ir}^{(1)}(\vk_1)}{\bar \rho_\ir} v_{\ir,m}^{(1)}(\vk_2)\right\}\,.
\end{eqnarray}
Note also that since $v^{\iT(2)}=V^{\iB(2)}-2\Psi^{(1)} V^{\iB(1)}$ for all species~\cite{Pitrou2008}, then 
\begin{eqnarray}
v^{(1)}_{0}(k)&=& V^{(1)}_0(k)=k^{(0)} V^{(1)}(k)=- k V^{(1)}(k)\,,\\
v^{(1)}_{\pm}(k)&=& V^{(1)}_\pm(k)=0\,,\\
v^{(2)}_{0}(k)&=& -k V^{(2)}(k) - 2 {\cal K}\left\{  \Psi^{(1)}(k_1) k^{(0)}_2 V^{(1)}(k_2)\right\}\,,\\
v^{(2)}_{\pm}(k)&=& V^{(2)}_\pm(k) - 2 {\cal K}\left\{  \Psi^{(1)}(k_1) k^{(\pm)}_2 V^{(1)}(k_2)\right\}\,.
\end{eqnarray}
\subsubsection{First order}
At first order we choose to align the Fourier mode considered $\gr{k}$ with the
direction with respect to which the moments are taken, and the dependence in $\gr{k}$ becomes only a
dependence in its magnitude $k$. The set of equations obtained at first order is (dropping the obvious dependence of
all quantities in $\eta$)
\bea\label{LI1}
\flamoi {\cal L}^{(1)}[\nohat{\cal I}]_\ell^0(k)&=&\nohat{\cal
  I}_\ell^{'0}(k)+k\left[\frac{\kPlmsn{\ell+1}{0}{0}{0}}{2\ell+3}\nohat{\cal
  I}_{\ell+1}^{0}(k)-\frac{\kMlmsn{\ell}{0}{0}{0}}{2\ell-1}\nohat{\cal
  I}_{\ell-1}^{0}(k) \right]\nonumber\\*
\flamoi&&-\delta_\ell^0 4 \Psi'(k)-\delta_\ell^1 4 k \nohat\Phi(k)\,,
\eea
\be\label{LE1}
\flamoi{\cal L}^{(1)}[\nohat{\cal E}]_\ell^0(k)=\nohat{\cal
  E}_\ell^{'0}(k)+k\left[\frac{\kPlmsn{\ell+1}{0}{2}{0}}{2\ell+3}\nohat{\cal
  E}_{\ell+1}^{0}(k)-\frac{\kMlmsn{\ell}{0}{2}{0}}{2\ell-1}\nohat{\cal
  E}_{\ell-1}^{0}(k) \right]\,,
\ee

\be\label{CI1}
\flamoi{\cal C}^{(1)}[\nohat{\cal I}]_\ell^0(k)=\bar{\tau'}\left\{-\nohat{\cal
  I}_\ell^0(k)+\delta_\ell^0 \nohat{\cal
  I}_0^0(k)+4 \delta_\ell^1  \nohat{v}_0(k)+\delta_\ell^2\frac{1}{10}\left[\nohat{\cal I}_2^0(k)-\sqrt{6}\nohat{\cal E}_2^0(k)\right] \right\}\,,
\ee

\be\label{CE1}
\flamoi {\cal C}^{(1)}[\nohat{\cal E}]_\ell^m(k)=\bar{\tau'}\left\{-\nohat{\cal E}_\ell^m(k)-\delta_\ell^2\frac{\sqrt{6}}{10}\left[\nohat{\cal I}_2^0(k)-\sqrt{6}\nohat{\cal E}_2^0(k)\right]\right\}\,.
\ee
The first order magnetic modes are not excited since
the first order vector and tensor modes are negligible, so we did not report them in
the above equations. Additionally, the intensity and electric multipoles are
only excited for $m=0$, that is the reason why we also only reported this case. 

\subsubsection{Second order}

At second order we can only align $\gr{k}$ with the azimuthal direction of the moments, but not $\gr{k}_1$
or $\gr{k}_2$ at the same time. However for a first order quantity the components $X_{\ell}^m$ for a mode in a given direction
$\nohat{\gr{k}}_1$ can be obtained by rotating the components $X_{\ell}^0$ obtained when
we had decided to align this direction with $\bar{\gr{e}}_3$. Namely this rotation
leads for a mode $\gr{k}$ to
\be\label{Rotatemode}
X_{\ell}^m(\gr{k})=\sqrt{\frac{4 \pi}{2 \ell +1}}Y^{\star \ell
  m}(\nohat{\gr{k}})X_{\ell}^0\left(k \right)\,,
\ee
and in particular for the first order velocity
\be
\nohat v_n(\gr{k})=-\hat k^{(n)} \nohat v_0(k)\,.
\ee

We finally obtain for the linear terms in the second order Liouville operator
\bea\label{LiouvilleNMI2}
\flamoi {\cal L}^{(2)}[\nohat{\cal I}^{(2)}]_\ell^m(k)&=&\nohat{\cal
  I}_\ell^{'m(2)}(k)+k\left[\frac{\kPlmsn{\ell+1}{m}{0}{0}}{2\ell+3}\nohat{\cal
  I}_{\ell+1}^{m(2)}(k)-\frac{\kMlmsn{\ell}{m}{0}{0}}{2\ell-1}\nohat{\cal
  I}_{\ell-1}^{m(2)}(k) \right]\nonumber\\*
\flamoi &&-\delta_\ell^0\delta_m^04 \Psi'^{(2)}(k)-\delta_\ell^1\delta_m^0 4 k
\nohat\Phi^{(2)}(k)\nonumber\\*
\flamoi && +\delta_\ell^2\delta_m^1\frac{4
  k}{\sqrt{3}}\nohat{\Phi}_1^{(2)}(k)+\delta_\ell^2\delta_m^{-1}\frac{4 k}{\sqrt{3}}\nohat{\Phi}_{-1}^{(2)}(k)\nonumber\\*
\flamoi &&+\delta_\ell^2\delta_m^2 4 \nohat{H}_1^{'(2)}+\delta_\ell^2\delta_m^{-2}4 \nohat{H}_{-1}^{'(2)}\,,
\eea

\bea
\flamoi {\cal L}^{(2)}[\nohat{\cal E}^{(2)}]_\ell^m(k)&=&\nohat{\cal
  E}_\ell^{'m(2)}(k)\\*
\flamoi &&+k\left[\frac{\kPlmsn{\ell+1}{m}{2}{0}}{2\ell+3}\nohat{\cal
  E}_{\ell+1}^{m(2)}(k)+\frac{2 m}{\ell(\ell+1)}\nohat{\cal
  B}_{\ell}^{m(2)}(k)-\frac{\kMlmsn{\ell}{m}{2}{0}}{2\ell-1}\nohat{\cal
  E}_{\ell-1}^{m(2)}(k) \right]\,,\nonumber
\eea

\bea
\flamoi {\cal L}^{(2)}[\nohat{\cal B}^{(2)}]_\ell^m(k)&=&\nohat{\cal
  B}_\ell^{'m(2)}(k)\\*
\flamoi &&+k\left[\frac{\kPlmsn{\ell+1}{m}{2}{0}}{2\ell+3}\nohat{\cal
  B}_{\ell+1}^{m(2)}(k)-\frac{2 m}{\ell(\ell+1)}\nohat{\cal
  E}_{\ell}^{m(2)}(k)-\frac{\kMlmsn{\ell}{m}{2}{0}}{2\ell-1}\nohat{\cal
  B}_{\ell-1}^{m(2)}(k) \right]\,.\nonumber
\eea
As for the quadratic terms, we obtain
\bea\label{EqLI11}
\flamoi {\cal L}^{(1)(1)}[\nohat{\cal I}]_\ell^m(k)&=&{\cal L}_{\rm time-delay}^{(1)(1)}[\nohat{\cal I}]_\ell^m(k)+{\cal L}_{\rm lensing}^{(1)(1)}[\nohat{\cal I}]_\ell^m(k)\nonumber\\*
\flamoi&&+2{\mathcal K}\left\{-\sum_{n=-1}^1\frac{\kPlmsn{\ell+1}{m}{0}{n}}{2\ell+3}\nohat{\cal
    I}_{\ell+1}^{m+n}(\gr{k}_2) 4 k_1^{(-n)}\nohat{\Phi}(k_1)\right.\nonumber\\*
\flamoi &&+\sum_{n=-1}^1
\frac{\kMlmsn{\ell}{m}{0}{n}}{2\ell-1}\nohat{\cal
  I}_{\ell-1}^{m-n}(\gr{k}_2) 4 k_1^{(n)}\nohat{\Phi}(k_1)- 4 \nohat{\Psi}'(k_1)\nohat{\cal I}_\ell^m(\gr{k}_2)\nonumber\\*
\flamoi && \left. +4 \delta_{\ell}^1\left[\nohat{\Psi}(k_1)-\nohat{\Phi}(k_1)\right]\nohat{\Phi}(k_2)k_2^{(m)} -8 \delta_{\ell}^0  \nohat{\Psi}(k_1)\nohat{\Psi}'(k_2)\right\}\,,
\eea
with
\bea
\flamoi {\cal L}_{\rm time-delay}^{(1)(1)}[\nohat{\cal I}]_\ell^m(k)&\equiv& -2{\mathcal K}\sum_{n=-1}^1\frac{\kPlmsn{\ell+1}{m}{0}{n}}{2\ell+3}\nohat{\cal
    I}_{\ell+1}^{m+n}(\gr{k}_2)k_2^{(-n)} \left[\nohat{\Phi}(k_1)+\nohat{\Psi}(k_1)\right]\nonumber\\*
\flamoi&&+2{\mathcal K}\sum_{n=-1}^1
\frac{\kMlmsn{\ell}{m}{0}{n}}{2\ell-1}\nohat{\cal
  I}_{\ell-1}^{m-n}(\gr{k}_2)k_2^{(n)}\left[\nohat{\Phi}(k_1)+\nohat{\Psi}(k_1)\right]\,,
\eea
\bea
\flamoi {\cal L}_{\rm lensing}^{(1)(1)}[\nohat{\cal I}]_\ell^m(k)&\equiv& 2{\mathcal K}\sum_{n=-1}^1\frac{\kPlmsn{\ell+1}{m}{0}{n}}{2\ell+3}\nohat{\cal I}_{\ell+1}^{m+n}(\gr{k}_2) (\ell+2)k_1^{(-n)}\left[\nohat{\Phi}(k_1)+\nohat{\Psi}(k_1)\right]\nonumber\\*
\flamoi&&+2{\mathcal K}\sum_{n=-1}^1 \frac{\kMlmsn{\ell}{m}{0}{n}}{2\ell-1}\nohat{\cal I}_{\ell-1}^{m-n}(\gr{k}_2)(\ell-1)k_1^{(n)}\left[\nohat{\Phi}(k_1)+\nohat{\Psi}(k_1)\right]\,.
\eea
For the electric-type polarization, we can also separate explicitely the lensing and time-delay contributions
\bea
\flamoi{\cal L}^{(1)(1)}[\nohat{\cal E}]_\ell^m(k)&=&{\cal L}_{\rm time-delay}^{(1)(1)}[\nohat{\cal E}]_\ell^m(k)+{\cal L}_{\rm lensing}^{(1)(1)}[\nohat{\cal E}]_\ell^m(k)\nonumber\\*
\flamoi&&+2{\mathcal K}\left\{-\sum_{n=-1}^1\frac{\kPlmsn{\ell+1}{m}{2}{n}}{2\ell+3}\nohat{\cal
    E}_{\ell+1}^{m+n}(\gr{k}_2) 4 k_1^{(-n)}\nohat{\Phi}(k_1)\right.\nonumber\\*
\flamoi &&\left.+\sum_{n=-1}^1
\frac{\kMlmsn{\ell}{m}{2}{n}}{2\ell-1}\nohat{\cal
  E}_{\ell-1}^{m-n}(\gr{k}_2) 4 k_1^{(n)}\nohat{\Phi}(k_1)-4 \nohat{\Psi}'(k_1)\nohat{\cal E}_\ell^m(\gr{k}_2)\right\}\,,
\eea
with
\bea
\flamoi {\cal L}_{\rm time-delay}^{(1)(1)}[\nohat{\cal E}]_\ell^m(k)&\equiv& -2{\mathcal K}\sum_{n=-1}^1\frac{\kPlmsn{\ell+1}{m}{2}{n}}{2\ell+3}\nohat{\cal
    E}_{\ell+1}^{m+n}(\gr{k}_2)k_2^{(-n)} \left[\nohat{\Phi}(k_1)+\nohat{\Psi}(k_1)\right]\nonumber\\*
\flamoi&&+2{\mathcal K}\sum_{n=-1}^1
\frac{\kMlmsn{\ell}{m}{2}{n}}{2\ell-1}\nohat{\cal
  E}_{\ell-1}^{m-n}(\gr{k}_2)k_2^{(n)}\left[\nohat{\Phi}(k_1)+\nohat{\Psi}(k_1)\right]\,,
\eea
\bea
\flamoi {\cal L}_{\rm lensing}^{(1)(1)}[\nohat{\cal E}]_\ell^m(k)&\equiv& 2{\mathcal K}\sum_{n=-1}^1\frac{\kPlmsn{\ell+1}{m}{2}{n}}{2\ell+3}\nohat{\cal E}_{\ell+1}^{m+n}(\gr{k}_2) (\ell+2)k_1^{(-n)}\left[\nohat{\Phi}(k_1)+\nohat{\Psi}(k_1)\right]\nonumber\\*
\flamoi&&+2{\mathcal K}\sum_{n=-1}^1
\frac{\kMlmsn{\ell}{m}{2}{n}}{2\ell-1}\nohat{\cal
  E}_{\ell-1}^{m-n}(\gr{k}_2)(\ell-1)k_1^{(n)}\left[\nohat{\Phi}(k_1)+\nohat{\Psi}(k_1)\right]\,.
\eea
For the magnetic-type polarization, we obtain (note that we have corrected for mistakes in Ref.~\cite{Pitrou2008} which were pointed out in Ref.~\cite{Fidler})
\bea
\flamoi {\cal L}^{(1)(1)}[\nohat{\cal B}]_\ell^m(k)&=&{\cal L}_{\rm time-delay}^{(1)(1)}[\nohat{\cal B}]_\ell^m(k)+{\cal L}_{\rm lensing}^{(1)(1)}[\nohat{\cal B}]_\ell^m(k)\nonumber\\
&&+{\mathcal K}\frac{2}{(\ell+1)}\sum_{n=-1}^1
  \llmn{\ell}{m}{n}\nohat{\cal E}_{\ell}^{m-n}(\gr{k}_2) k_1^{(n)} 8\nohat{\Phi}(k_1)
\eea
\be
\flamoi {\cal L}_{\rm time-delay}^{(1)(1)}[\nohat{\cal B}]_\ell^m(k)\equiv 2{\mathcal K}\frac{2}{(\ell+1)}\sum_{n=-1}^1
  \llmn{\ell}{m}{n}\nohat{\cal E}_{\ell}^{m-n}(\gr{k}_2)k_2^{(n)}\left[\nohat{\Phi}(k_1)+\nohat{\Psi}(k_1)\right] 
\ee
\be
\flamoi {\cal L}_{\rm lensing}^{(1)(1)}[\nohat{\cal B}]_\ell^m(k)\equiv -2{\mathcal K}\frac{2}{(\ell+1)}\sum_{n=-1}^1
  \llmn{\ell}{m}{n}\nohat{\cal E}_{\ell}^{m-n}(\gr{k}_2)k_1^{(n)}\left[\nohat{\Phi}(k_1)+\nohat{\Psi}(k_1)\right] 
\ee
Note that the lensing terms clearly differs from Eq.~(44) of Ref.~\cite{Nitta2009}, and this affects the results presented 
in Eq.~(52) of Ref.~\cite{Nitta2009} (and subsequently this error spreads as well into Eq.~(450) of  Ref.~\cite{Bartolo2010revue}).
 Indeed the harmonic expansion  of the lensing term~(\ref{lensingterm}) should be treated as in section 
 IV.A of Ref.~\cite{Hu2000}. There, the screen projector $S^\iT_\jT=\delta^\iT_\jT-n^\iT
n_\jT$ which is implied in our notation $\nabla^{n^\iT}$ (see Ref.~\cite{Pitrou2008} for details) is not split 
into $\delta^\iT_\jT$ and $n^\iT n_\jT$ before performing the harmonic decomposition 
as done in Ref.~\cite{Nitta2009}, but is instead used to ensure that the object manipulated 
in the lensing term remains tangent to a two-dimensional sphere whose radius is parameterized 
by $n^\iT$. This enables to work with spinned spherical harmonics as in Ref.~\cite{Hu2000} and to 
derive the result almost immediately. Additionally, the multipoles of the radiation brightness ($\Delta_\ell(\vk)$ in the notation 
of Ref.~\cite{Nitta2009}) are not rotated correctly when the azimuthal direction, with respect to which the spherical 
harmonics are defined, is shifted from the axis $\vk_1$ or $\vk_2$ of the first order calculation 
to the axis $\vk$ of the second order calculation. Indeed, this change of reference axis leads to a transformation 
given by Eq.~(\ref{Rotatemode}). Though this is taken into account correctly for the baryons 
velocity (for instance in the eleventh line of Eq.~(44) of Ref.~\cite{Nitta2009}), it is clearly forgotten 
for the brightness moments in the last term of the fourth line or in the last line. 
These two reasons explain why our treatment of quadratic terms is substantially different from Ref.~\cite{Nitta2009}.

Using a similar method applied to the collision term leads to
\be\label{C2a}
\flamoi{\cal C}^{(2)}[\nohat{\cal I}]_\ell^m(k)=\bar{\tau'}\left[-\nohat{\cal
  I}_\ell^{m(2)}(k)+\delta_\ell^0 \delta_m^0 \nohat{\cal
  I}_0^{0(2)}(k)+4 \delta_\ell^1 \nohat{v}_m^{(2)}(k)+\delta_\ell^2 \nohat P^{m(2)}(k) \right]\,,
\ee
where $\nohat P^{m(2)}(k)$ is non-vanishing only if $-2 \le m\le 2 $ and is defined in
that case by
\be
\nohat P^{m(2)}(k)=\frac{1}{10}\left[\nohat{\cal
      I}_2^{m(2)}(k)-\sqrt{6}\nohat{\cal E}_2^{m(2)}(k)\right]\,.
\ee
For the electric and magnetic type collision terms, their second order linear components read
\be
{\cal C}^{(2)}[\nohat{\cal E}]_\ell^m(k)=\bar{\tau'}\left[-\nohat{\cal
    E}_\ell^{m(2)}(k)-\delta_\ell^2 \sqrt{6}\nohat P^{m(2)}(k)\right]\,,
\ee
\be
{\cal C}^{(2)}[\nohat{\cal B}]_\ell^m(k)=-\bar{\tau'}\nohat{\cal B}_\ell^{m(2)}(k)\,.
\ee
The quadratic terms are then given by
\bea
\flamoi && {\cal C}^{(1)(1)}[\nohat{\cal I}^{(1)}]_\ell^m(k)=\\*
\flamoi && 2 \bar{\tau'}{\cal  K}\left\{-\sum_{n=-1}^1\frac{\kPlmsn{\ell+1}{m}{0}{n}}{2\ell+3}\nohat{\cal I}_{\ell+1}^{m+n}(\gr{k}_2)\nohat{v}_{-n}(\gr{k}_1)+\sum_{n=-1}^1
\frac{\kMlmsn{\ell}{m}{0}{n}}{2\ell-1}\nohat{\cal
    I}_{\ell-1}^{m-n}(\gr{k}_2)\nohat{v}_n(\gr{k}_1)\right.\nonumber\\*
\flamoi &&\qquad\qquad+\delta_{\ell}^0\left[-\frac{4}{3}\sum_{n=-1}^{1}(-1)^n\nohat{v}_n(\gr{k}_1)\nohat{v}_{-n}(\gr{k}_2)+\sum_{n=-1}^12 \frac{\kPlmsn{1}{0}{0}{n}}{3} \,\nohat{\cal I}_{1}^{n}(\gr{k}_2)\nohat{v}_{-n}(\gr{k}_1)\right]\nonumber\\*
\flamoi &&\qquad\qquad+\delta_{\ell}^13\sum_{n=-1}^1
\kMlmsn{1}{m}{0}{n}\nohat{\cal I}_{0}^{m-n}(\gr{k}_2)\nohat{v}_n(\gr{k}_1)\nonumber\\*
\flamoi &&\qquad\qquad+\delta_{\ell}^2\sum_{n=-1}^1
\frac{\kMlmsn{2}{m}{0}{n}}{3}\left[-\frac{1}{2}\nohat{\cal
    I}_{1}^{m-n}(\gr{k}_2)+7 \nohat{v}_{m-n}(\gr{k}_2)\right]\nohat{v}_n(\gr{k}_1)\nonumber\\*
\flamoi &&\qquad\qquad\left.+\delta_{\ell}^3\frac{1}{2}\sum_{n=-1}^1\frac{\kMlmsn{3}{m}{0}{n}}{5}\left[\nohat{\cal
      I}_{2}^{m-n}(\gr{k}_2)-\sqrt{6}\nohat{\cal
      E}_{2}^{m-n}(\gr{k}_2)\right]\nohat{v}_n(\gr{k}_1) \right\}\nonumber\\*
\flamoi &&+2 {\cal K}\left\{\left[\nohat{\tau'}^{(1)}(\gr{k}_1)+\Phi^{(1)}(\gr{k}_1)\right]{\cal C}^{(1)}[\nohat{\cal I}^{(1)}]_\ell^m(\gr{k}_2)\right\}\,,\nonumber
\eea

\bea
\flamoi&&{\cal C}^{(1)(1)}[\nohat{\cal E}^{(1)}]_\ell^m(k)=\\
\flamoi && 2 \bar{\tau'}{\cal
  K}\left\{-\sum_{n=-1}^1\frac{\kPlmsn{\ell+1}{m}{2}{n}}{2\ell+3}\nohat{\cal
      E}_{\ell+1}^{m+n}(\gr{k}_2)\nohat{v}_{-n}(\gr{k}_1)+\sum_{n=-1}^1
\frac{\kMlmsn{\ell}{m}{2}{n}}{2\ell-1}\nohat{\cal
    E}_{\ell-1}^{m-n}(\gr{k}_2)\nohat{v}_n(\gr{k}_1)\right.\nonumber\\*
\flamoi &&\qquad \qquad +\delta_{\ell}^2\sum_{n=-1}^1
\frac{\kMlmsn{2}{m}{0}{n}}{3}\left[\frac{\sqrt{6}}{2}\nohat{\cal
    I}_{1}^{m-n}(\gr{k}_2)-\sqrt{6} \nohat{v}_{m-n}(\gr{k}_2)\right]\nohat{v}_n(\gr{k}_1)\nonumber\\*
\flamoi &&\qquad\qquad \left.+\delta_{\ell}^3\frac{1}{2}\sum_{n=-1}^1\frac{\kMlmsn{3}{m}{2}{n}}{5}\left[-\sqrt{6}\nohat{\cal I}_{2}^{m-n}(\gr{k}_2)+6
    \nohat{\cal E}_{2}^{m-n}(\gr{k}_2)\right]\nohat{v}_n(\gr{k}_1)\right\}\nonumber\\*
\flamoi &&+2 {\cal K}\left\{\left[\nohat{\tau'}^{(1)}(\gr{k}_1)+\Phi^{(1)}(\gr{k}_1)\right] {\cal C}^{(1)}[\nohat{\cal E}^{(1)}]_\ell^m(\gr{k}_2)\right\}\,,\nonumber
\eea
\bea\label{CollisionNMB11}
\flamoi{\cal C}^{(1)(1)}[\nohat{\cal B}^{(1)}]_\ell^m(k)&=& 2 \bar{\tau'}{\cal K}\left\{\frac{-2 }{(\ell+1)}\sum_{n=-1}^1
  \llmn{\ell}{m}{n} \nohat{v}_n(\gr{k}_1) \nohat{\cal E}_{\ell}^{m-n}(\gr{k}_2)\right.\\*
\flamoi&&\qquad\left.-\delta_{\ell}^2\sum_{n=-1}^1 \llmn{2}{m}{n} \nohat{v}_n(\gr{k}_1) \left[\frac{4}{5}\nohat{\cal E}_{2}^{m-n}(\gr{k}_2)-\frac{2}{15}\sqrt{6}\nohat{\cal I}_{2}^{m-n}(\gr{k}_2)\right] \right\}\,.\nonumber
\eea

\subsection{Fluid equations}

For baryons and cold dark matter we only need the two first moments, that is the continuity and Euler equations, since there is no anisotropic stress. The continuity equations are
\be
\flamoi \left[\frac{\nohat \rho^{(1)}(k)}{\bar \rho}\right]'+k
\nohat{v}_{0}^{(1)}(k)-3 \nohat \Psi^{'(1)}(k)=0\,,\label{ContinuiteBaryons1}
\ee

\bea\label{ContinuiteBaryons2}
\flamoi \left[\frac{\nohat \rho^{(2)}(k)}{\bar \rho}\right]'+k
\nohat{v}_{0}^{(2)}(k)-3 \nohat \Psi^{'(2)}(k)\nonumber\\*
\flamoi +2{\cal  K}\left\{-\sum_{n=-1}^1(-1)^n\left[\HH\nohat{v}_n(\gr{k}_1)\nohat{v}_{-n}(\gr{k}_2)+\frac{\nohat
      \rho(\gr{k}_1)}{\bar \rho}k_1^{(n)} \nohat{v}_{-n}(\gr{k}_2)+ 2 \nohat{v}_n(\gr{k}_1)\nohat{v}'_{-n}(\gr{k}_2)\right]\right.\nonumber\\*
\flamoi \qquad\quad+\left[\frac{\nohat \rho(k_1)}{\bar \rho}+\nohat \Phi(k_1)+\nohat
  \Psi(k_1)\right] k_2 \nohat v_0(k_2)-3 \nohat \Psi'(k_1)\frac{\nohat \rho(k_2)}{\bar
  \rho}-6\nohat \Psi(k_1)\nohat \Psi'(k_2)\nonumber\\*
\flamoi \qquad\quad\left.-2 \sum_{n=-1}^1(-1)^n \left[\nohat
    \Phi(k_2)-\nohat \Psi(k_2)\right]k_2^{(n)}\nohat
  v_{-n}(\gr{k}_1)\right\}\nonumber\\*
\flamoi =-\frac{2 \bar{\tau'}}{R}{\cal  K}\left\{\sum_{n=-1}^1(-1)^{n}\left[\frac{1}{4}\nohat{\cal
      I}_{1}^{n}(\gr{k}_2)\nohat{v}_{-n}(\gr{k}_1)-\nohat{v}_n(\gr{k}_1)\nohat{v}_{-n}(\gr{k}_2)\right]\right\}\,.
\eea

The Euler equation reads
\be\label{EulerBaryons1}
\flamoi \nohat v^{'(1)}_m(k)+\HH \nohat v^{(1)}_m(k)-\delta_m^0 k \nohat
\Phi^{(1)}(k)=-\frac{\bar{\tau'}}{3 R}\left[-\nohat{\cal
  I}_1^{m(1)}(k) + 4 \nohat{v}_m^{(1)}(k)\right]\,,
\ee

\bea\label{EulerNM2}
\flamoi&& \nohat v^{'(2)}_m(k)+\HH \nohat v^{(2)}_m(k)- \delta_m^0 k \nohat
\Phi^{(2)}(k)\\*
\flamoi&&+2{\cal K}\left\{\left[\frac{\nohat \rho(k_1)}{\bar \rho}-\nohat \Phi(k_1)
  \right] \left[\nohat v'_m(\gr{k}_2)+\HH\nohat v_m(\gr{k}_2) \right]+\left[\frac{\nohat \rho(k_1)}{\bar \rho}+\nohat \Psi(k_1)
  \right]k_2^{(m)}\nohat \Phi(k_2)\right.\nonumber\\*
\flamoi&&\left.+\nohat v_m(\gr{k}_2)\left[\left(\frac{\nohat \rho(k_1)}{\bar \rho}\right)'-4\nohat \Psi'(k_1)\right]+\delta_m^0 k\nohat \Phi(k_1)\nohat
  \Phi(k_2) + k \nohat v_m(\gr{k}_2)v_0(\gr{k}_1) 
\right\}\nonumber\\*
\flamoi&&=-\frac{\bar{\tau'}}{4 R}\left[-\nohat{\cal
  I}_1^{m(2)}(k) + 4 \nohat{v}_m^{(2)}(k)\right]-\frac{1}{2 R}{\cal K}\left\{\nohat{\tau'}^{(1)}(\gr{k}_1)\left[-\nohat{\cal  I}_1^{m(1)}(\gr{k_2}) + 4 \nohat{v}_m^{(1)}(\gr{k}_2)\right]\right\}\nonumber\\*
\flamoi&&\quad -\frac{1}{2 R} \bar{\tau'} {\cal K}\left\{4 \nohat{\cal
      I}_{0}^{0}(\gr{k}_2)\nohat{v}_{m}(\gr{k}_1)-\sum_{n=-1}^1\frac{\kPlmsn{2}{m}{0}{n}}{5}\nohat{\cal
      I}_{2}^{m+n}(\gr{k}_2)\nohat{v}_{-n}(\gr{k}_1)\right\}\,.\nonumber
\eea
In the case of cold dark matter, the right hand side of all these equations vanishes since it is collisionless.

\subsection{Perturbed recombination}

\begin{figure}[htb]
\center
\includegraphics[width=8cm]{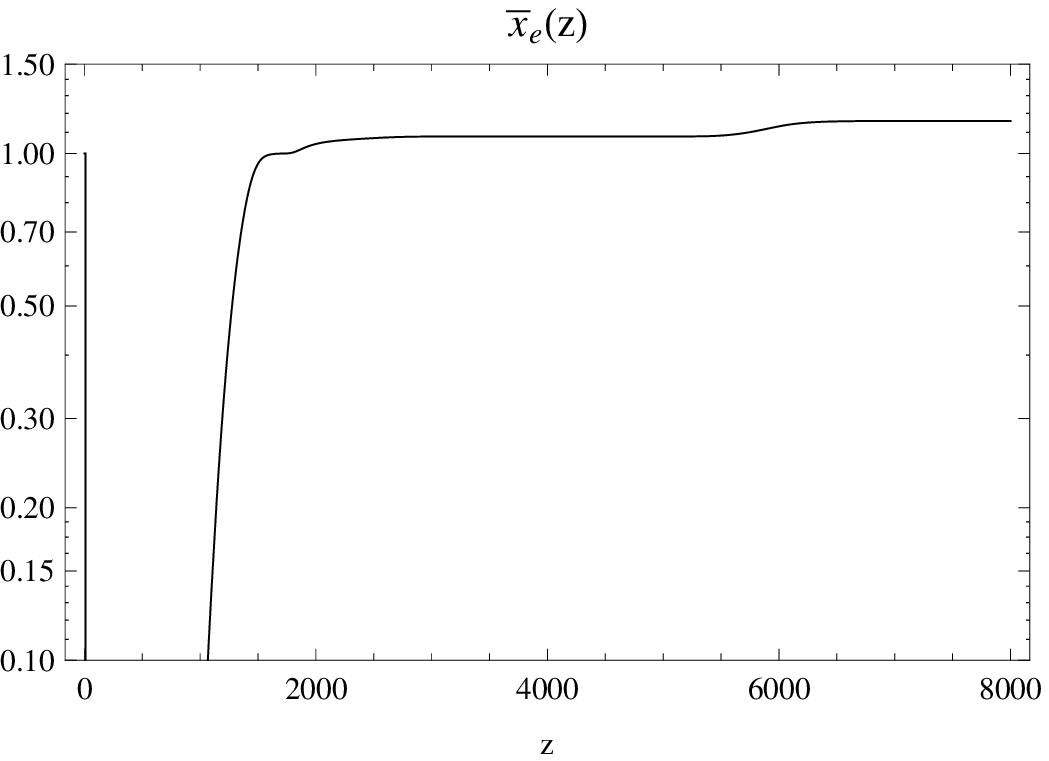}
\caption{Evolution of the free electrons fraction as a function of the redshift. We can see the transition of HeII and HeI at approximately z=6000 and z=2500 respectively, and then the recombination of Hydrogen around z=1000.}
\label{Figxe}
\end{figure}

The fraction of free electron at the background level is obtained from a Boltzmann equation where the collision term encodes all the relevant processes in the recombination. We shall not detail any of it here and it can be found in 
Refs.~\cite{Senatore2008a,Ma1995,Peebles1970,Lewis2007} (see also the different method of Ref.~\cite{Novosyadlyj2006}).
Since the interaction rate is given by $\tau' \equiv a x_\ie n_{\rm e} \st$, where $n_{\ie}$ is the number density of electrons (free and non-free), $\st$ the Thomson cross-section, and $x_\ie$ the fraction of free electrons, its perturbation is given by
\be\label{EqDeftauprime1}
\tau'^{(1)}=\bar \tau'\left(\frac{n_\ie^{(1)}}{\bar n_\ie}+\frac{x_\ie^{(1)}}{\bar x_\ie}\right)\,.
\ee
Since electrons and baryons are tightly coupled through Compton interactions and collisions~\cite{Senatore2008a},
\be
\frac{n_\ie^{(1)}}{\bar n_\ie}=\frac{n_\ib^{(1)}}{\bar n_\ib}\,.
\ee
As for the fraction of free electrons, at the background level it obeys 
\be
\bar x_\ie' + 3 H \bar x_\ie = \bar Q\,,
\ee
where the expression of $Q$ can be found in Ref.~\cite{Senatore2008a} (more precisely what is called $Q$ in this reference is equivalent to $n_\ie Q$ in here). It is in general a function of $(x_\ie,n_\ie,T,H)$. $T$ is the temperature of baryons and photons since it is approximately the same around recombination~\cite{Senatore2008a}. $H$ is the local divergence of baryons ($H \equiv \nabla_\mu u_\ib^\mu /3$) which at the background is the Hubble factor (in cosmic time). $\bar Q$ is obtained by taking the background value of all its arguments, that is 
\be
\bar Q=Q (\bar x_\ie,\bar n_\ie,\bar T,\bar H)\,.
\ee
We report on Fig.~\ref{Figxe} the numerical results obtained for the background fraction of free electrons. This has to be compared to the results obtained with RECFAST~\cite{Seager1999a,Wong2007} since we have implemented the correction to the three-level atom approximation implemented in this code, that is using its fudge factor of $1.14$ for the last stages of the Hydrogen recombination and the fudge factor $0.86$ 
for the HeI recombination.
 
\begin{figure}[htb]
\center
\includegraphics[width=6cm]{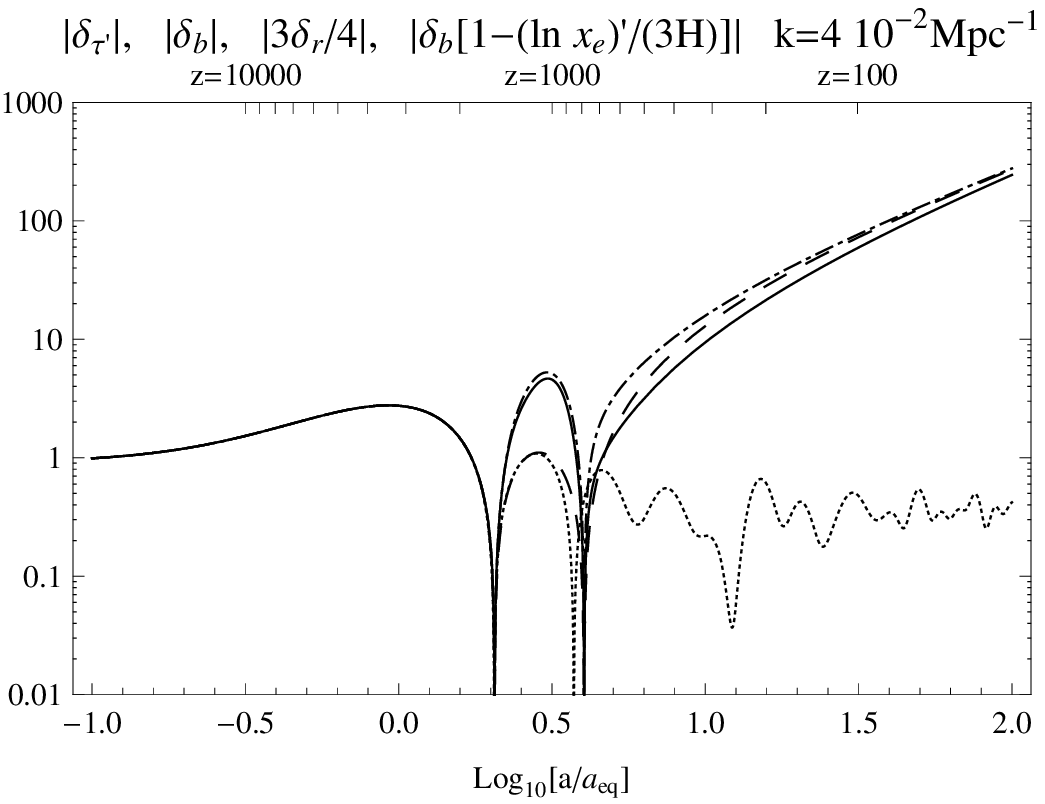}
\includegraphics[width=6cm]{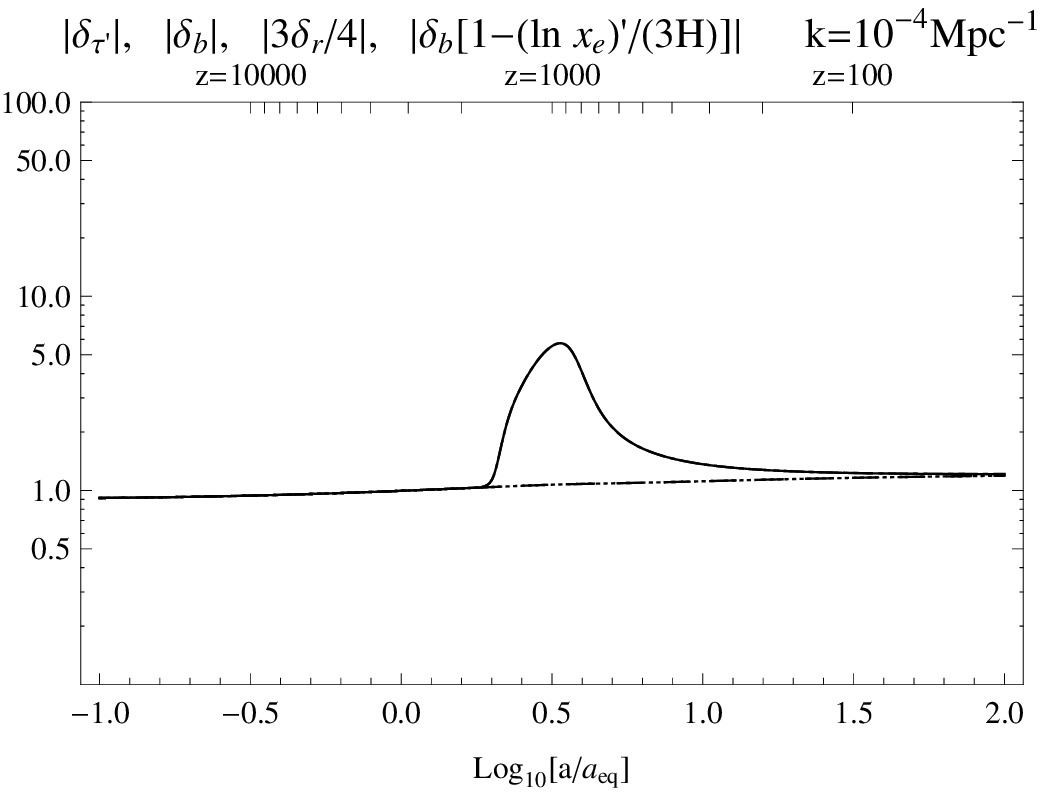}
\caption{Enhancement of the interaction rate due to the perturbations in the fraction of free electrons. 
$\delta_{\tau'}$,  $\delta_\ib$ and $\delta_\ir$ are respectively the relative perturbation of $\tau'$, $n_\ie$ and $\rho_\ir$.
We plot $\delta_{\tau'}$ in continuous line, and $\delta_\ib$ in dashed line. According to Eq.~(\ref{EqDeftauprime1}), the difference between the two is the enhancement due to perturbed recombination. This has to be compared to the large scale limit of the delayed recombination $\delta_\ib[1-1/3 (\ln x_\ie)'/\HH]$, which is plotted in dashed-dotted line.  We also plot in dotted line $3\delta_\ir/4$ to show the departure from tight coupling around recombination. We have considered the two cases $k=10^{-4} \,{\rm Mpc}^{-1}$ (right) and $k=0.04 \,{\rm Mpc}^{-1}$ (left). }
\label{fig2}
\end{figure}

At first order, the evolution of the fraction of free electrons reads~\cite{Senatore2008a} simply
\be
\left(\frac{x_\ie^{(1)}}{\bar x_\ie}\right)'=\Phi^{(1)} \bar Q + Q^{(1)}\,.
\ee
The perturbation $Q^{(1)}$ is obained by taking the first order of $Q(x_\ie,n_\ie,T,H)$ which is obtained by partial differentiation with respect to all the arguments of $Q$. The quantity $H^{(1)}$ is given by
\be
a H^{(1)}=-\HH \Phi^{(1)} -\frac{1}{3}\left(\frac{n_\ib^{(1)}}{\bar n_\ib}\right)'\,.
\ee
The enhancement due to the perturbations in the fraction of free electrons is plotted in Fig.~\ref{fig2}.

\section{Emitting sources}\label{AppEmittingSources}

In the line of sight approach, we need to define the emitting sources. Instead of working with the brightness ${\cal I}$ and its corresponding electric and magnetic type polarizations ${\cal E}$ and ${\cal B}$, we work with ${\Theta}$ (defined in Eq.~(\ref{defTbrightness})), $\tilde {\cal E} \equiv {\cal E}/(4\bar{\cal I})$ and $\tilde {\cal B} \equiv {\cal B}/(4\bar{\cal I})$.
They are expanded similarly to Eq.~(\ref{DecL}).
We obtain at first order
\begin{equation}\label{Slm1_expression1}
S_{\Theta,\ell}^{0(1)}(k)\equiv\delta_\ell^0 \Psi'(k)+\delta_\ell^1 k \nohat\Phi(k)+\frac{1}{4}\left[{\cal C}^{(1)}[\nohat{\cal I}]_\ell^0(k)+\bar{\tau'} \nohat{\cal I}_\ell^{0(1)}(k)\right]\,,
\end{equation}
\begin{equation}
S_{\tilde {\cal E},\ell}^{0(1)}(k)\equiv\frac{1}{4}{\cal C}^{(1)}[\nohat{\cal E}]_\ell^0(k)\,.
\end{equation}
However, if we want to have all effects but the late ISW localized on the last scattering surface, we need to integrate by parts in the line of sight integral solution. This boils down to using instead
\begin{equation}\label{Slm1_expression2}
\flamoi S_{\Theta,\ell}^{0(1)}(k)\equiv\delta_\ell^0 \left[\Psi'(k)+\Phi'(k)\right]+\delta_\ell^0 \bar \tau' \Phi(k)\left[{\cal C}^{(1)}[\nohat{\cal I}]_\ell^0(k)+\bar{\tau'} \nohat{\cal I}_\ell^{0(1)}(k)\right]\,.
\end{equation}

At second order the sources are
\begin{eqnarray}
S_{\Theta,\ell}^{m(2)}(k)&\equiv&\delta_\ell^0\delta_m^0 \left[\Phi'^{(2)}+\Psi'^{(2)}(k)\right]+\delta_\ell^0\delta_m^0 \bar \tau' \nohat\Phi^{(2)}(k) \\*
&&-\delta_\ell^2\delta_m^1\frac{k}{\sqrt{3}}\nohat{\Phi}_1^{(2)}(k)-\delta_\ell^2\delta_m^{-1}\frac{k}{\sqrt{3}}\nohat{\Phi}_{-1}^{(2)}(k)-\delta_\ell^2\delta_m^2  \nohat{H}_1^{'(2)}\nonumber\\*
&&-\delta_\ell^2\delta_m^{-2} \nohat{H}_{-1}^{'(2)}+\frac{1}{4}\left[{\cal C}^{(2)}[\nohat{\cal I}]_\ell^m(k)+\bar \tau' {\cal I}_\ell^{m(2)}(k) \right]\nonumber\,,
\end{eqnarray}
\begin{equation}
S_{\tilde {\cal E},\ell}^{m(2)}(k)\equiv \frac{1}{4}\left[{\cal C}^{(2)}[\nohat{\cal E}]_\ell^m(k)+\bar \tau' {\cal E}_\ell^{m(2)}(k) \right]
\end{equation}
\begin{equation}
S_{\tilde {\cal B},\ell}^{m(2)}(k)\equiv\frac{1}{4}\left[{\cal C}^{(2)}[\nohat{\cal B}]_\ell^m(k)+\bar \tau' {\cal B}_\ell^{m(2)}(k) \right]\,,
\end{equation}

\begin{equation}
S_{\Theta,\ell}^{m(1)(1)}(k)\equiv\frac{1}{4}{\cal C}^{(1)(1)}[\nohat{\cal I}]_\ell^m(k)-\frac{1}{4}{\cal L}^{(1)(1)}[\nohat{\cal I}]_\ell^m(k)\,,
\end{equation}
\begin{equation}
S_{\tilde {\cal E},\ell}^{m(1)(1)}(k)\equiv\frac{1}{4}{\cal C}^{(1)(1)}[\nohat{\cal E}]_\ell^m(k)-\frac{1}{4}{\cal L}^{(1)(1)}[\nohat{\cal E}]_\ell^m(k)\,,
\end{equation}
\begin{equation}
S_{\tilde {\cal B},\ell}^{m(1)(1)}(k)\equiv\frac{1}{4}{\cal C}^{(1)(1)}[\nohat{\cal B}]_\ell^m(k)-\frac{1}{4}{\cal L}^{(1)(1)}[\nohat{\cal B}]_\ell^m(k)\,.
\end{equation}

\section{Quadrupole initial conditions} \label{App_quadrupoleconditions}

Since $\HH \sim 1/\eta$ in the radiation era, we need to determine the initial conditions for ${\cal N}_2^0$ up to terms in $(k \eta_\init)^2$, and thus to know the initial conditions for ${\cal N}_1^m$ up to terms of order $k \eta_\init$. In fact, deep in the radiation dominated era, ${\cal N}_\ell^{0(1)} \sim (k \eta)^\ell$.
\begin{equation}
{\cal I}_{1}^{0(1)}(k) = {\cal N}_{1}^{0(1)}(k) =\frac{2 k }{\HH} \Phi^{(1)}(k)\,,
\end{equation}
\begin{equation}
{\cal I}_{1}^{0(2)}(k) = {\cal N}_{1}^{0(2)}(k) =\frac{2 k }{\HH} \Phi^{(2)}(k)-{\cal K}\left\{\frac{8 k}{\HH}\Phi^{(1)}(\vk_1)\Phi^{(1)}(\vk_2)\right\}\,.
\end{equation}
Using this, we deduce from the Boltzmann equation that at first order~\cite{Ma1995}
\begin{equation}
{\cal N}_2^{0(1)}(k)=\frac{2 k^2}{3 \HH^2}\Phi^{(1)} \,.
\end{equation}
At second order we obtain the following initial condition
\begin{equation}
{\cal N}_2^{0(2)}(k)=\frac{2 k^2}{3 \HH^2}\Phi^{(2)}(k)+S_5
\end{equation}
with
\begin{eqnarray}
\flamoi S_5&=&-{\cal K}\left\{\frac{8 k^2}{3\HH^2}\Phi^{(1)}(\vk_1)\Phi^{(1)}(\vk_2) \right\}-\frac{1}{2 \HH} {\cal L}^{\hashamoi(1)(1)}[\nohat{\cal N}]_2^0(k)\,\\
\flamoi &=&-{\cal K}\left\{\frac{8 k^2}{3\HH^2}\Phi^{(1)}(\vk_1)\Phi^{(1)}(\vk_2) +\frac{2 k_1^2}{3
    \HH^2}\Phi^{(1)}(\vk_1)\Phi^{(1)}(\vk_2)\right.\nonumber\\
\flamoi &&\left.\qquad -\sum_{n=-1}^1
  \frac{\kMlmsn{2}{0}{0}{n}}{6 \HH}{\cal N}_1^{-n(1)}(\vk_2)\left[5 k_1^{(n)}\Phi^{(1)}(\vk_1)+(k_2^{(n)}+k_1^{(n)})\Psi^{(1)}(\vk_1) \right] \right\}\nonumber\,.
\end{eqnarray}
As for radiation, though the anisotropic stress is vanishing at initial time due to the high collision rate, the moment for $\ell=2$ is not vanishing since the bulk motion of radiation has a quadrupole~\cite{Pitrou2008,Pitrou2007,Bartolo2007}. We deduce that at initial time 
\begin{eqnarray}
{\cal I}_2^{0(2)}(k) &=& {\cal K} \left\{\frac{20}{3} \sum_{n=-1}^1 \kMlmsn{2}{0}{0}{n} \frac{{\cal I}_1^{n(1)}(\vk_1)}{4} \frac{{\cal I}_1^{-n(1)}(\vk_2)}{4}\right\}\nonumber\\
 &=& {\cal K} \left\{\frac{5}{3 \HH^2} \sum_{n=-1}^1 \kMlmsn{2}{0}{0}{n} \,k_1^{n} k_2^{-n} \Phi^{(1)}(k_1) \Phi^{(1)}(k_2)\right\}\,.
\end{eqnarray}

\section{Closure relations for vector and tensor modes} \label{App_closure}

We reproduce here the closure relations needed to integrate the
Boltzmann equation for the vector and tensor modes. They
are derived in Eqs.~(5.4.48) and~(5.4.49) of Ref.~\cite{Riazuelo} which is not easily available.
They correspond to the equations satisfied by the last multipoles in brightness and 
polarization kept in the truncated Boltzmann hierarchy.
At first order we need only to retain the scalar closure relation ($m=0$). However at second order we have 
to consider also the vector and tensor parts ($m=1,2$). Note that the presence of quadratic terms 
at second order implies that these closure relations are not necessarily accurate as they were derived for the linear Boltzmann equation. 
For the brightness part, the closure relation reads
\be
{{\cal I}_\ell^m}'=k\left[ \sqrt{\frac{\ell+|m|}{\ell-|m|}} \frac{(2 \ell+1)}{(2 \ell-1)}{\cal I}_{\ell-1}^m-\frac{\ell+1+|m|}{k \eta}{\cal I}_{\ell}^m\right]\,.
\ee
For the polarization part, the closure relation for ${\cal E}_\ell^m$ and ${\cal B}_\ell^m$ is deduced by taking the real and imaginary part of
\bea
\flamoi \left({\cal E}_\ell^m+\ii{\cal B}_\ell^m\right)'&=&k\left[ \sqrt{1-\frac{m^2}{\ell^2}}\sqrt{\frac{\ell+2}{\ell-2}} \frac{(2 \ell+1)}{(2 \ell-1)}\left({\cal E}_{\ell-1}^m+\ii{\cal B}_{\ell-1}^m\right)-\frac{\ell+3}{k \eta}\left({\cal E}_\ell^m+\ii{\cal B}_\ell^m\right)\right.\nonumber\\
\flamoi &&\quad \left.+\ii\frac{m}{\ell} \left({\cal E}_\ell^m+\ii{\cal B}_\ell^m\right) \right]\,.
\eea

\ifcqg
\section*{References}
\else
\fi
\ifcqg
\bibliographystyle{h-physrev}
\else
\bibliographystyle{apsrev}
\fi

\bibliography{comptonsanstitres}


\end{document}